# The Galactic Center Massive Black Hole and Nuclear Star Cluster

Reinhard Genzel[1], Frank Eisenhauer & Stefan Gillessen
*Max-Planck Institut für Extraterrestrische Physik, Garching, FRG*

The Galactic Center is an excellent laboratory for studying phenomena and physical processes that may be occurring in many other galactic nuclei. The Center of our Milky Way is by far the closest galactic nucleus, and observations with exquisite resolution and sensitivity cover 18 orders of magnitude in energy of electromagnetic radiation. Theoretical simulations have become increasingly more powerful in explaining these measurements. This review summarizes the recent progress in observational and theoretical work on the central parsec, with a strong emphasis on the current empirical evidence for a central massive black hole and on the processes in the surrounding dense nuclear star cluster. We present the current evidence, from the analysis of the orbits of more than two dozen stars and from the measurements of the size and motion of the central compact radio source, Sgr A*, that this radio source must be a massive black hole of about $4.4 \times 10^6$ $M_\odot$, beyond any reasonable doubt. We report what is known about the structure and evolution of the dense nuclear star cluster surrounding this black hole, including the astounding fact that stars have been forming in the vicinity of Sgr A* recently, apparently with a top-heavy stellar mass function. We discuss a dense concentration of fainter stars centered in the immediate vicinity of the massive black hole, three of which have orbital peri-bothroi of less than one light day. This 'S-star cluster' appears to consist mainly of young early-type stars, in contrast to the predicted properties of an equilibrium 'stellar cusp' around a black hole. This constitutes a remarkable and presently not fully understood 'paradox of youth'. We also summarize what is known about the emission properties of the accreting gas onto Sgr A* and how this emission is beginning to delineate the physical properties in the hot accretion zone around the event horizon.

---

[1] Also Department of Physics, University of California, Berkeley

# Table of Contents

# 1. Introduction

## 1.1 Massive black holes

Following the discovery of distant luminous quasars (QSOs) in the early 1960s (Schmidt 1963) considerations of energetics suggested that the enormous luminosities and energy densities of QSOs can be explained most plausibly by the conversion of gravitational energy into radiation when matter accretes onto massive black holes (Lynden-Bell 1969, Rees 1984). Lynden-Bell (1969) and Lynden-Bell & Rees (1971) proposed that most galactic nuclei, including the Center of the Milky Way, might harbor such massive black holes, which in most cases are much less active than in QSOs. An unambiguous proof of the existence of a massive black hole as defined by General Relativity requires the determination of the gravitational potential to the scale of the event horizon. This proof can in principle be obtained from spatially resolved measurements of the motions of test particles (interstellar gas or stars) in close orbit around the nucleus. In practice it is not possible (yet) to probe the scale of an event horizon of any black hole candidate (stellar as well as massive black holes) with spatially resolved dynamical measurements. A more modest goal then is to show that the gravitational potential of a galaxy nucleus is dominated by a compact non-stellar mass and that this central mass concentration cannot be anything but a black hole because all other conceivable configurations are more extended, are not stable, or produce more light. Even this test cannot be conducted yet in distant QSOs from dynamical measurements. It has become feasible in nearby galaxy nuclei, however, including the Galactic Center.

Solid evidence for central 'dark' (i.e. non-stellar) mass concentrations in about 50 nearby galaxies has emerged over the past two decades (e.g. Kormendy 2004, Gültekin et al. 2009) from optical/infrared imaging and spectroscopy on the Hubble Space Telescope (HST) and large ground-based telescopes, as well as from very long baseline radio interferometry (VLBI). The first really convincing case that such a dark mass concentration cannot just be a dense nuclear cluster of white dwarfs, neutron stars and perhaps stellar black holes emerged in the mid 1990s from spectacular VLBI observations of the nucleus of NGC 4258, a mildly active galaxy at a distance of 7 Mpc (Miyoshi et al. 1995). The VLBI observations showed that the galaxy nucleus contains a thin, slightly warped disk of $H_2O$ masers in Keplerian rotation around an unresolved mass of 40 million solar masses. The inferred density of this mass exceeds a few $10^9$ solar masses $pc^{-3}$ and thus cannot be a long-lived cluster of 'dark' astrophysical objects of the type mentioned above. As we will discuss in this review, a still more compelling case can be made in the case of the Galactic Center.

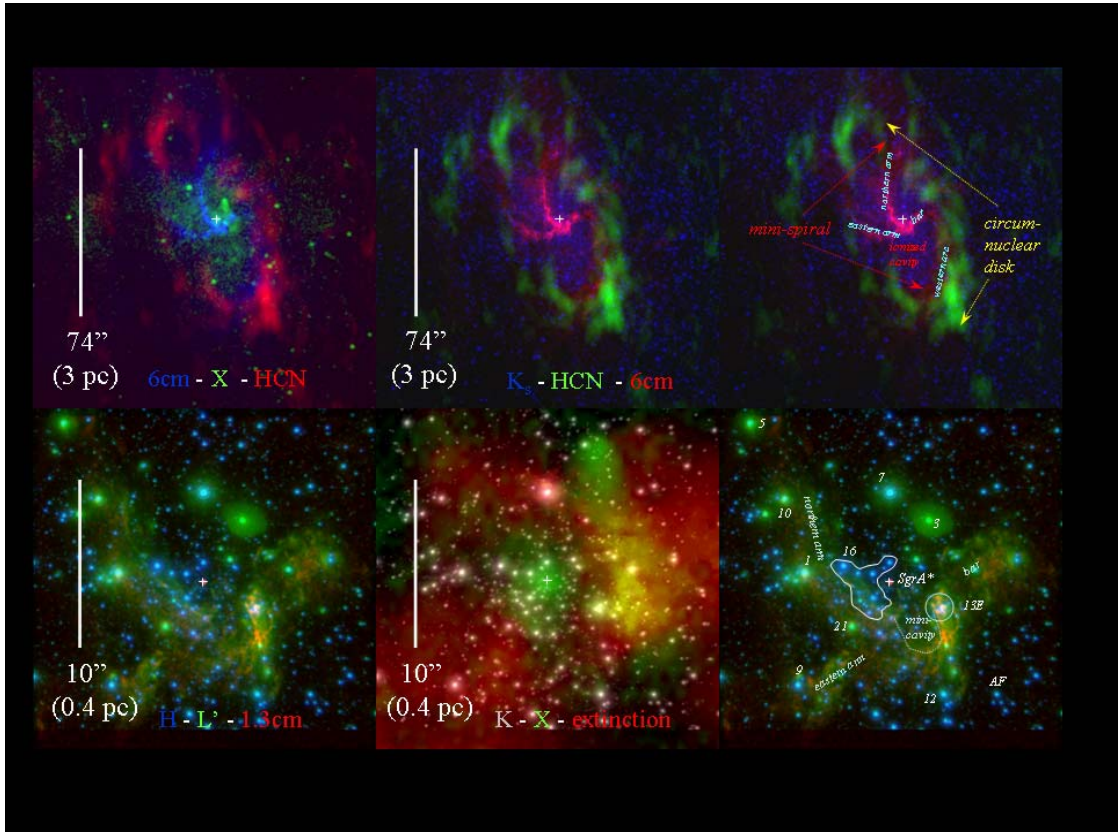

**Figure 1.1.** Multi-wavelength overview of the central few parsecs of the Milky Way. Top three panels: larger scale view. All three top images are on the same scale (at the distance of the Galactic Center (8.3 kpc) 1' = 2.41 pc). The cross marks the position of the compact radio source Sgr A*. Top left: Color composite of radio 6cm emission (blue: VLA, Yusef-Zadeh et al. 1986, Roberts & Goss 1993), HCN 1-0 emission (red: OVRO, Christopher et al. 2005) and X-ray emission (green: Chandra, Baganoff et al. 2003). The Galactic plane runs at position angle 32° southwest-northeast across the image. Top middle: 6 cm VLA radio image (pink), HCN emission (green) and 2.2 μm (K-band) image (blue, VLT-ISAAC, Schödel et al. 2007a). Top right: Copy of the central panel with the key interstellar features marked. Bottom: zoom into the central region. All three insets are on the same scale. Bottom left: Color composite near-IR adaptive optics image (blue: H-band (1.6 μm), green L'-band (3.8 μm), VLT-NACO, Genzel et al. 2003a), and the 1.3 cm VLA radio continuum (red, Zhao & Goss 1998). Bottom center: K-band (white, VLT-NACO, Genzel et al. 2003a), derived dust extinction (red-yellow, Schödel et al. 2007) and X-ray emission (green, Chandra, Baganoff et al. 2003). In addition to X-ray emission from Sgr A*, IRS 13E and a diffuse component, the elongated structure 10" north-west of Sgr A* is the pulsar-wind nebula 259.95−0.04, which may also be associated with the HESS TeV source J1745−290. Bottom right: Copy of left panel with the 'IRS' names of the stars, as well as some of the interstellar features marked. Here and in all other images of the Galactic Center region, North is up and East is to the left.

## 1.2 The Galactic Center Laboratory

The central few parsecs of our Milky Way contain a dense and luminous star cluster, as well as several components of neutral, ionized and extremely hot gas (Figure 1.1, Becklin & Neugebauer 1968, Rieke & Rieke 1988, Genzel & Townes 1987, Genzel, Hollenbach & Townes 1994, Mezger, Duschl & Zylka 1996). The central parsec diameter region is mostly ionized. It consists of the HII region Sgr A West and a concentration of ~ $10^6$ K hot, X-ray emitting gas (Baganoff et al. 2001, 2003; Muno et al. 2004). This low density, ionized 'central cavity' is pervaded by a set of orbiting ionized gas filaments (the 'mini-spiral', Lo & Claussen 1983), which in turn are surrounded by orbiting, dense molecular cloud streamers at $R \sim 1.5 - 4$ pc (the 'circum-nuclear disk', CND, Becklin et al. 1982, Güsten et al. 1987, Jackson et al. 1993, Christopher et al. 2005). Beyond its outer edge the CND is bordered by a young supernova remnant, Sgr A East, and surrounded by a number of dense molecular clouds on a scale of $5 - 100$ pc (Güsten & Downes 1980; Mezger, Duschl & Zylka 1996).

The stellar density in the nuclear cluster increases inward from a scale of tens of parsecs to within the central parsec (Becklin & Neugebauer 1968, Catchpole, Whitelock & Glass 1990). At its center is a very compact radio source, Sgr A* (Figure 1.1, Balick & Brown 1974, Lo et al. 1985, Yusef-Zadeh et al. 1986). Short-wavelength centimeter and millimeter VLBI observations have established that its intrinsic size is a mere $3 - 10$ light minutes (Krichbaum et al. 1993, Rogers et al. 1994, Doeleman et al. 2001, 2008, Bower et al. 2004, Shen et al. 2005). Sgr A* thus is the most likely candidate for the location of a possible central black hole. However, most of the time Sgr A* is disappointingly faint in all bands other than the radio to submillimeter region (Falcke et al. 1998). For this reason, many considered the case for a massive black hole in the Galactic Center initially fairly unconvincing (e.g. Rees 1982, Allen & Sanders 1986).

This conclusion changed when increasingly detailed dynamical measurements of the mass distribution became available. At the time of writing this review, there are precise determinations of the orbits of about thirty stars in the immediate vicinity of SgrA*, including one complete orbit. Several of these stars approach SgrA* to within tens of light hours, moving there with a speed of several $10^3$ km/s. Combined with evidence for little motion of SgrA* itself from VLBI data, these measurements now provide firm and compelling evidence that Sgr A* must be massive black hole of about four million solar masses.

Because of its proximity - the distance to the Galactic Center is about $10^5$ times closer than the nearest quasars - high-resolution observations of the Milky Way nucleus yield much more detail and specific information than possible in any other galaxy nucleus, to a linear scale comparable to the radius of the Earth's orbit around the Sun. As such, the Galactic Center is a unique laboratory for testing the massive black hole paradigm and for exploring the impact of a massive black hole on its stellar and interstellar environment. Since the nucleus is highly obscured by interstellar dust particles in the plane of the Galactic disk, however, observations in the optical waveband are not possible. Measurements need to be carried out at longer wavelengths, in the infrared and microwave bands, or at shorter wavelengths, at hard X-rays and γ-rays, where the veil of interstellar gas and dust is more transparent. The dramatic progress in our knowledge of the Galactic Center over the past two decades is a direct consequence of the development of novel facilities, instruments and techniques across the whole range of the electromagnetic spectrum (Figure 1.1).

There have been a number of review articles on the Galactic Center region. Oort's 1977 review covered many of the pioneering results: the first detailed radio and infrared maps, the discovery of Sgr A* and the properties of the ionized and molecular gas and stars on scales of 1 pc to 1 kpc. Ten years later Genzel & Townes (1987) emphasized the phenomena in the central few parsecs and discussed the emerging dynamical evidence for a central mass, mainly based at that time on observations of gas motions. Blitz et al. (1993) discussed the evidence that the central bulge of the Milky Way actually is bar-shaped and what that means for the gas dynamics in the central few hundred parsecs. Genzel, Hollenbach & Townes (1994) summarized the then newly discovered young massive stars in the central parsec, analyzed the luminosity production there and the importance of stellar winds, and gave an update of the measurements of the mass distribution, including the first stellar dynamics work. Morris & Serabyn (1996) focussed on interstellar phenomena, magnetic fields and star formation in the larger Galactic Center zone and discussed the evidence for a limit cycle of activity. Mezger, Duschl & Zylka (1996) reviewed how Sgr A*, the central ionized 'cavity' and the nuclear star cluster relate to the dense and massive molecular clouds in the central ten parsecs, including the CND and the far-infrared/submillimeter dust emission. Melia & Falcke (2001) emphasized the most important theoretical aspects of accretion onto black holes and the puzzling faint emission of Sgr A* at different wavelengths. Alexander (2005) and Merritt (2006) gave in-depth discussions of the rich physics of star clusters in the vicinity of massive black holes, with the Galactic Center nuclear star cluster and cusp as the most powerful testing ground for a number of predictions and explanations of the recent theoretical work. Genzel (2007) summarized the status of the proof of the black-hole paradigm with special reference to the instrumental progress across the electromagnetic spectrum. Finally Reid (2009) gave a concise summary of the evidence for the black hole paradigm in its most recent form, including the stellar orbit evidence and the beautiful VLBI work on Sgr A*. Two entire books are devoted to the Galactic Center story, a textbook written by A. Eckart, R. Schödel, & C. Straubmeier (2005) and a more popular account by F. Melia (2003).

# 2. The nuclear star cluster

Over the last one and a half decades diffraction limited near-infrared imaging and spectroscopy, especially with the advent of adaptive optics and integral field spectroscopy, have revolutionized the exploration of the nuclear star cluster. Compared to earlier work, the angular resolution has improved by more than one order of magnitude (to ~ 50 mas, corresponding to a linear scale of $2 \times 10^{-3}$ pc) and the sensitivity by 3 to 5 magnitudes to $K_s \sim 16$ and 18, for spectroscopy and imaging observations, respectively (Figures 1.1 & 2.1.1). The depth of the best current imaging observations is limited by crowding and confusion throughout the central ten arcseconds (Genzel et al. 2003a, Schödel et al. 2007a, Ghez et al. 2005a, 2008, Gillessen et al. 2009b, Fritz et al. 2010a). As a result of these improvements all supergiants, giants and Wolf-Rayet stars, plus the main-sequence to ~ A0 stars (birth mass ~ 3 $M_\odot$) can now be studied throughout the cluster (Figure 2.1.1.).

One of the surprises of early observations was the fairly large number of bright stars, a number of which were already apparent on the discovery infrared images of Becklin & Neugebauer (1975) and Becklin et al. (1978): IRS 7, 13, 16, see Figure 1.1. Infrared spectroscopy reveals that many of these bright stars are somewhat older, late-type (red) giants, supergiants (IRS 7) and asymptotic giant branch (AGB) stars, which are naturally expected in an old nuclear star cluster. Starting with the discovery of the 'Allen-Forrest (AF)'-star (Figure 1.1, Allen, Hyland & Hillier 1990, Forrest et al. 1987), however, an ever increasing number of these bright stars were found to be hot early-type stars (Krabbe et al. 1991, 1995, Tamblyn & Rieke 1993, Blum et al. 1995, Libonate et al. 1995, Tamblyn et al. 1996, Genzel et al. 1996, Paumard et al. 2001, 2003, 2006, Tanner et al. 2006). Stellar atmosphere modeling shows that these 'HeI-stars' are post main-sequence, blue supergiants and Wolf-Rayet (WR) stars with ages of 2 to 8 Myrs and zero age main-sequence (ZAMS) masses $30 - 100$ $M_\odot$ (Najarro et al. 1994, 1997, Martins et al. 2007). A particularly impressive concentration of three O/WR stars is found in the IRS 13E region (Figures 1.1 & 5.4.1), which is also a dense and dusty concentration in the HII region and an X-ray emission source (Figure 1.1 bottom right). A remarkable concentration of mainly B-stars, the so-called 'S-star cluster'[2] is found in the central arcsecond, centered on Sgr A$^*$ (Figure 2.4.1).

These findings are highly surprising, especially if there is a massive black hole at the center of the nuclear cluster. How did these young stars get there? Did they form in situ, or were the transported into the center from further out? The most recent count from deep SINFONI integral field spectroscopy yields 177 O/WR/B-stars with good spectral identification (Bartko et al. 2010), including the entire plethora of known types of luminous blue supergiants and Wolf-Rayet stars (WN, WC, Ofpe), as well as dwarf and giant, main-sequence O- and B-stars (Paumard et al. 2006, Bartko et al. 2009, 2010). The nuclear star cluster is one of the richest concentrations of young massive stars in the entire Milky Way.

---

[2] The naming of the 'S'-stars originated in Eckart & Genzel (1996) to denote those remarkably fast moving stars in the 'Sgr A*(IR)-cluster' that were known at that time. Since that time the number of S-stars has grown to over 200 (Ghez et al. 1998, Gillessen et al. 2009b); unfortunately, the MPE and UCLA groups have been using different nomenclature.

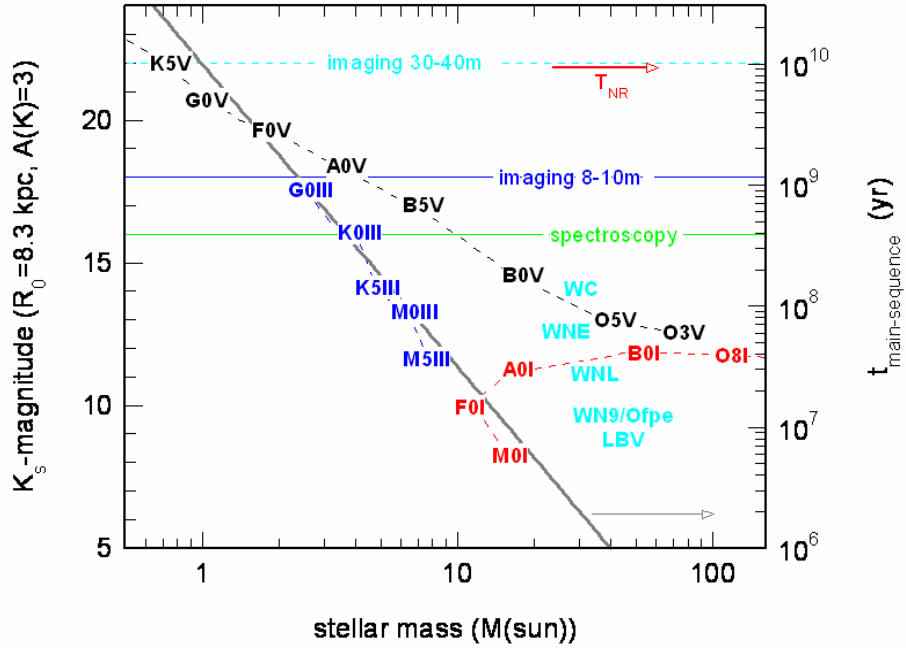

**Figure 2.1.1.** $K_s$-band magnitude as a function of stellar mass: black denotes main-sequence stars of luminosity class V (and mass is then the zero age main-sequence (ZAMS) mass), giants (III, blue), supergiants (I, red) and various Wolf-Rayet stars (cyan). The apparent $K_s$-band magnitudes are given for a $R_0 = 8.3$ kpc (Ghez et al. 2008, Gillessen et al. 2009b) and $K_s$-band extinction of $A_{Ks} = 3$ mag. The thick grey curve shows the main-sequence lifetime as a function of mass. The two-body relaxation time in the Galactic Center, $T_{NR} \sim 1-20 \times 10^{10}$ yr (Alexander 2005, Merritt 2010, Figure 2.3.3, red arrow), the current confusion limited (40 − 70% depending on location in cluster, Genzel et al. 2003a, Schödel, Merritt, Eckart 2009) imaging detection limit in the $K_s$-band with 8 − 10 m class telescopes in the central few arcseconds ($K_s \sim 18$), the adaptive optics spectroscopic detection limit ($K_s \sim 15.5 - 16$), as well as the confusion limit for imaging with a future 30 − 40 m-class telescope are marked by horizontal dotted lines.

## 2.1 The nuclear cluster of cool, old stars

96% of the currently observed stars in the central parsec (to $K_s \sim 18$, M(K) ~ +1) are old (> 1 Gyr) late-type giant stars and helium-burning stars on the horizontal branch/ 'red clump', which have low to intermediate masses ($m_\Box \sim 0.5 - 4$ M$_\odot$) and temperatures of ~ 3500 − 3700 K. In addition there are a few dozen very bright and extremely cool ($T \sim 2700$ K) asymptotic giant branch stars (AGB), which are in brief thermally pulsing phases accompanied by large mass loss (AGB-TP). Finally there are a few more massive, red supergiants stars in the central region (two within $R \leq 0.5$ pc, and 15 within 2.5 pc, Blum et al. 2003).

Observations of the 2D (3D) stellar motions (two proper motions on the sky and a radial velocity) are now available for ~ 6000 (660) of these stars (Trippe et al. 2008,

Schödel et al. 2009). These show that the motions and distribution are largely random and isotropic, with no detectable phase space fluctuations greater than a few percent (Trippe et al. 2008). The old stellar cluster exhibits slow, solid-body rotation in the sense of the rotation of the Galaxy and the entire Galactic Bulge, with an amplitude of ~ 1.4 km/s/arcsec (McGinn et al. 1989, Genzel et al. 1996, Trippe et al. 2008, Schödel et al. 2009).

## 2.2 The disk(s) of young massive stars

The radial velocities of the massive young stars are mostly blue-shifted north, and red-shifted south of Sgr A*, exactly opposite to the rotation of the Milky Way and the old stars in the nuclear star cluster (Genzel et al. 1996, 2000, Tanner et al. 2006). This pattern suggests a large-scale coherent velocity field, such as rotation. When good quality proper motions were becoming available in the late 1990s it became clear that most of the bright early-type stars in the central few arcseconds (especially in the IRS 16 complex) exhibit a clockwise motion pattern ($j_z > 0$)[3] on the sky, consistent with an inclined disk in Keplerian rotation around Sgr A* (Genzel et al. 2000). However, there clearly were also some bright stars in counter-clockwise motion ($j_z < 0$: e.g. IRS 16NE, IRS 16NW). A single rotating disk model thus cannot fit the motions of all stars known in 2000. By only considering the clockwise stars Levin and Beloborodov (2003) then were able to demonstrate that the 3D velocities of ten out of 13 stars in the 2000-sample are consistent with a ***thin, rotating disk***. By including more and better velocity data Genzel et al. (2003a), Tanner et al. (2006) and Paumard et al. (2006) proposed that there are two rotating star disks or rings, at very large angles with respect to each other, and not coinciding with any other known orientation in and around the Galactic Center (Table 2.2.1). The clockwise system is much better defined, however, and appears to be located inside most of the counter-clockwise stars. Paumard et al. (2006) were able to detect for the first time also 40 main-sequence and giant O-stars (through absorption lines) and thus substantially expanded the sample of massive stars. More than two thirds of these have $j_z > 0$.

---

[3] $j_z = (x\, v_y - y\, v_x) / (\{x^2 + y^2\} \{v_x^2 + v_y^2\})^{1/2}$ is the projected, normalized specific angular momentum of the motion on the sky. If $j_z > 0$ the motion is clockwise, and if $j_z < 0$ it is counter-clockwise. Orbits with tangential motion in 3D also have projected tangential motion on the sky ($|j_z| = 1$). Orbits with projected radial motion on the sky ($j_z = 0$) can be due to either true 3D radial orbits or tangential orbits, or a mixture of such orbits (Genzel et al. 2000).

**Table 2.2.1** Orientations of planar structures in the central parsec

| Name | inclination $i^a$ | angle of line of nodes on sky $\Omega^b$ | method/reference |
|---|---|---|---|
| rotation axis of Galaxy | 90° | 31.4° | |
| northern arm | 45° ± 10° | 15° ± 15° | Paumard et al. 2004 Zhao et al. 2009 |
| bar | 76° | 115° | Liszt 2003 |
| eastern arm | 58° ± 5° | 132° ± 11° | Zhao et al. 2009 |
| circum-nuclear molecular disk/western arc | 66° ± 6° | 25° ± 5° | Lacy, Achtermann & Selabyn 1991 Jackson et al. 1993 Zhao et al. 2009 |
| clockwise system (at inner edge) | 122° ± 7° | 99° ± 3° | Lu et al. 2009 Bartko et al. 2009 |
| counter-clockwise disk (at $p = 7''$) | 60° ± 15° | 243° ± 14° | Bartko et al. 2010 |

[a] angle between orbital and sky plane (clockwise: 0° − 90°, counter-clockwise: 90° − 180°): $\theta_n = 180° - i$
[b] angle between North and the line of ascending (= receding) nodes (= intersection of disk and sky plane), increasing east of north, $\phi_n = -\Omega$

## *2.2.1 The clockwise stellar disk*

Lu et al. (2006, 2009) achieved the next substantial step forward by deriving constraints for the individual orbits of the O/WR-stars from a Monte Carlo analysis, including the spatial coordinates of the stars, as well as the 3D space velocities and limits to (and in one case, a detection of) their accelerations. For this analysis they used much improved proper motion data in the central few arcseconds, as well as (for their 'extended sample') the Paumard et al. (2006) data set. Because of the lack of accelerations, this technique still does not yield individual orbits but gives much better defined probability distributions of the orbital parameters. The top graph in Figure 2.2.1 shows the overall angular momentum distribution projected on the sky derived by Lu et al. (2009) for all O/WR-stars in their extended sample. There is a highly significant over-density of the orbital angular momentum vector distributions toward the direction of the clockwise disk found in the previous studies.

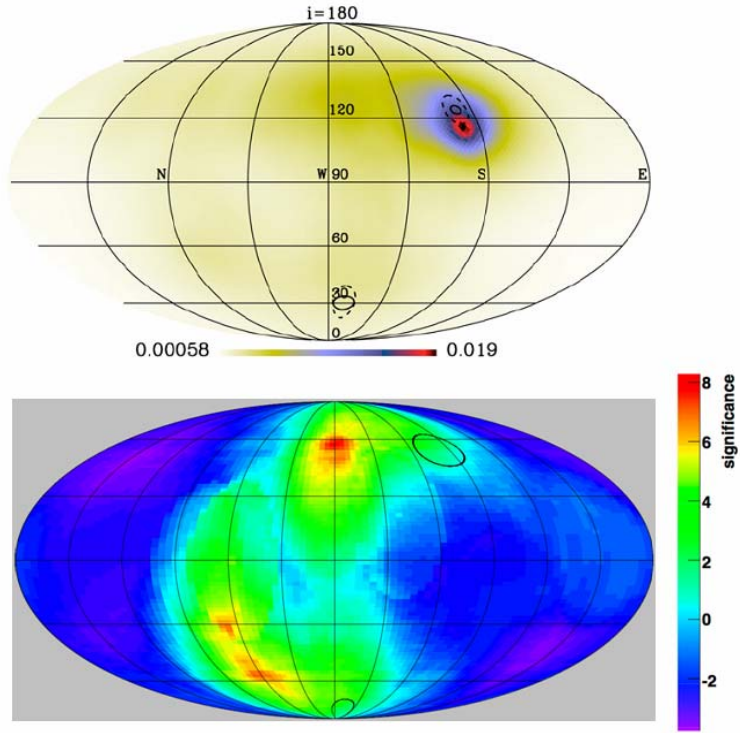

**Figure 2.2.1.** Distribution of orbital normal vectors on the sky of the O/WR-stars in the central cluster. Top: distribution of normal vectors for the extended sample of Lu et al. (2009), including 73 stars with projected radii from 0.8" to 12". Densities are indicated in colors (stars deg$^{-2}$) on a linear scale and the peak indicates an overdensity of stars with similar orbital planes. Overplotted in black are the candidate orbital planes as proposed by Levin & Beloborodov (2003) and Genzel et al. (2003a) with updated values from Paumard et al. (2006) for the candidate plane normal vector and uncertainties (solid black) and the disk thickness (dashed black) shown as solid angles of 0.05 sr and 0.09 sr for the clockwise and counterclockwise disks, respectively (adapted from Lu et al. 2009). Bottom: Sky projection of the distribution of significance in the sky (25° aperture) for 82 bona fide early-type stars ($K_s < 14$) with projected distances between 3.5" and 15". The disk positions of Paumard et al. (2006) are marked with full black circles. There are two extended excesses visible for clockwise and counter-clockwise stars, with maximum (pre-trial) significances of 8.2σ and 7.1σ, respectively. The upper black circle in the bottom panel corresponds to the peak in the top panel (adapted from Bartko et al. 2010).

    Bartko et al. (2009, 2010) carried out a similar Monte Carlo orbit construction analysis based on a still larger data set of about 150 robustly identified early-type stars with proper motions and radial velocities, including more than 30 B-dwarfs. For the clockwise stars their findings are in good agreement with Lu et al. and confirm the highly significant detection and average location of the clockwise system making up about 55 − 60% of all the O/WR-stars in the central parsec. Bartko et al. (2009) further find that the location of the normal vector of the clockwise disk changes with separation from Sgr A*, putting on firm ground an earlier trend already seen in the analysis of the velocity vectors only (Figure 2.2.2, see also Löckmann & Baumgardt 2009). The change in angle between the inner edge at ~ 1" and the outermost stars at > 10" corresponds to 64° ± 6°. The angular momentum direction changes in both angles on the sky. In the framework of a single disk structure the clockwise disk thus

appears to be highly warped. Alternatively, the clockwise system may consist of a system of streamers with different orientations at different radii. Taken together, the various studies discussed above provide a fairly consistent picture of the properties of the clockwise star disk.

The location of the O stars in the Hertzsprung-Russell diagram and the number ratios of different sub-types of Wolf-Rayet stars show that most of the O/WR-stars in the central $R \leq 0.5$ pc are coeval. The massive stars appear to have formed in a well-defined star formation episode, of short duration ($t_{OW} \sim 6 \pm 2$ Myr, Paumard et al. 2006). There is no discernable age difference between clockwise and counter-clockwise orbiting stars, or for stars at different radii from Sgr A* (Bartko et al. 2009).

The stellar surface density distribution of the stars in the clockwise disk is fairly steep and scales as $\Sigma(R) \sim R^{-\beta}$, with $\beta$ ranging from 1.5 to 2.3 in the analyses of Paumard et al. (2006), Lu et al. (2009) and Bartko et al. (2009, 2010). The inner cutoff of the clockwise system is remarkably sharp. There are six O/WR-stars projected between 0.96" and 1.5" but none inside of 0.96".

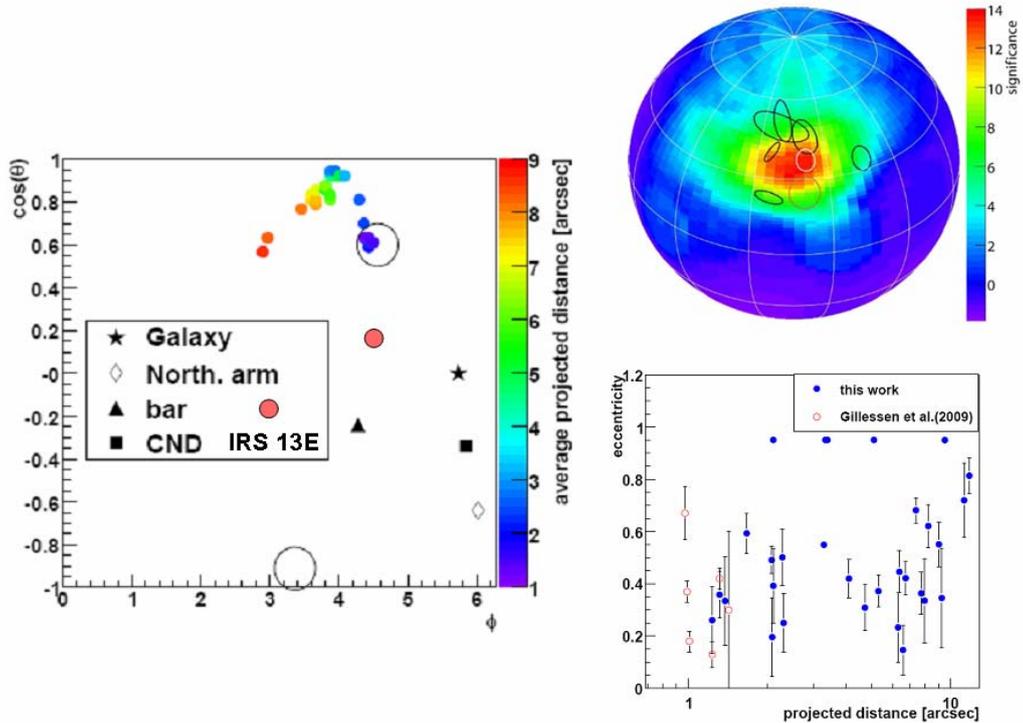

**Figure 2.2.2**. Left: Cylindrical equal area projection of the local average stellar angular momentum direction for the clockwise stars as a function of the average projected distance. The points are correlated through a sliding window technique. The average angular momenta for the innermost stars agree well with the orientation of the clockwise disk of Paumard et al. (2006), shown by the black circle. The asterisk shows the Galactic pole (Reid & Brunthaler 2004), the diamond indicates the normal vector to the northern arm of the mini-spiral (Paumard et al. 2004), the triangle indicates the normal vector to the bar of the mini-spiral (Liszt 2003), the square shows the rotation axis of the circum-nuclear disk (CND, Jackson et al. 1993). The red filled circle marks the position of the angular momentum vector of the orbit of IRS 13E if it is located in the bar, about 10" behind the sky plane (Fritz et al. 2010b). Top right: Orthographic projection (seen from $\varphi_0 = 254°$, $\theta_0 = 54°$) of the significance sky map in the interval of projected distances 0.8" − 3.5". It is overlaid with the 2σ contours (black lines) for the direction of the orbital angular momentum vectors of the six early-type stars (S66, S67, S83, S87, S96 and S97) with $0.8'' \leq p \leq 1.4''$ for which Gillessen et al. (2009b) were able to derive individual orbital solutions. All of these stars seem compatible with being members of the clockwise system. Still the orbital angular momenta of all of these stars are offset from the local angular momentum direction of the clockwise system at a confidence level beyond 90%. The white ellipse shows the 2σ contour of the clockwise system as determined by Paumard et al. (2006) and the brown one the 2σ contour of Lu et al. (2009). Bottom right: Reconstructed eccentricity as a function of projected distance for the 30 clockwise moving WR/O stars (blue points), which have a minimum angular distance below 10° from the (local) average angular momentum direction of the clockwise system. Red circles show the six early-type stars S66, S67, S83, S87, S96 and S97. Error bars denote the RMS of the reconstructed eccentricities (adapted from Bartko et al. 2009).

The disk has a significant geometric (z-) thickness. The local orbital inclinations relative to the disk's midplane exhibit a 1σ dispersion of between 7° and 14° in the different analyses, with a mean of ~ 10°. This opening angle corresponds to a ratio of z-scale height to radius of ~ 0.1. For six of the disk stars near the inner edge, Gillessen et al. (2009b) derive individual stellar orbits. The locations of their angular momentum vectors on the sky are plotted in the upper right inset of Figure 2.2.2. The spread in angles of these six stars agrees with a Gaussian model of dispersion 10°.

On average the stellar orbits are not circular. Beloborodov et al. (2006) find an average eccentricity of 0.3. Lu et al. (2009) conclude that a number of candidate stars have an eccentricity of at least 0.2. The mean ellipticity inferred by Bartko et al. (2009) is 0.37, with an uncertainty of the mean of ± 0.07 (lower right inset of Figure 2.2.2). The average eccentricity of the six stars with individual orbits (Gillessen et al. 2009b) is also 0.36 (lower right inset of Figure 2.2.2). This significant eccentricity of the orbits provides strong constraints for the star formation processes and initial conditions of the orbits (§ 6.3).

The surface brightness distribution of near-infrared light is not symmetric with respect to Sgr A*. Most of the bright stars near Sgr A* are in the IRS 16 'cluster' located 1" − 2" east of Sgr A* (Figure 1.1). Most of these stars are members of the clockwise disk. Since the orbital times at this radius ($1 - 2 \times 10^3$ yr) are a small fraction of the ages of the stars, the distribution of stars at the inner edge of the clockwise disk should be well phase mixed. Is the strong asymmetry of light reflected in an azimuthal asymmetry of the density of stars and does the asymmetry perhaps signal the presence of a gravitationally bound structure (Lu et al. 2006)? Once projection effects and apparent crowding along the line of nodes are taken into account, as well as the presence of additional, slightly fainter O-stars west of Sgr A*, both Paumard et al. (2006) and Lu et al. (2009) find that any remaining anisotropy is not significant at the > 3σ level. In addition there appears to be somewhat higher extinction west of Sgr A*, compared to east of Sgr A* (central bottom inset of Figure 1.1, Schödel et al. 2007a), which could be the result of additional dust in the orbiting gas streamers of the mini-spiral (Vollmer & Duschl 2000, Paumard et al. 2004). While this issue is probably not completely closed it may thus not be necessary to contemplate the alternative that the IRS 16 complex is self-gravitating, perhaps because of a central intermediate mass black hole (e.g. Lu et al. 2006).

*2.2.2 More than one disk?*

The analyses of Lu et al. (2009) and Bartko et al. (2009) show that only slightly more than half of the stars are members of the well-defined clockwise disk. One question is whether the other ~ 40 − 45% can be assigned to a second disk-like system, as proposed by Genzel et al. (2003a) and Paumard et al. (2006), or whether they are distributed more isotropically. Lu et al. (2009) do not find a second significant overdensity (at a level > 1σ above the background) at or near the position of the counter-clockwise disk system proposed by Genzel et al. (2003) and Paumard et al. (2006, left panel in Figure 2.2.1). Lu et al. (2009) conclude that there is only one disk of young stars.

The most recent work of Bartko et al. (2009, 2010), however, adds many new O/WR-and B-stars in the range of 5" ≤ $p$ ≤ 15" and does show a peak in the angular momentum distribution of the counter-clockwise stars (right panel of Figure 2.2.1). The significance of the peak of counter-clockwise stars in Figure 2.2.1 is 7σ, and is unlikely due to a statistical fluctuation in an underlying isotropic distribution. The difference in the conclusions of Lu et al. and Bartko et al. can be plausibly understood from the fact that the former study concentrates on the innermost few arcseconds, which has few counter-clockwise stars, while the latter (especially Bartko et al. 2010) has many more stars at $p$ > 5". It thus seems probable that there is indeed a second counter-clockwise system, co-eval with the clockwise star disk but oriented almost orthogonally to it.

As in the case of the clockwise system, Bartko et al. (2009, 2010) find evidence for a large (50° − 70°) change in the direction or the angular momentum direction with distance from Sgr A*. In addition perhaps 20% of the counter-clockwise O/WR -stars, and a larger fraction of the fainter counter-clockwise B-stars do appear to be distributed more isotropically, as proposed by Lu et al. (2009). The counter-clockwise disk or streamer structure may be in a more disrupted state than the clockwise disk. ***The growing complexity of the dynamics of the ~ 200 early-type stars in the central parsec probably indicates that the two-disk model may be a simplified approximate description***. In reality the stellar distribution may be more irregular (Kocsis & Tremaine 2010).

Cuadra, Armitage & Alexander (2008b) conclude from N-body simulations that the current configuration of O/WR-stars cannot be the result of the evolution of one single, initially very thin circular disk over 6 Myrs. They find that the known perturbers within the central parsec, such as massive stars, stellar black holes and even putative intermediate mass black holes are not sufficient to explain the thickness, ellipticity or warping. Interactions of the disk stars with stars and especially with remnants in the stellar cusp near Sgr A* help in increasing eccentricities and inclinations more rapidly (Perets et al. 2009, Löckmann, Baumgardt & Kroupa 2009) but probably are still not sufficient for explaining the current configuration from a single initial disk. If there was only one disk initially the large range of inclinations and large eccentricities of some of the stars must have been created during the formation process. The required large perturbing potential may result from the circum-nuclear disk if it is massive enough (Šubr, Schovancova & Kroupa 2009), an intermediate mass black hole or star cluster (Yu, Lu & Lin 2007), or a second disk (Nayakshin et al. 2006). However, these simulations treat the background stellar cluster as a smooth gravitational potential. Using an analytical model based on the growth of normal modes in Laplace-Lagrange theory, Kocsis & Tremaine (2010) find that when the 'graininess' of the cluster is taken into account, a strong warp of an initially thin disk arises naturally and inevitably through vector resonant relaxation within the influence of a central massive black hole (Figure 2.3.3). It is not clear yet from the work by Kocsis & Tremaine (2010), however, whether the specific warp structure of the clockwise disk, and/or the counter-clockwise system inferred by Bartko et al. (2009, 2010) can be explained by this mechanism.

Nayakshin (2005a) and Nayakshin et al. (2006) have pointed out that the interaction between two disks or rings of stars induces mutual orbital precession and thickens initially thin disks over time. The precession frequency is proportional to the mass of the ring/disk. The induced precession is fastest at radii inside of the ring and slowest at large radii. Nayakshin et al. (2006) find that two disks of mass a few $10^3$ to $10^4$ $M_\odot$ induce then a thickness of $h_z/R \sim 0.1$ over $4-6$ Myrs, in good agreement with the observed values of $10^4$ $M_\odot$ and $5 \times 10^3$ $M_\odot$ for the clockwise and counter-clockwise systems (Bartko et al. 2010). Löckmann, Baumgardt & Kroupa (2009) find that the interaction between two disks can easily explain the large range of orbital orientations and warping of the O/WR-stars observed by Lu et al. (2009) and Bartko et al. (2009). Resonant relaxation processes of the disk(s)-cusp system may be rapid enough to then also explain the ellipticities in the disks (Löckman, Baumgardt & Kroupa 2009).

### *2.2.3 Massive binaries in the disk(s)*

Another important issue is the abundance of binary stars in the central parsec. Because of the large velocity dispersion, binaries can only be long-lived if they are tightly bound. Hopman (2009) finds that the binary fraction can be a few percent outside 0.1 pc, but has to smaller further in. Within that radius, binaries can only exist on highly eccentric orbits. For the O/WR-stars residing in the disks, binarity could open up a new route for investigating the mode of star formation that has occured ~ 6 Myrs ago. The binary fraction is an indicator of the cooling time scale of the disks (Alexander et al. 2008b). Also for the B-stars (§ 2.3) the binary fraction may be a signature of their formation (Perets et al. 2008a, Perets 2009). For an in-situ formation scenario a rather normal binary fraction would be expected, whereas for a binary capture scenario (§ 6.3), a low binary fraction should result.

Observationally, relatively little is known currently about binaries in the Galactic Center. Ott et al. (1999) and Martins et al. (2006) have demonstrated from lightcurves and line profile modulations that IRS16SW, one of the brightest O/WR-stars, is an eclipsing contact binary with a $\geq 50$ $M_\odot$ mass of each member. Rafelski et al. (2007) list a few more candidate stars, one of which is probably spectroscopically confirmed (O. Pfuhl, priv. comm.). The number of O/WR-stars explored for binarity through lightcurves or spectral variations is ~ 15. For the B-stars, no evidence for binarity has been reported so far.

## 2.3 The central S-star cluster and the distribution of B-stars

While the O/WR-stars appear to reside in one or two non-equilibrium planar structures still reflecting their formation process a few million years ago, the somewhat lighter (~ 3.5 − 20 $M_\odot$), and potentially older, B-stars trace a very different distribution. We will discuss in § 2.4 that the overall radial surface density distribution of the B-stars in the current data sets scales as $p^{-1.4}$ (Figure 2.4.2)[4], similar to the O/WR-stars but continuing all the way to Sgr A*, without the sharp inner cutoff observed for the O/WR-stars. Of the 31 $K_s \leq 16$ stars that reside in projection in the central $p \leq 1$" around Sgr A*, 16 (52%) are B-stars, three are O-stars and twelve are late-type stars. The three O-stars are found at $p \geq 0.9$" and are the innermost members of the clockwise disk discussed in the last section. If the 3D-space velocity of each star is used to eliminate probable fore- and background interlopers, there remain 17 B-stars (74%) and six late-type stars (26%) with $K_s \leq 16$ that reside in the innermost cusp. A still more robust accounting can be made for 28 stars for which Gillessen et al. (2009b) were able to derive individual orbits. Of the 19 stars with a semi-major axis $a \leq 1$", 16 (84%) are B-stars, three (16%) are late-type stars. Clearly, the brightest members of the central cusp are B-stars.

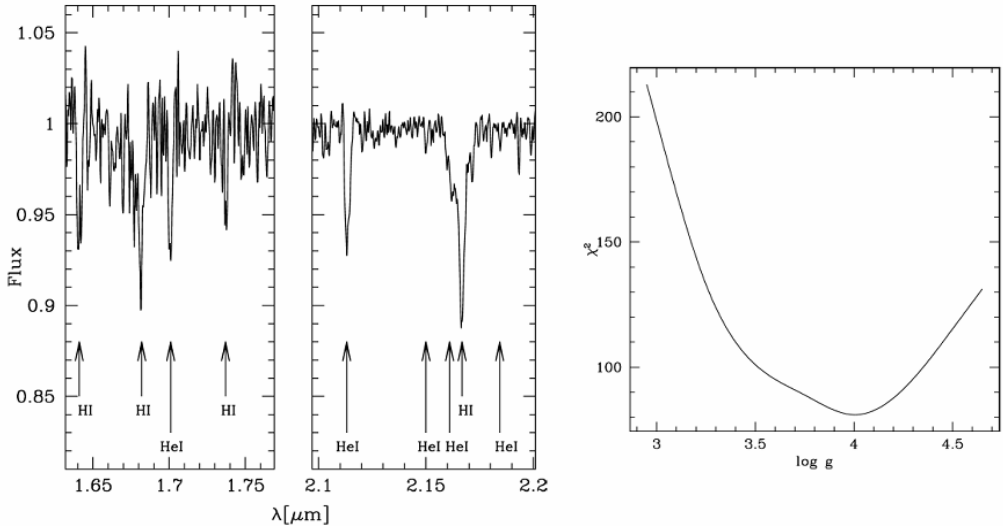

**Figure 2.3.1.** Spectral properties of S2/S02, the brightest member of the B-star cluster around Sgr A*, which was the one identified first (Ghez et al. 2003) and for which a full orbit is now available (Figure 4.3.1). The left panel shows its H/K-band spectrum from the co-addition of SINFONI spectra taken between 2004 and 2007 (23.5 hours total integration time). The various spectral features are marked. After a full modeling of the line equivalent widths, line shapes and the absolute $K_s$-band magnitude of −2.75, a quantitative comparison with stellar

---

[4] In the following *p* denotes projected angular distance, while *R* refers to the 3D physical distance.

models shows that the star must be a true main-sequence B0-2.5V-star ($\log g = 4$) of temperature $T \sim 22{,}000$ K, $\log L/L_\odot = 4.45$, $m = 19.5$ $M_\odot$, $r = 11.6$ $R_\odot$, with low rotation velocity of $100 \pm 30$ km/s typical of solar neighborhood B-stars, low mass loss rate ($dm/dt < 3 \times 10^{-7}$ $M_\odot$/yr) and somewhat enriched He-abundance (He/H = 0.3 − 0.5). Adopted from Martins et al. (2008).

Martins et al. (2008) have analyzed a high-quality, combined multi-epoch SINFONI spectrum of the brightest member of the B-star cusp, S2/S02, with model spectra and standard spectral atlas data (Figure 2.3.1). S2/S02 appears to be a main-sequence (dwarf) B0-2.5 V star of zero age main-sequence (ZAMS) mass of $m_{ZAMS} \sim 19.5$ $M_\odot$ and with a low rotation velocity typical of solar neighborhood BV stars. The data exclude the possibility that the star is the He-rich core of a stripped $m_{ZAMS} < 8$ $M_\odot$ AGB star, as proposed by Davies & King (2005). The He/H abundance of S2/S02 is relatively high (0.25 − 0.5), making the star a member of the so-called 'He-rich' subclass of main-sequence B-stars.

Eisenhauer et al. (2005) co-added the spectra of several of the fainter members of the S-star cluster and showed that they too have spectral properties (equivalent widths of HI Br-$\gamma$ and He 2.1 µm, line profiles, absolute magnitudes, rotation velocities) consistent with solar neighborhood B2-9 V stars. Given these spectral identifications, the B-stars in the S-star cluster could have ages between 6 and 400 Myrs, given by the main-sequence lifetimes. Interestingly, the lower limit to the age of S2/S02 is 6 Myrs, similar to the age of the O/WR-star disk(s).

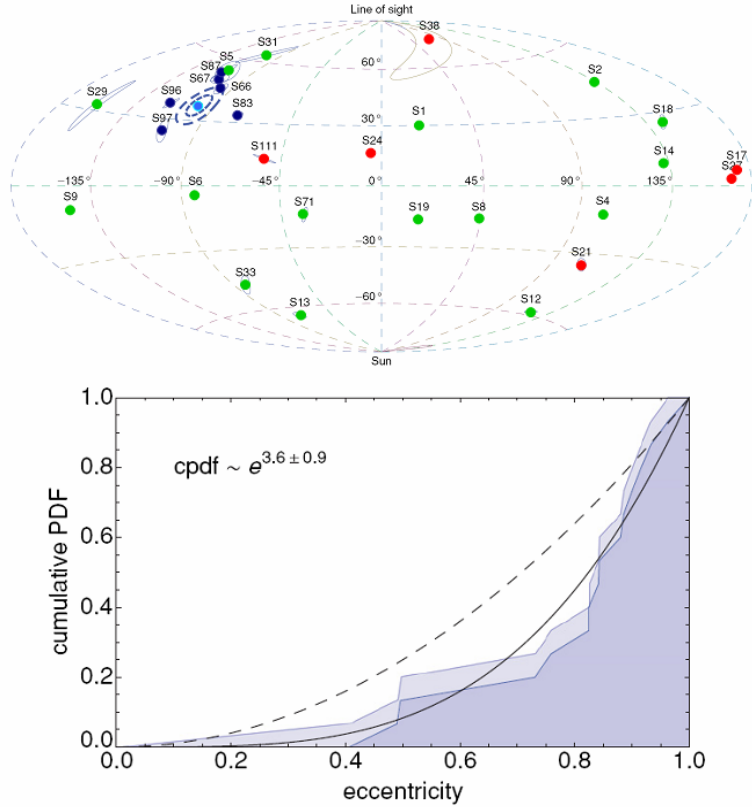

**Figure 2.3.2.** Properties of the orbits of the 'S-stars' in the central cusp around Sgr A*. Top: Orientation of the orbital planes of S-stars with individually determined orbits. The orientation of the orbits in space is described by the orbital angular momentum vector,

corresponding to a position in this all sky plot, in which the vertical dimension corresponds to the inclination *i* of the orbit and the horizontal dimension to the longitude of the ascending node *Ω*. A star in a face-on, clockwise orbit relative to the line of sight, for instance, would be located at the top of the graph, while a star with an edge-on seen orbit would be located on the equator of the plot. The error ellipses correspond to the statistical 1$\sigma$ fit errors only, thus the area covered by each is 39% of the probability density function. Stars with an ambiguous inclination have been plotted at their more likely position. The stars S66, S67, S83, S87, S96 and S97, which were suspected to be part of the clockwise stellar disk by Paumard et al. (2006) and Bartko et al. (2009) at ($\Omega = 98°$, $i = 129°$), actually are found very close to the position of the disk. The latter is marked by the thick light blue dot and the dashed lines, indicating a disk thickness of 14° ± 4°. Red dots mark late-type stars, green dots B-stars and dark blue dots O-stars. The orbits of the other stars are oriented randomly. Bottom: Cumulative probability distribution function (pdf) for the eccentricities of the early-type B-stars. The two curves correspond to the two ways to plot a cumulative pdf, with values ranging either from 0 to $(N-1)/N$ or from $1/N$ to 1. The distribution is only marginally compatible with a thermal isotropic pdf $n(e) \sim e$ (dashed line), the best fit is $n(e) \sim e^{2.6 \pm 0.9}$ (corresponding to a cumulative pdf $\sim e^{3.6 \pm 0.9}$). Adapted from Gillessen et al. (2009b).

From the studies by Schödel et al. (2003), Ghez et al. (2005b), Eisenhauer et al. (2005) and Gillessen et al. (2009b) it is now possible to constrain the orbital parameters of the B-stars. The upper panel of Figure 2.3.2 gives the orientations of the angular momentum vectors of all S-stars with individual orbits. If only the 22 B-stars and late-type stars outside the clockwise disk are considered, the probability density distribution of the angular momenta are in very good agreement with an ***isotropic distribution***. A Rayleigh test yields a probability of 73% that the sample is drawn from a random distribution. The bottom panel of Figure 2.3.2 shows the cumulative distribution of the eccentricities of the B-stars. The best fit differential distribution of eccentricities is $n(e)\, de \sim e^{2.6 \pm 0.9}\, de$. This favors somewhat higher eccentricities than, but is still marginally consistent with, a thermal isotropic distribution: $n(e)\, de \sim e\, de$ (Schödel et al. 2003, Alexander 2005). The fact that the observed eccentricity distribution shows a larger fraction of highly eccentric orbits than in a relaxed, thermal distribution may give a first interesting hint on the origin of the S-stars. This is because the normal, two-body relaxation time scale, $T_{NR}$, (involving long distance, random interactions between two stars) is several Gyrs or more, much longer than the lifetime of all S-stars (Figure 2.3.3). Hence, the eccentricity distribution points towards a mechanism producing highly eccentric orbits.

However, in the sphere of influence of the central massive black hole stellar orbits are near-Keplerian, precession time scales are much longer than orbital time scales, and the interaction between stars builds up coherently (resonantly) over many revolutions (Rauch & Tremaine 1996). In this regime a change in the value of the angular momentum (and thus the eccentricity of the orbit) occurs on the 'scalar' resonant relaxation time scale $T^s_{RR} \ll T_{NR}$. Changes of the orientation of the orbital angular momentum (but not its magnitude) can occur still faster, on the 'vector' resonant relaxation time scale $T^v_{RR} \ll T^s_{RR}$ (e.g. Hopman & Alexander 2006b, Alexander 2007). Figure 2.3.3 indicates that $T^v_{RR}$ is sufficiently short in the central cusp that a complete randomization of all S-star orbital momentum orientations could have occurred during the entire range of their possible time of residence there (6 Myrs $< t < t_{ms}$). For the innermost S-stars Lense-Thirring precession around the spin axis of the black hole may also be important (Levin & Beloborodov 2003). However, the scalar relaxation time scale is too long for a thermalization of all but the lower mass S-stars, even if they have resided there for their entire main-sequence lifetime. Unless there are additional perturbers or still faster relaxation processes, these considerations would suggest that the observed eccentricity distribution still somewhat reflects the initial eccentricity distribution. We will return to these issues in § 6.3 when we discuss the origin of the S-stars in more detail.

Gillessen et al. (2009b) also derive the distribution of the volume density distribution of the cusp's B-stars and find for the 15 stars with a semi-major axis ≤ 0.5", $\Sigma_B(R) \sim R^{-1.1 \pm 0.3}$. That slope appears to differ somewhat from the overall B-star distribution, which shows a surface density profile out to $p \sim 15$" of $\Sigma_B(p) \sim p^{-1.5 \pm 0.2}$ (Figure 3.3.2, Bartko et al. 2010). This would suggest that the S-star cluster is a distinct component and not the inward continuation of an overall larger cusp of B-stars. Still, the fainter B-stars ($K_s > 15$) at $p > 0.8$" may, reside in a more isotropic distribution than the O/WR-stars, and resemble the B-stars in the S-star cluster in this respect (Bartko et al. 2010).

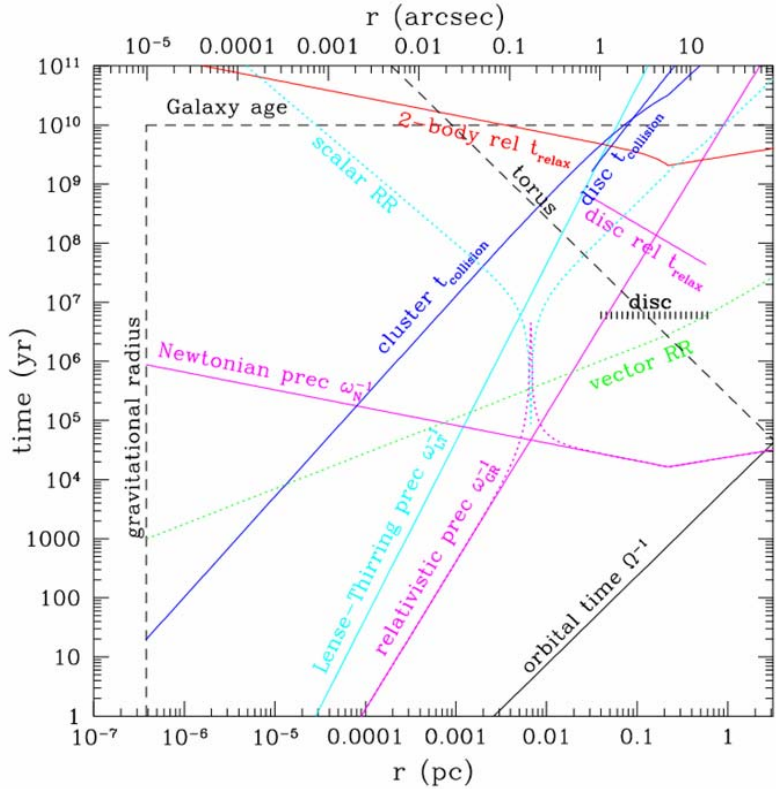

**Figure 2.3.3.** Time-scales in the central parsec of the Galaxy. The dashed black lines show the gravitational radius of the central black hole (vertical) and the age of the Galaxy, which is an upper limit to the age of the nuclear cluster $T_{cl}$ (horizontal) The horizontal thick hatched black line shows the age and radial extent of the disks(s) of WR/O-stars. The slanted black line at the lower right shows the orbital time. The magenta lines show the apsidal precession times due to the Schwarzschild term of General Relativity (GR) and the Newtonian retrograde precessiondue to the stellar cluster (N). The dotted thin magenta curve shows the combined apsidal precession time $(\omega_{GR} + \omega_N)^{-1}$. The solid cyan line slanting up to the right shows the Lense–Thirring precession time due to frame dragging of circular orbits around a maximally spinning black hole. The red line shows the two-body relaxation time $T_{NR}$ for the stellar cluster for an average stellar mass of 1 $M_\odot$. The down-sloping magenta line shows the two-

body relaxation time within the disk, assuming an average stellar mass of 20 M$_\odot$ and a disk mass of 5000 M$_\odot$. The cyan and green, dotted lines show the scalar ($T^s{}_{RR}$) and vector ($T^v{}_{RR}$) resonant relaxation time-scales, for 20 M$_\odot$ stars. The solid blue lines show the collision time in the stellar cluster for stars with mass $m$ = 1 M$_\odot$ and radius $r_\star$ = 1 R$_\odot$ and in the disk for $m$ = 20 M$_\odot$ and radius $r_\star$ = 10 R$_\odot$. Finally, the slanted dashed black line shows the precession time due to the circum-nuclear disk (CND) if the high-mass estimate of 10$^6$ M$_\odot$ of Christopher et al. (2005) is applicable. For more modest CND masses the time scale scales inversely to that mass (§ 3.2). Adopted from Kocsis & Tremaine (2010).

## 2.4 Is there an 'equilibrium' stellar cusp?

Depending on the mass function and radial distribution of the nuclear stars, including the fraction of stellar remnants, estimates of the two-body relaxation time scale $T_{NR}$ in the central parsec range between 1 and 20 Gyrs, close to or somewhat smaller than the Hubble time $T_H$ (Figure 2.3.3, Alexander 2005, Merritt 2006, 2010, Kocsis & Tremaine 2010). In an equilibrium cluster (age >> $T_{NR}$) around a massive black hole a ***dense stellar cusp of old stars and remnants*** is expected to form (Bahcall & Wolf 1976, 1977, Cohn & Kulsrud 1978, Young 1980). The Galactic center nuclear cluster (age $T_{cl} \sim T_{NR}$) probably approaches but may not fully reach this equilibrium state. The equilibrium stellar distribution of a spherically symmetric single mass star cluster is a singular power-law of slope *ζ = 7/4 (ρ ~ R$^{-ζ}$*, Bahcall & Wolf 1976). In a multi-component stellar cluster mass-segregation results in the more massive stars being more concentrated than the less massive ones. For a mass-segregated, multi-mass cluster with a range of masses from $m_{min}$ to $m_{max}$, the equilibrium density distribution of stars with mass $m_{min} \leq m \leq m_{max}$ is independent of the initial conditions and scales as

$$\rho(R,m) \propto R^{-\zeta(m)} \text{ with } \zeta(m) = 1.5 + \left(\frac{m}{4m_{max}}\right) \qquad (1)$$

(Bahcall & Wolf 1977). More recent analytical work and numerical simulations of multi-component nuclear clusters broadly confirm this classical analytical result (Hopman & Alexander 2006a, Freitag, Amaro-Seoane & Kalogera 2006). For the parameters of the Galactic Center, the Monte Carlo simulations of Freitag et al. (2006) predict 560, 2.4 × 10$^4$ and 2.1 × 10$^5$ M$_\odot$ in 10 M$_\odot$ stellar black holes within 0.01, 0.1 and 1 pc from the center, with a slope approaching *ζ ~ 7/4*. Within the same radii Freitag et al. find 180, 6500 and 3.4 × 10$^5$ M$_\odot$ in main-sequence stars (and giants), 30, 2000 and 1.2 × 10$^5$ M$_\odot$ in white dwarfs and ~ 0, 600 and 1.5 × 10$^5$ M$_\odot$ in neutron stars. The lighter stars attain a slope of *ζ* = 1.3 − 1.4, close to but somewhat shallower than the low mass limit of Bahcall & Wolf.

If the 'heavy' stars (mass $m_h$, number $N_h$) are rare while the 'light' stars (mass $m_l$, number $N_l$) dominate such that the self-relaxation parameter
$\Delta = 4N_h m_h{}^2 /( N_l m_l{}^2$ [$3+m_h/m_l$]) is much less than unity (the Bahcall & Wolf solution is reached when *Δ > 1*), much stronger relaxation can occur (Alexander 2007). The slope of the most massive stars can then be significantly steeper than *ζ* = 7/4. This limit would be applicable to stellar black holes, with *ζ* = 2 to 11/4. Massive stars are too short-lived to reach relaxation. Alexander & Hopman (2009) showed that old or continuously star-forming systems with a Salpeter-Kroupa-IMF have *Δ << 1*. The strong segregation solution might thus be relevant for most galaxies, if they have old relaxed cusps with a standard IMF. If physical collisions are frequent in the densest, innermost part of the cusp (at stellar densities > $10^8$ M$_\odot$pc$^{-3}$) the cusp slope can become shallower than $\zeta = 1.5$ and in principle can reach a minimum value of $\zeta = 1/2$ (Murphy, Cohn and Durisen 1991). Deviations from the Bahcall & Wolf solution also would generally be expected if the cusp has not reached equilibrium, following a strong disturbance, such as a merger or star formation event, or the in-spiral of an intermediate mass black hole (Merritt 2006, Baumgardt et al. 2006).

Is such an equilibrium stellar cusp observed in the Galactic Center? Speckle imaging with ~ 0.15" resolution in the early 1990s showed that the star counts to K ≤ 14 are consistent with a power law density distribution ($\zeta \sim 2$) to ~ 0.15 pc with a flattening to a core inside (Eckart et al. 1993, Allen & Burton 1994). Following the availability of high quality adaptive optics imaging at 8−10 m-class telescopes Genzel et al. (2003a) reported the first significant detection of an inward increase of the K ≤ 17 stellar number counts to scales of ≤ 0.04 pc from Sgr A*. A stellar maximum of faint stars centered on Sgr A* is detected and is centered on Sgr A* to within ± 0.2" (Genzel et al. 2003a, Schödel et al. 2007a). The azimuthally averaged, projected stellar surface density distribution can be described by a broken-power law (Genzel et al. 2003a, Schödel et al. 2007a). The inferred power-law slope of the cusp is $\zeta = 1.3 \pm 0.1$ within $R_b = 6.0" \pm 1.0"$ (0.22 ± 0.04 pc), outside of which the slope steepens to $\zeta = 1.8 \pm 0.1$, in agreement with earlier studies of the surface brightness distribution by Becklin & Neugebauer (1968), Catchpole, Whitelock & Glass (1990), Eckart et al. (1993), Haller et al. (1996), Genzel et al. (2000), or Launhardt et al. (2002) on larger scales. Refining these works, Graham & Spitler (2009) subtract off the contribution from bulge stars to the light profile. They show that the nuclear cluster of the Milky Way is reasonably well represented by a Sersic function with index $n = 3$ and an effective half-light radius of 80". That profile is steeper in the outer parts than the previous power-law describptions due to the omission of the bulge stars. Schödel (2010) confirms this finding.

In contrast to the surface density distribution of the faint stars, the surface brightness distribution of the 'diffuse' infrared light between bright stars is somewhat flatter and does not show a maximum toward Sgr A* (Scoville et al. 2003). Following up on similar results in earlier work with seeing limited images by Rieke & Rieke (1988) and Allen & Burton (1994), Scoville et al. (2003) find from NICMOS-HST 1.9 μm imaging that the diffuse light corrected for extinction is flat outside about 10", slowly increases inward to about 1" but then has a core or even a dip toward Sgr A*. Because of the spillover of light from bright stars it is not clear how well this diffuse light traces the intrinsic distribution of the underlying faint stars (Schödel et al. 2007a). The confusion corrected stellar surface density distribution thus is arguably the best tracer of the older giants and the B/A main-sequence population in the nuclear cluster. Neither the near-infrared light nor the star counts, however, are sensitive to main-sequence stars less than a few solar masses, to white dwarfs or stellar remnants.

### *2.4.1 Radial distribution of different stellar components*
We will show in this section that the concentration of faint stars near Sgr A* probably does not represent a classical Bahcall & Wolf cusp, once the nuclear stars are assigned to different stellar types based on their spectral properties. Instead the *superposition of a fairly flat, large core of (relaxed?) old giants and a compact concentration of unrelaxed young massive stars toward Sgr A\* conspire to create a cusp in the number counts of faint stars*.

It has been clear for almost two decades that the mix of stars changes with radius as one approaches the nucleus from the bulge. The relative fraction of bright stars ($M_{bol} \sim -5$ to $-8$) increases significantly from a few parsecs to the center (Rieke & Rieke 1988). While the light of the old, low mass stars dominates the outer star cluster, the central few arcseconds are dominated by massive early-type stars (Figure 2.4.1, Krabbe et al. 1991, Burton & Allen 1992). There is a central depression in the surface brightness or equivalent width of the light of the red giants and bright AGB stars (Figure 2.4.1, Sellgren et al. 1990, Allen & Burton 1994, Genzel et al. 1996, Haller et al. 1996). The key question is whether this inferred depression of late-type stars is a true physical effect, or whether it is the result of the strong increase in the density of early-type stars.

Increasingly more sensitive and higher angular resolution imaging spectroscopy over the past few years now makes a compelling case that the main effect is the steep increase in the density (and brightness) in the density of bright early-type stars (e.g. Genzel et al. 2003a, Eisenhauer et al. 2005, Schödel et al. 2007a). This conclusion is demonstrated qualitatively in Figure 2.4.1, which shows the spatial distribution of early and late-type stars, colored blue and red, respectively, as well as quantitatively in Figure 2.4.2, which gives the binned surface density distributions of different types of stars averaged over annuli centered on Sgr A\* (corrected when necessary for confusion, incomplete coverage etc., see Bartko et al. 2010, Do et al. 2009a). The data shown in Figure 2.4.2 are from SINFONI/VLT integral field spectroscopy (Bartko et al. 2010) but OSIRIS/Keck spectroscopy by Do et al. (2009a) and narrow-band NACO/VLT spectro-photometry by Buchholz, Schödel & Eckart (2009) yield very similar results, with some differences mainly between the spectroscopic and narrow-band photometric data due to erroneous stellar identifications in the narrow-band method. Because of a combination of confusion/completeness and reliability of spectroscopic features, the limiting magnitude of the spectroscopic data currently reaches a $40 - 50\%$ completeness at $M(K) \leq -1.5$ ($K \leq 15.5$, Do et al. 2009a, Bartko et al. 2010), with the exception of the central $1" - 2"$ where the spectroscopic data are somewhat deeper ($K_s \sim 16.5$) and more complete (Bartko et al. 2010). The spectroscopy thus currently samples all O/WR and AGB stars, B-stars to $\sim$ B3V, and red giants to K0III, including the peak of the red clump.

The surface density of the O- and Wolf-Rayet stars, representing the most massive component of the star cluster rises steeply *($\Sigma_{O/WR} \sim p^{-1.4 \pm 0.2}$* for the 'clockwise disk' stars) from $\sim 15"$ to a few arcseconds (black points in Figure 2.4.2). There is a sharp inner cutoff of the O/WR-population at $p \leq 1"$ and no O/WR-stars have so far been discovered for $p > 13"$ (0.5 pc, Paumard et al. 2006, Lu et al. 2009, Bartko et al. 2009, 2010). Likewise the surface density of B-stars ($14 \leq K_s \leq 15.5$) also sharply increases inward (blue points in Figure 2.4.2), with a similar slope as that of the O/WR stars *($\Sigma_B \sim p^{-1.5 \pm 0.2}$*). All of these early-type stars have an age much less than $T_{NR}$. The central concentration of B-stars centered on Sgr A\* thus cannot be an equilibrium Bahcall-Wolf cusp.

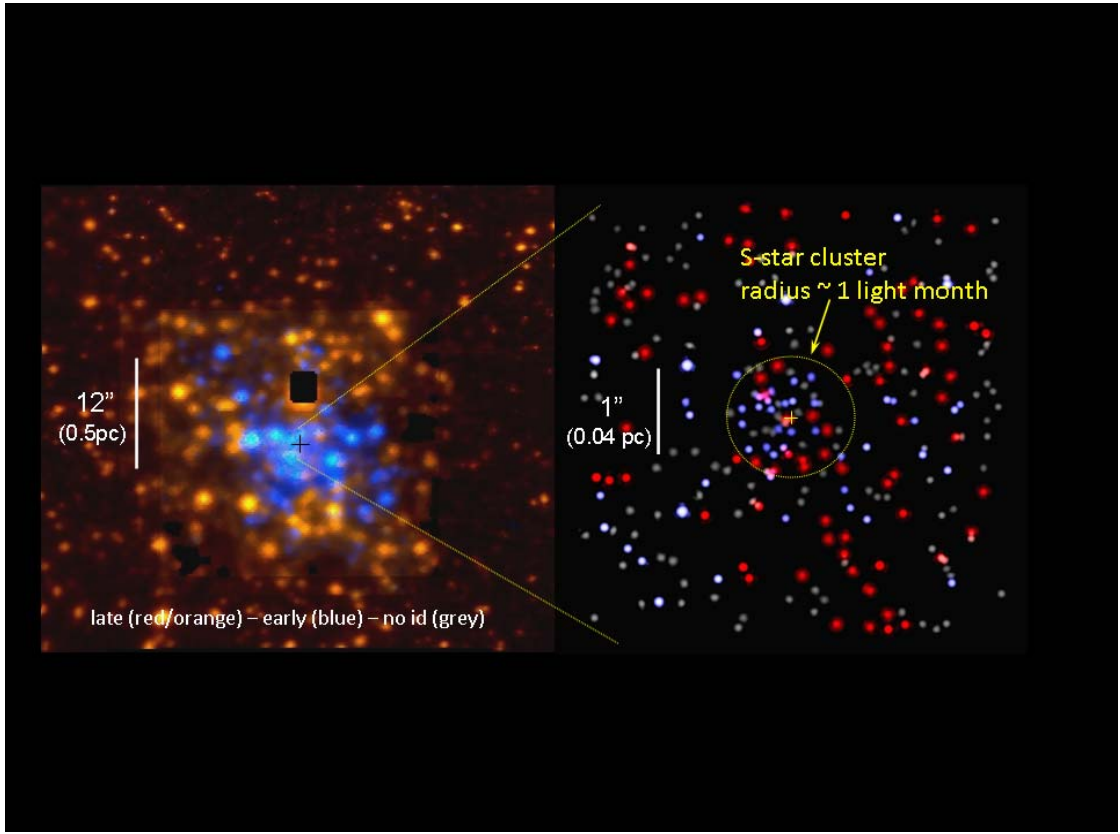

**Figure 2.4.1.** Distribution of early-type stars (blue) and late-type stars (orange/red), as obtained from SINFONI integral field spectroscopy in the central $p \sim 1$ pc (left) and central $p \sim 0.08$ pc (right). The blanked out region in the left image is the location of the bright red supergiant IRS 7. The image is a deconvolved NACO image of the central $p \sim 0.08$ pc region with SINFONI spectroscopic identifications marked as colors. Grey stars lack a spectroscopic identification.

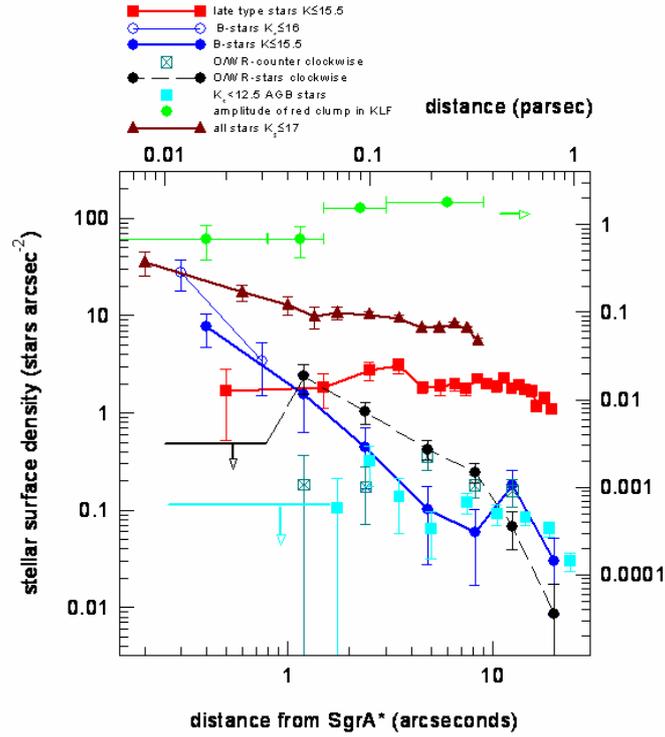

**Figure 2.4.2.** Radial surface density distributions of different stellar components in the Galactic Center, as a function of projected distance from Sgr A* (from NACO and SINFONI observations on the VLT). Counts are corrected for spatially variable extinction, incompleteness of coverage and local confusion with nearby bright stars. Filled black circles and dark green crossed squares mark the distribution of spectroscopic O/WR-stars in the clockwise and counter-clockwise systems ($K_s \leq 14$). Filled and open blue circles denote spectroscopic B-stars with $14.5 \leq K_s \leq 15.5$ and $K_s \leq 16$. Filled red and cyan squares mark spectrosopic late-type stars to $K_s \leq 15.5$ and $K_s \leq 12.5$. Open green circles denote the ratio of the red clump feature at $K_s \sim 15 - 16$, relative to the underlying KLF, marking the relative fraction of the oldest stars currently accessible to study (right ordinate). Filled brown triangles represent the overall distribution of all stars to $K_s \leq 17$ (adapted from Bartko et al. 2010).

In contrast to the early-type stars, the density of late-type giants does not peak on/near Sgr A*. There are only a handful of bright giants ($K_s \leq 12.5$) in the central few arcseconds and no bright TP-AGB stars ($K_s \leq 10.5$) in the central 7" (cyan points in Figure 2.4.2, Genzel et al. 1996). The fainter K-giants (later than K0III: $K_s \leq 15.5$) exhibit a flat surface brightness distribution in the central 10" − 12", perhaps even with a central depression (red points in Figure 2.4.2, Buchholz et al. 2009, Do et al. 2009a, Bartko et al. 2010). There is also a lack of horizontal branch/red clump stars toward the central few arcseconds, which sample the oldest ($\geq 1$ Gyr) and lowest mass ($\leq 2$ M$_\odot$) stars accessible to the current spectroscopic studies (open green circles in Figure 2.4.2, Genzel et al. 2003a, Schödel et al. 2007a). The central S-star cluster is dominated by main-sequence B-stars to the current spectroscopic limit (Eisenhauer et al. 2005, Gillessen et al. 2009b, Do et al. 2009a). There definitely are a few late-type stars within the central S-star cluster, however. Gillessen et al. (2009b) find that four of the 17 $K_s \leq 16.7$ spectroscopically identified late-type stars in the central arcsecond have orbital periods ≤ 120 years, the remainder may just be projected toward the S-star cluster. However, from a Monte Carlo analysis Do et al. (2009a) conclude that the intrinsic density distribution of giants and red clumps stars to $K_s \leq 15.5$ has a power law slope that is flatter than $\zeta = 1$ at the 99.7% probability level. All these results lead to the robust conclusion that the old, late-type stars have a large central core ($\zeta \sim 0$), a central hole ($\zeta < 0$) or at most a very flat cusp ($\zeta < 1$).

As discussed already above, analytical and numerical work cannot explain cusp slopes with $\zeta \leq 1$ solely by mass segregation in an equilibrium, multi-mass system. One possible explanation for the flat or perhaps even centrally depleted density distribution of giants in the central arcseconds are physical collisions with other stars and remnants. They may result in a long lasting or permanent loss of the giant's envelope (Lacy, Townes & Hollenbach 1982, Phinney 1989, Sellgren et al. 1990, Genzel et al. 1996). The efficacy of this process is quite uncertain, however, despite increasingly more sophisticated simulations of collisions. From analytical considerations Alexander (1999) concluded that the depletion of moderately bright giants in the central ~ 1" could be accounted for by such collisions, while Bailey & Davies (1999) found from 'smoothed particle hydrodynamics' (SPH) simulations that the high velocity collisions in the innermost arcseconds are not effective enough in most instances to remove enough mass to destroy the envelope permanently. Adopting a background stellar cusp model from Freitag et al. (2006), Dale et al. (2009) carried out more detailed SPH and Monte Carlo simulations to investigate collisional mass removal and envelope destruction as a function of mass and evolutionary stage of the giant, as a function of distance and for different collision partners. Dale et al. find that penetrating collisions indeed can eject several tens of percent of the giant's mass that then subsequently permanently removes the star as a bright K-band object. 10 $M_\odot$ black holes and main-sequence solar mass stars contribute approximately equally to the collisional destructions. For a population of a few $10^4$ stellar black holes in the central 0.1 pc the cumulative effects of these collisions are able to account for a significant depletion of moderately bright ($10 < K_s < 12$) giants within the central ~ 1" around Sgr A*. The depletion zone is more spatially confined for the more massive bright AGB stars. Dale et al. also conclude that collisional depletion is probably irrelevant for faint giants and red clump stars ($K_s < 15$), because of the extreme mass loss required. Physical collisions with a standard stellar mass function thus do not appear to be able to explain the large observed core of late-type stars. Very recent work with a top-heavy IMF suggests that the collisionally depleted zone may then be significantly larger (Davies et al. 2010a) but the issue needs to be investigated in more detail before a final verdict is possible.

Another option is that an additional massive or intermediate mass black hole has gouged out a core in the old star cluster, on its inward spiral due to dynamical friction (Milosavljevic & Merritt 2001, Baumgardt et al. 2006, Merritt 2010). Baumgardt et al. (2006) find that the in-spiral of a $3 \times 10^3 - 10^4$ $M_\odot$ intermediate mass black hole would create a central flat core of radius $R \sim 0.2$ pc, broadly consistent with the density distributions shown in Figure 2.4.2 (Buchholz et al. 2009, Do et al. 2009a). Finally a minor merger or major star formation event may also strongly perturb the central cluster (Merritt 2006, 2010). However, Trippe et al. (2008) find that the dynamics of the old cluster is well described by a uniform, slowly rotating and relaxed system, with no evidence for any large-scale disturbance. Likewise the dynamics of the S-stars and of Sgr A* also strongly restrict the parameter space available for an intermediate mass black hole (Figure 4.6.1).

If the flat spatial distribution of the $K_s \leq 15.5$ late-type stars is respresentative of most of the stellar mass the two-body relaxation time scale at $R \ll 1$ pc would be significantly longer than the age of the Galactic Center star cluster (and the Hubble time, Merritt 2010). The observed large core may then reflect the initial conditions of the nuclear cluster at formation (Merritt 2010). Note, however, that the relaxation time depends sensitively on the mass function, abundance of remnants, radial distribution and other parameters (Figure 2.3.3, Kocsis & Tremaine 2010). In addition Preto & Amaro-Seoane (2010) argue that analytical estimates might overestimate the relaxation time by a factor 4 to 10. In addition the existence of the central S-star cluster implies that relaxation processes much faster than the standard two-body rate are relevant in the nuclear star cluster (see §6.3, Alexander 2007). Another important consequence of the flat core and long relaxation time scale, if representative for the overall central star cluster, would be *reduced extreme-mass ratio in-spiral rates, and thus smaller event rates for the planned gravitational wave observatory LISA*, when compared to most current simulations (Hopman & Alexander 2006a,b, Freitag et al. 2006, Alexander 2007, Merritt 2010, Preto and Amaro-Seoane 2010).

A top-heavy IMF (Bartko et al. 2010, § 2.5), if applicable throughout the evolution of the central nuclear cluster, would also lead to a lack of old, low mass giants in the very center, which would then be completely dominated by stellar remnants (but see Löckmann et al. 2010). To explain the flat or even inverted radial slope (Do et al. 2009; Buchholz et al. 2009), the IMF probably would have to depend strongly on radius. Finally, it is perhaps interesting to note that the mass of the missing late-type stars in the cusp between Sgr A* and $R \sim 10$" (subtracting the observed core from a $n(R) \sim R^{-1.5}$ cusp) is $\sim 3 - 5 \times 10^3$ $M_\odot$ for $K_s \leq 17 - 18$. This is comparable to the stellar mass in the cusp of early-type stars (~1500 $M_\odot$ for $14.5 \leq K_s \leq 17.5$, and substantially larger if any of the O/WR-stars between $p \sim 0.8$" $- 10$" are considered as well). Could it be that non-equilibrium effects due to episodic star formation (with a top-heavy IMF) and capture of B-stars in the central S-star cluster might be capable of driving out a large number of giants from the innermost cusp region?

Muno et al. (2005) have reported the detection of four (of seven total) X-ray transients in the central parsec from multiple Chandra observations over a five-year period (Figure 2.4.3), representing a 20-fold excess of such transients with respect to the greater Galactic Center environment on 10 pc scales. These sources each vary in luminosity by more than a factor of 10 and have peak X-ray luminosities greater than $5 \times 10^{33}$ erg s$^{-1}$, suggesting that they are accreting black holes or neutron stars. The peak luminosities of the transients are intermediate between those typically considered outburst and quiescence for X-ray binaries. The excess transient X-ray sources may be low-mass X-ray binaries that were produced by three-body interactions between binary star systems and either black holes or neutron stars that have migrated into the central parsec as a result of dynamical friction. Alternatively, they could be high-mass X-ray binaries that formed among the young stars that are present in the central parsec. The Muno et al. transients thus may represent direct evidence for a concentrated population of remnants in the central cusp.

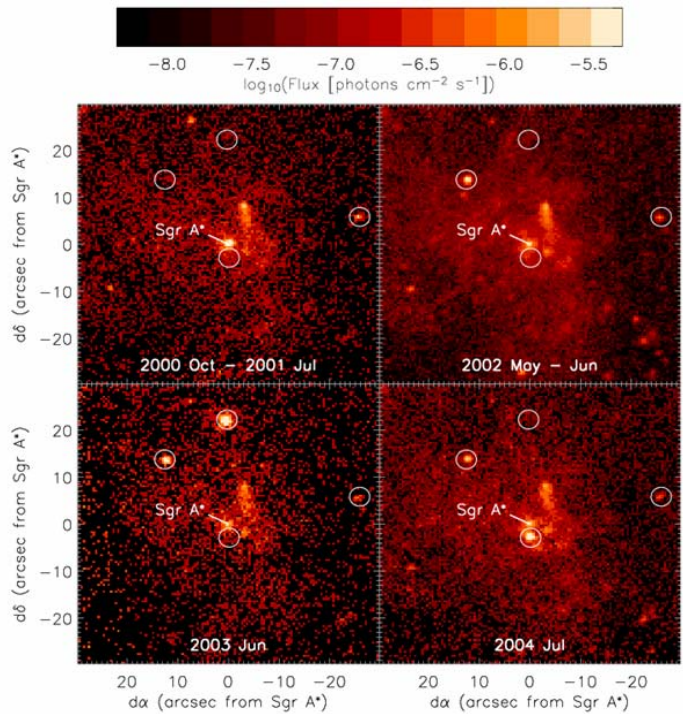

**Figure 2.4.3.** Multi-epoch Chandra observations of the Galactic Center showing the detection of four transient X-ray sources in five years (white circles). Adapted from Muno et al. (2005).

## 2.5 Stellar mass function

The distribution of stellar masses at birth, the so-called initial mass function (IMF), is well described in most Galactic environments by a power law ($dN(m_*) \sim m_*^{-\gamma} dm_*$) with exponent $\gamma = 2.3$ from $m_* \sim 1$ to $\sim 120$ M$_\odot$. Below this mass the IMF turns over and flattens (Salpeter 1955, Kroupa 2002, Chabrier 2001). It is of obvious general interest to ask whether in the unique environment around a massive black hole this 'universal' IMF holds as well.

There are several constraints on the IMF in the central parsec. The first (and strongest) comes from the number counts of young, massive stars as a function of K-band magnitude (the K-band luminosity function: KLF). Generally, the 'present day' stellar mass function (PMF) and its associated KLF depend on the IMF, as well as on the star formation history and the mapping of each star onto the KLF through stellar atmosphere modeling. The IMF is a reasonable approximation for the PMF inferred from the KLF when modeling (1) very recently born stars, or (2) a decaying, sharp star formation burst (for masses below a suitably adjusted turn-off mass), or (3) when modeling the top-most edge of the PMF in an old system with continuous star formation. The latter is the case because very massive stars have quite similar lifetimes. The rate at which stars 'accumulate' across the PMF due to the continuous star formation is then roughly independent of mass. The star disk(s) discussed above have a common and reasonably well-defined age of $6 \pm 2$ Myrs and thus match conditions (1) and (2) very well. Figure 2.5.1 shows different estimates of the KLF of the early-type stars in three different radial bins: the S-star cluster ($p \leq 0.8$"), the zone of the O/WR-star disks (0.8" < $p$ ≤ 12") and the region $p$ > 12" (all from Bartko et al. 2010). These KLFs and the corresponding best fit model IMFs (shown as histograms, under the assumption of in situ formation) rely on a deep spectroscopic survey with SINFONI, which has an average 50% spectrosopic and photometric completeness at K ~ 15. In the S-star cluster and a few deeper fields the spectroscopic data reach K ~ 16.5. The KLFs in Figure 2.5.1 were corrected for the spectroscopic, photometric and area incompleteness.

The KLF in the young star disk(s) is remarkably flat with a best-fit power-law PMF/IMF of slope γ = 0.45 ± 0.3, almost 2 dex flatter than a standard IMF. The Bartko et al. results confirm and put on a robust spectroscopic footing the earlier finding of Paumard et al. (2006) that *the PMF/IMF in the O/WR-star disk(s) is top-heavy*. The disk(s) contain fewer B- than O/WR-stars, while a Salpeter-Kroupa-Chabrier IMF would have 2.6 times as many $K_s$ ≤ 16 B-stars than O/WR-stars. The inferred top-heavy IMF in the O/WR-star disk(s) is in strong contrast to the S-star cluster ($p$ ≤ 0.8") and probably also the region outside the disk(s), where the KLF is well fit by a standard IMF for a single burst population or a continuous star formation history with a range of ages between 6 and a few tens of Myrs (Figure 2.5.1).

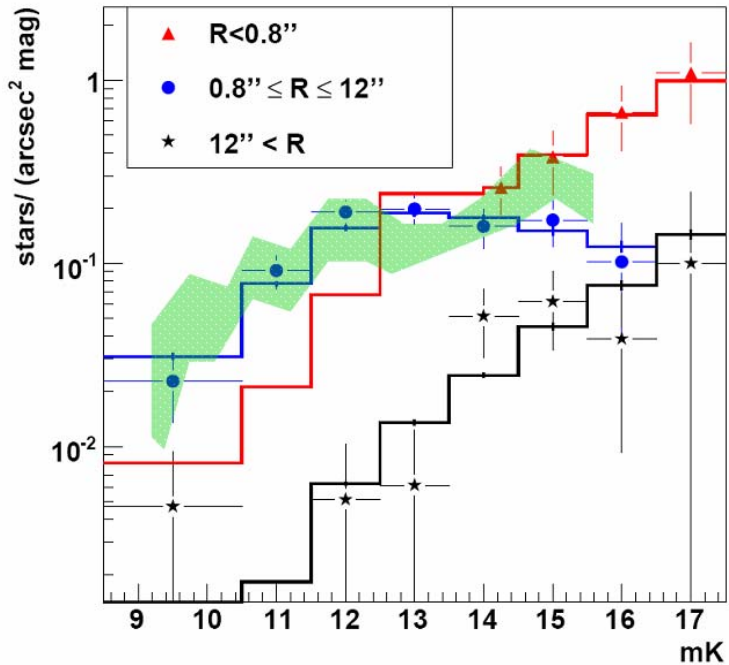

**Figure 2.5.1** Completeness corrected K-band luminosity functions of spectroscopic early-type stars in three radial intervals $p$ < 0.8″ (red triangles, scaled by a factor 0.05), 0.8″ ≤ $p$ ≤ 12″ (blue points) and 12″ < $p$ < 25″ (black asterisks). For comparison, the green shaded distribution is the KLF for p < 7" derived from the narrow-band spectro-photometric technique of Buchholz, Schödel & Eckart (2009, normalized to the spectroscopic data). This technique allows full aerial coverage of the central region and sets an upper limit to the number of fainter early-type stars. For a 6 Myr population the best fitting inferred IMF in the radial interval 0.8″ ≤ $p$ ≤ 12″ (blue histogram), where the disks of early-type stars are most prominent, is extremely top-heavy and clearly different from the IMFs of the S-stars (red histogram) and the field stars beyond 12" (black histogram). It can be fitted by a power-law IMF with a slope of 0.45 ± 0.3 or the IMF proposed by Bonnell & Rice (2008). The IMF of the field stars beyond 12" as well as the S-stars within 0.8" can be fitted by a power-law IMF with a slope of 2.15 ± 0.3, consistent with a standard Salpeter/Kroupa IMF. These latter KLFs can also be fitted by somewhat older populations with a continuous star formation history, which would then, however, predict lower numbers of K ≤ 14 stars (relevant for S-star cluster). All IMFs use a mass range of 0.1 to 120 $M_\odot$ (adapted from Bartko et al. 2010).

There are three main uncertainties affecting this conclusion. One is the mapping of the KLF into the PMF at the bright end ($K_s \leq 13$), where most stars are post-main sequence supergiants. In this range there is no unique relation between $K_s$-band magnitude and stellar mass and the transformation relies on population synthesis modeling. It is encouraging that the modeled mapping between KLF and PMF has been verified by more detailed spectral atmosphere modeling for about 20 of the O/WR-stars in Martins et al. (2007, see § 6.4). The second issue is the reliability of the area and spectroscopic completeness correction at the faint end ($K_s \sim 14.5$ to 16). The area coverage of the Bartko et al. (2010) data, as well as the probability and reliability of correct spectral identification both decrease with increasing $K_s$-magnitude. This may suggest that the $K_s > 14.5$ KLF-points underestimate the true distribution. This conclusion is supported qualitatively by the photometric analyses of Paumard et al. (2006) and Bucholz, Schödel & Eckart (2009), both of which indicate a larger number of faint early-type stellar candidates (green shaded distribution in Figure 2.5.1) than obtained in the spectroscopic technique of Bartko et al. (2010). However, the resulting possible increase in faint-end counts is modest (factor ≤ 2, Figure 2.5.1). In addition, a number of the $K_s \sim 15$ early-type candidates identified with the Bucholz et al. technique were targeted in the spectroscopic observations of Bartko et al. (2010) and turned out to be late-type stars. The green shaded KLF in Figure 2.5.1 thus represents an upper bound to the number of fainter early-type stars in the star disk(s). Finally some of the B-stars counted in the faintest bins ($K_s \geq 15$) of Figure 2.5.1 appear to not be situated within the disks but in a more isotropic distribution (Bartko et al. 2010). Taken together this suggests that possible upward corrections of the B-star counts of Bartko et al. (2010) may result in a somewhat steeper slope of the PMF but will remain significantly flatter than the standard IMF. The final and probably most important uncertainty is the relation between the PMF and the IMF in case the stars in the disk(s) have not formed in their current locations. We will return to this issue at the end of this section and in § 6.2.2.

The second constraint on the IMF in the Galactic Center comes from the X-ray emission in the central region. Nayakshin & Sunyaev (2005) compare the diffuse X-ray emission observed with Chandra in Sgr A* with that of the Orion nebula cluster. The IMF in Orion is known to be close to a Kroupa-Chabrier IMF. If the Galactic Center has a standard IMF its X-ray emission should then be larger than that in Orion by the ratio of OB stars in these two regions. The predicted diffuse X-ray flux in the central parsec exceeds the value observed by Baganoff et al. (2003) and Muno et al. (2004) by a factor of 20 to100. Since the X-ray emission of a young star cluster is dominated by low-mass T-Tauri stars, Nayakshin & Sunyaev conclude that the ratio of OB stars to low mass stars must be at least an order of magnitude larger in the Galactic Center than in Orion. This implies a very flat IMF, similar to the arguments based on the KLF in the last section. However, since T-Tau stars have a mass of

~ 1 $M_\odot$, the lack of X-ray emission could also be consistent with a more canonical IMF that is truncated below 1 − 2 $M_\odot$.

Could the PMF be different from the IMF because the massive young stars did not form in their present location but migrated inwards in a mass-dependent manner? The 'in-spiraling cluster' scenario for the origin of the star disks discussed in more detail in § 6.2 indeed predicts a fairly flat PMF for the innermost deposited stars although the assumed intrinsic IMF of the cluster is a standard one ($\gamma_{PMF}$ ~ 1.1, Gürkan & Rasio 2005). This is because mass segregation within the cluster leads to a differential shedding of stars of different masses when the cluster is tidally stripped during in-spiral. Lower mass stars are deposited further out, while the most massive stars sink the farthest in, especially if the cluster contains a central intermediate mass black hole (Gürkan & Rasio 2005). The finding by Bartko et al. (2010) of a flat PMF/KLF in the O/WR-disks, and a steeper KLF further out, is qualitatively consistent with such a model. However, the differential shedding of different stellar masses combined with the predicted flat surface density of the deposited cluster stars ($\Sigma \sim p^{-0.75}$, Berukoff & Hansen 2006) would require a large 'sea' of B-stars outside the location of the O/WR-disk(s). The ratio of $K_s$ < 16 B stars (8 − 20 $M_\odot$) to O stars (> 20 $M_\odot$) is 2.6 for a Kroupa/Chabrier IMF, but 10 times smaller (0.25) for $\gamma$ = 0.45. The ratio of the number of B-stars inferred from the radial surface distribution in Figure 3.3.2 ($\Sigma_B \sim p^{-1.4 \pm 0.2}$) to lie between $p$ ~ 12" and 24" to that observed in the region of the disks ($p \leq 12$") is 0.6 ± 0.2, by far not sufficient to make up the lack of B-stars for a normal IMF. In addition, the angular momentum distribution of B-stars at $p$ > 12" appears to be more isotropic. This means that the observed number of B-stars in the disk planes outside $p$ > 12" is still smaller. Finally in the simulations of Gürkan & Rasio (2005) the differential shedding is a fairly slow function of distance from the center. Unless there is a very much larger surface density of B-stars outside the region probed by the current observations, a top-heavy PMF at $p \leq 12$" probably is a reasonable estimate of the in situ IMF.

The third constraint on the IMF comes from the relatively small z-thickness of the disks if there are indeed two separate star disks. Nayakshin et al. (2006) have shown from numerical simulations that in this case, the mutual interaction induces warping and precession of their stellar orbits, leading to thickening and eventual destruction. If the disks were initially very thin the z-thickness of each disk depends on the time since formation and is proportional to the mass of the other disk. Nayakshin et al. infer that the observed thickness requires that the masses of the clockwise and counter-clockwise systems do not exceed 1 and 0.5 × $10^4$ $M_\odot$, in excellent agreement with the values estimated by Bartko et al. (2010) from direct integration of the observed KLFs. Since Nayakshin et al. adopted an age of 4 Myrs (instead of 6 Myrs) these upper limits should be quite conservative. Nayakshin et al. conclude that this consideration further supports the evidence in favor of a top-heavy IMF.

Löckmann et al. (2010) have pointed out that star formation in the nuclear cluster cannot always have been characterized by a strongly top heavy IMF throughout the last 10 Gyrs, at least not for R ≥ 1 pc. This is because otherwise the number of stellar remnants is too large and the predicted ratio of stellar mass (including remnants) to the faint-end K-band luminosity is significantly larger than the observed ratio, as inferred from the dynamical mass of the stellar cluster (Figure 5.1.1, equation 5) and the diffuse K-band light (Schödel et al. 2009). Pfuhl et al. (in prep.) agree with this conclusion, based on the observed number counts of faint stars (and especially the number of A-stars) in the central parsec.

Weaker evidence exists for a somewhat flatter than standard IMF in the young 'Arches' and 'Quintuplet' star clusters located ~ 30 pc from Sgr A* ($\tau$ ~ 2−4 Myrs, Morris & Serabyn 1996, Figer et al. 1999). Figer et al. (1999), Stolte et al. (2005) and Kim et al. (2006) find that the slope of the integrated PMF in these clusters ranges between $\gamma$ = 1.7 and 2 in the mass range 1 to 50 $M_\odot$, perhaps with a tendency of flattening at higher mass. The corresponding IMF slope then is $\gamma$ = 2, close to but perhaps somewhat flatter than the standard IMF (Kim et al. 2006). From another analysis of data on the Arches cluster, however, including a treatment of the effect of strong differential reddening on the inferred instrinsic photometry, Espinoza, Selman & Melnick (2009) find $\gamma$ = 2.1 ± 0.2, consistent with a Salpeter IMF to within the uncertainties.

In contrast to the picture we just have developed for the region outside the central S-star cluster and within $p \leq 12$", the KLF in the **S-star cluster and at p > 12"** appears to be very different (Figure 2.5.1). In these regions the observed KLF is consistent with the KLF of a young stellar component with a **normal IMF**. Normal IMFs with older ages but a continuous star formation history are also possible but for $t >$ 50 Myrs the more massive B-stars move off the main-sequence and the KLF becomes too steep (Bartko et al. 2010). This finding is of great interest for understanding the origin of the B-stars in the central cusp. The steep KLF would appear to speak against a scenario where the B-stars formed in the star disk(s) 6 Myrs ago and then migrated into the central cusp since that time. Instead it appears more likely that the B-stars in the cusp were formed over larger regions and longer times, perhaps with a close to universal IMF outside the sphere of influence of the massive black hole and then migrated into the central cusp.

In summary of this section the **evidence for a top-heavy IMF in the disk(s) of young O/WR-stars has become fairly strong**. While in most astrophysical environments where sufficiently detailed information is available, the concept of a universal IMF seems to hold true (Kroupa 2002), the sphere of influence of a massive black hole may favor the formation of O-stars. We come back to this issue when we discuss the theoretical concepts of star formation in this environment in § 6.

## 2.6 Chemical abundances

The Galactic Center is also an interesting laboratory for studying the evolution and enrichment of heavy elements by nucleosynthesis. For several decades there has been evidence for an increase of the interstellar oxygen to hydrogen abundance ratio with decreasing Galactocentric radius, from the outer to the inner Galaxy. Electron temperature determinations with radio recombination lines, optical forbidden line ratios and infrared fine structure line ratios all suggest an increase of α-element abundances (O/H, Ar/H, Ne/H) by a factor of 1.5 to 3 between the solar neighborhood and the Galactic Center region, albeit with large uncertainties and substantial scatter (left panel of Figure 2.6.1, Shaver et al. 1983, Lester et al. 1987, Simpson et al. 1995, Afflerbach, Churchwell & Werner 1997, Giveon et al. 2002). The central parsec fits in with this general picture. Lacy et al. (1980) were the first to conclude that Ne and Ar are enriched by a factor of 1.5 to 2 in the mini-spiral streamers of the Sgr A East

HII region. This evidence for elevated α-abundances in the ionized gas of Sgr A West has been confirmed by more recent observations and analyses, although Shields and Ferland (1994) find twice the solar values for N/O and Ar/H but solar abundance for Ne/H. The most recent analysis by Giveon et al. (2002) includes the most recent photoionization and stellar models and yields Ar/H and Ne/H abundances in the central parsec and in the Arches cluster about 1.5 solar (left panel of Figure 2.6.1). Is this somewhat super-solar abundance in the ionized interstellar gas also present in the Galactic Center stars?

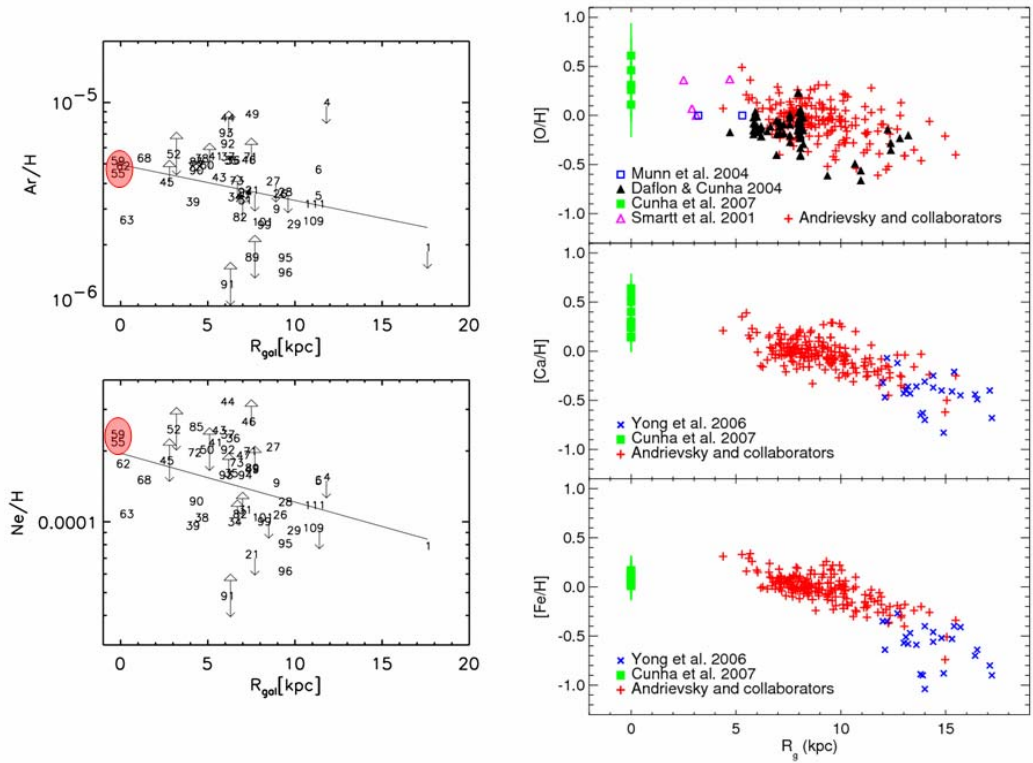

**Figure 2.6.1.** Chemical abundances in the Galactic Center. Left panel: α-element abundances in interstellar HII regions, as a function of Galactocentric distance, obtained from mid-infrared fine structure line ratios obtained with the ISO SWS and analyzed with the most recent generation of stellar atmosphere models. Each number refers to a single HII region. The red oval marks the Galactic Center (central parsec and the Arches region 30pc from Sgr A*, adapted from Giveon et al. 2002). Right panel: stellar α-element and Fe-abundances as a function of Galactocentric distance. Galactic Center stars (red supergiants and AGB stars in the central parsec, as well as a supergiant in the Quintuplet cluster 30 pc from Sgr A*) are marked green, along with the measurement uncertainties. While the Fe-abundance is solar throughout the central 10 kpc, the α-element abundances (and the α/Fe-ratio) appears to be super-solar by ~ 0.2 − 0.3 dex, perhaps consistent with the interstellar abundances (adapted from Cunha et al. 2007).

There is ample evidence for the release of heavy element (O, N, C, Si, Mg) nucleosynthesis products in the intense winds of the massive young stars throughout the central parsec and the Arches and Quintuplet clusters (Najarro et al. 1994, 1997, 2004, 2009, Figer et al. 1999, Martins et al. 2007, 2008a). Determination of the *initial* (ZAMS) element abundances in the hot stars is more difficult. Najarro et al. (2004) and Martins et al. (2008a) infer the initial O/H abundances of hot young stars in the Arches cluster from the maximum, asymptotic nitrogen abundances of the cluster's most evolved Wolf-Rayet stars. Both papers compare the spectroscopic data to the latest stellar evolutionary models and infer solar (Najarro et al.), or marginally super-solar ($Z/Z_\odot = 1.35 \pm 0.2$, Martins et al. 2007, 2008a), initial metal abundances. Martins et al. point out that this technique in essence gives only a lower limit to the initial metallicity since most or all of the stars in the sample may not have reached the asymptotic limit. In a study of two Luminous Blue Variables in the Quintuplet cluster Najarro et al. (2009) deduce Fe- and α-element (Si, Mg) abundances. These authors find Fe to hydrogen to be approximately solar, while Si and Mg appear to be super-solar by $0.3 \pm 0.2$ dex. Geballe et al. (2006) study the star IRS 8 in the northern arm of the mini-spiral in the central parsec. They conclude that the 2.116 μm feature, which is a blend of C III, N III and O III transitions, is sensitive to the oxygen abundance, and infer an O/H abundance that is about 2σ above the Asplund et al. (2005) solar abundance.

Maeda et al. (2002) and Sakano et al. (2004) carried out X-ray spectroscopy of the Sgr A East supernova remnant and also found strong evidence for α-enrichment (four times solar), perhaps not surprising, as this is a remnant of a recent (< a few $10^4$ yr) core-collapse supernova in the central few parsecs.

There have been a series of studies of heavy element abundances in red supergiants and AGB stars in the Galactic Center region, based on high-quality, high-resolution absorption line H/K-band spectra. Carr, Sellgren & Balachandran (2000) were the first to determine abundances in IRS 7, the brightest red supergiant in the central parsec. They found approximately solar values for Fe to within their quoted uncertainties, with a depletion of oxygen, probably as a result of internal CNO processing. Ramirez et al. (2000) analyze three bright red giants/AGB stars and six red supergiants in the central 2.5 pc (including IRS 7), as well as the red supergiant VR 5-7 in the Quintuplet cluster 30 pc north of Sgr A*. Compared to a solar neighborhood comparison sample the Fe-abundance in the Galactic Center sample is fully consistent with the solar value ($\Delta$(Fe/H) = $0.1 \pm 0.2$). Cunha et al. (2007) re-analyzed the Ramirez et al. (2000) sample after adding H-band spectra to the dataset. They confirm the solar Fe-abundances but also conclude that α-element abundances and thus, α/Fe-ratios (O/Fe, Ca/Fe) are $0.2 - 0.3$ dex above the solar values (right panel of Figure 2.6.1). The significance of this α/Fe-overabundance is ~ 2σ. Davies et al. (2009) analyzed an independent H-band data set of IRS 7 and VR 5-7. Again, they confirm solar Fe-abundances, and also find slight ($0.1 - 0.2$ dex) super-solar abundances of Si, Ca and Ti but solar values for Mg in both stars. The situation with respect to oxygen is puzzling. Oxygen is super-solar in VR 5-7 but substantially sub-solar in IRS 7 (see also Carr, Sellgren & Balachandran 2000), perhaps due to fast rotation and enhanced mass loss in this evolved supergiant. Davies et al. (2009) conclude that once uncertainties in data and modeling are properly taken into account, there is no evidence for super-solar α- and α/Fe-abundances in the two Galactic Center stars.

The emerging picture for the stellar abundances is somewhat confusing but reflects the scatter of the observational data. The subject is obviously tricky, given systematic uncertainties in data and modeling. If a global average of all stars, data sets and analyses are able to average out the substantial scatter from star to star and the influence of different analyses, there appears to be an overall trend toward a modest (factor 1.5 ± 0.25) enhancement of α-elements, but a solar iron abundance in the Galactic Center stars. The α/Fe-enhancement is very similar to that seen in the ionized interstellar gas. However, the comparison of the different papers we have discussed clearly shows that this result is overall not (yet) compelling, certainly when considering a specific star or HII region.

How would a super-solar α/Fe-abundance ratio have to be interpreted? The most likely possibilities lie in the properties and evolution of star formation in the Galactic Center. A super-solar α/Fe-abundance could have been created by a recent burst of star-formation creating massive stars, and in turn core-collapse supernovae. This explanation would work locally near OB-star forming sites. Yet Cunha et al. (2007) find the same O/Ca-overabundances in three older red giants. Further, the star formation history of the Galactic Center region is probably best described by relatively continuous star formation over the last few hundred Myrs (see § 6.1, Blum et al. 2003, Figer et al. 2004, Maness et al. 2007, Pfuhl et al. 2010).

An alternative explanation involves a top-heavy IMF, as discussed in the last section (Cunha et al. 2007). In this case there is a continuing enhanced rate of core collapse supernovae, pushing the α/Fe-ratio above that of the Galactic disk. Another aspect is the source of the gas that is collapsing to form new stars. If there is little influx of fresh gas from outside the Galactic Center, a significant fraction of the star forming gas is affected by winds from red giants/AGB stars, which tend to have high α/Fe ratios. Thus the natal material was already α-enriched before it arrived in various regions in the Galactic Center (Morris & Serabyn 1996).

# 3. Observed properties of the nuclear interstellar matter

The properties of the interstellar matter in the central parsec(s) have been previously discussed in some detail in the reviews of Genzel, Hollenbach & Townes (1994), Mezger, Duschl & Zylka (1996) and Serabyn & Morris (1996) to which we refer here for details. In the following we will concentrate on updates since these reviews. Figure 1.1 summarizes the spatial distribution and nomenclature of the gas components discussed below. The global properties of the interstellar material in the central parsecs can be described by a central ~ 1 − 1.5 pc radius '*ionized cavity*' devoid of much dense gas and surrounded by a set of streamers of dense molecular gas and warm dust, the ***circum-nuclear disk*** (CND), extending from 1.5 to ~ 4 pc (Becklin, Gatley & Werner 1982, Güsten et al. 1987, Jackson et al. 1993, Christopher et al. 2005). There is a sharp transition at ~ 1.5 pc between the ionized cavity and the CND. Outside of the CND there are several massive molecular clouds (the so-called '+20 km/s' and '+50 km/s'-clouds, Mezger, Duschl & Zylka 1996 and references therein).

## 3.1 Ionized gas in Sgr A West

To first order, the morphology and dynamics of the ionized gas in Sgr A West (the '*mini-spiral*') is well described by a system of ionized, clumpy streamers or filaments orbiting Sgr A* (Figure 1.1, Figure 3.1.1, Lo & Claussen 1983, Ekers et al. 1983, Serabyn & Lacy 1985, Serabyn et al. 1988, Schwarz, Bregman & van Gorkom 1989, Lacy, Achtermann & Serabyn 1991, Roberts & Goss 1993, Herbst et al. 1993, Roberts, Yusef-Zadeh & Goss 1996, Vollmer & Duschl 2002, Liszt 2003, Paumard, Maillard & Morris 2004, Zhao et al. 2009). Some of these streamers (e.g. the 'western arc') may be on mostly circular orbits, similar to the neutral gas in the CND (see below), while others (the 'northern' and 'eastern' arms and the 'bar') penetrate deep into the central ionized cavity, to within a few arcseconds from Sgr A*.

These streamers are almost certainly photoionized by the combined ultraviolet radiation from the massive stars in the young star disk(s) (Shields & Ferland 1994, Martins et al. 2007, § 6.4). The detailed morphology of the mini-spiral probably results from the combination of the physical distribution of the streamers with their illumination by the ionizing radiation (Paumard, Maillard & Morris 2004). Radio and infrared recombination line fluxes (and their ratios to the radio continuum flux) are well fit by electron temperatures between 5000 and 7500 K (Roberts & Goss 1993, Rieke, Rieke & Paul 1989, Maloney, Hollenbach & Townes 1992). The total mass of ionized gas in the central cavity is fairly small and amounts to only about ~ 25 $(n_e/10^4$ cm$^{-3})^{-1}$ M$_\odot$. In addition there is about 300 M$_\odot$ of neutral atomic gas (Jackson et al. 1993) and a few M$_\odot$ of warm dust (Davidson et al. 1992). Some or most of the central gas/dust may be associated with the mini-spiral filaments and the ionized gas may in part be the ionized surfaces/rims of the more massive gas clouds (Paumard, Maillard & Morris 2004). The average gas density in the central parsec thus is much lower than in the surrounding CND. This central gas cavity may be the result of the young massive star disks being in the post main-sequence 'wind' phase (Genzel, Hollenbach & Townes 1994, Morris & Serabyn 1996). The mini-spiral streamers represent an apparent mass inflow rate of ~ $10^{-3}$ $M_\odot \text{yr}^{-1}$ into the central few arcseconds (Genzel, Hollenbach & Townes 1994).

With the availability of proper motions from multi-epoch VLA observations and including H96α spectroscopic data, it is now possible to derive unique orbital solutions for the main gas streamers. The most recent work of Zhao et al. (2009) confirms earlier inferences from near- and mid-infrared imaging spectroscopy that the western arc is the inner ionized edge of the neutral circum-nuclear disk and has an orbit that is close to circular, while the northern arm and eastern arm (+ bar) are on highly elliptical orbits. Zhao et al. also strengthen the earlier conclusions of Vollmer & Duschl (2000), Liszt (2003) and Paumard, Maillard & Morris (2004) that all these features and the CND form a bundle of orbits with the same inclination relative to the plane of the sky. The northern arm and western arc (+ CND) also may be in the same plane, which prompted Lacy et al. (1991) to propose that both features are part of a single gas flow with dissipative loss of angular momentum along its length (perhaps as a result of frictional losses). The most recent data of Zhao et al. probably still permit this model but favor two distinct kinematic features. Northern arm and eastern arm gas may collide in the region of the bar (several arcseconds behind the plane of the sky). There are significant deviations from the simple single Keplerian orbit fits. Given the compelling evidence from infrared polarization and radio Zeeman effect measurements for an ordered mG-magnetic field in the central few parsecs (Aitken et al. 1991, Hildebrand et al. 1993, Morris & Serabyn 1996 and references therein), these deviations and the morphology of some of the filaments (e.g. braided structures, Zhao et al. 2009) may make a good case in favor of a significant impact of magnetic fields on the ionized gas dynamics. We refer to Morris & Serabyn (1996) for a detailed discussion on the role of magnetic fields in the Galactic Center.

While the dynamics of most of the mass of the ionized gas in the mini-spiral may be accounted for by ***tidally stretched gas filaments orbiting*** in the central gravitational potential, some of the ionized gas shows much larger motions (up to ~ 700 km/s) and complex patterns (Yusef-Zadeh, Roberts & Biretta 1998, Zhao & Goss 1998, Zhao et al. 2009). These peculiar motions may be the result of the ionized gas interacting with the orbiting stars and their stellar winds, or of cloud-cloud collisions (Morris & Yusef-Zadeh 1987, Genzel, Hollenbach & Townes 1994). Star-wind-gas interactions are documented in about a dozen cases as bow shocks in gas (and warm dust) distributions pointing opposite to the motion of bright stars centered on/near the gas clumps (Serabyn, Lacy & Achtermann 1991, Yusef-Zadeh & Melia 1992, Tanner et al. 2002, 2005, Paumard, Maillard & Morris 2004, Genzel et al. 2003a, Geballe et al. 2004, 2006, Viehmann et al. 2006, Perger et al. 2008, Zhao et al. 2009). If the plowing star is a Wolf-Rayet star with a powerful wind, these interactions can create bright radio and mid-infrared dust emission sources (IRS 1, 5, 8, 10, 13E, 21, 33: Zhao & Goss 1998, Tanner et al. 2002, 2005, Paumard, Maillard & Morris 2004, Muzic et al. 2008, Zhao et al. 2009). The case of IRS 13E is especially dramatic as this star-wind interaction region also is a prominent source of X-ray emission (Baganoff et al. 2003), which can be modeled by a wind-wind collision region (Coker, Pittard & Kastner 2002).

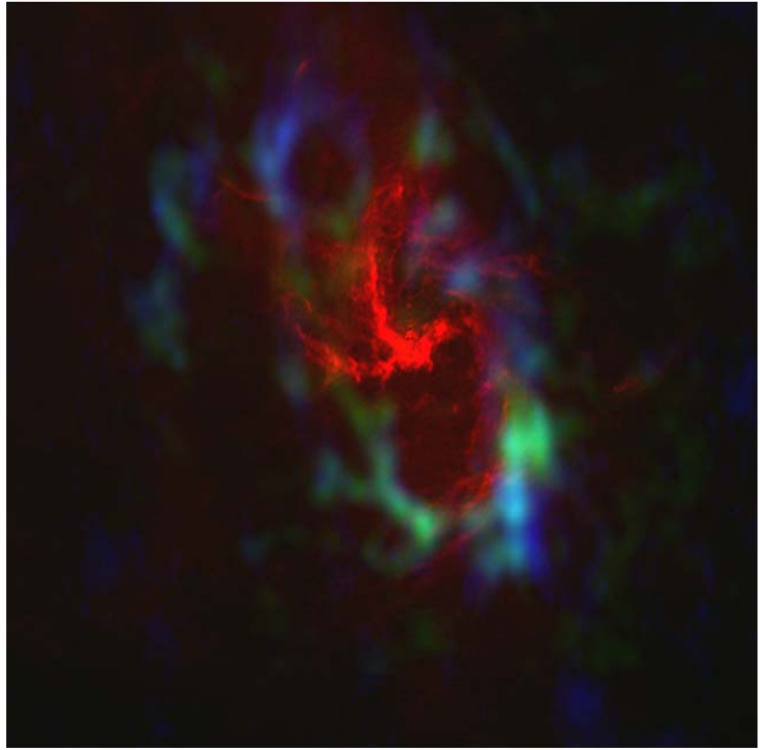

**Figure 3.1.1.** Superposition of integrated line emission in HCN 4-3 (green: SMA, Montero-Castano, Herrnstein & Ho 2009) and HCN 1-0 (blue: OVRO, Christopher et al. 2005), both at a resultion of 3" × 5", as well as 6 cm radio continuum emission (red: Lo & Claussen 1983, Ekers et al. 1983, Yusef-Zadeh, priv. comm., resolution ~ 1").

### 3.2 Neutral gas

The central 5 parsecs of the Galaxy contain a concentration of dense and warm, molecular and atomic gas and dust that is commonly referred to as the 'circum-nuclear disk' (CND: Figures 1.1 and 3.1.1, Becklin, Gatley & Werner 1982, Liszt et al. 1983, Genzel et al. 1985, Serabyn et al. 1986, Güsten et al. 1987, Jackson et al. 1993, Christopher et al. 2005, Montero-Castano, Herrnstein & Ho 2009). The most prominent feature is a rotating disk or set of filaments at inclination 60° - 70° (inclined 20° − 30° relative to the Galactic Plane) that can be followed for a good fraction of the circumference at $R \sim 1.5$ pc. The western arc (Figure 3.1.1) appears to be the ionized inner surface of the CND (Serabyn and Lacy 1985, Genzel et al 1985). The HCN clumps are also bright in vibrationally excited near-infrared $H_2$ line emission, suggesting that they are illuminated by far-ultraviolet radiation and/or are the sites of dissipative shocks (Gatley et al. 1986, Yusef-Zadeh et al. 2001). Figure 3.1.1 shows that the distributions of ionized and molecular gas are spatially anticorrelated. The northern arm appears to penetrate into the central cavity from outside through a gap in the inner edge of the CND. Some clumps and streamers may have large non-circular motions.

The CND clumps have large velocity dispersions ($\sigma = 17 \pm 8$ km/s), are very dense ($n(H_2) \sim 10^{6...8}$ cm$^{-3}$) and have moderately high gas and dust temperatures ($T_{gas}$ ~ 50 − 200 K, $T_{dust}$ ~ 50 − 70 K, Becklin, Gatley & Werner 1982, Güsten et al. 1987, Davidson et al. 1992, Jackson et al. 1993, Mezger, Duschl & Zylka 1996, Telesco et al. 1996, Latvakoski et al. 1999, Christopher et al. 2005, Montero-Castano, Herrnstein & Ho 2009). It is not clear whether the large line widths are caused by gravity induced, virialized motions, or by a transient turbulent cascade, as in Galactic disk GMCs (McKee & Ostriker 2007, Krumholz & McKee 2005). In the former case the clump densities would exceed the local Roche density at $R$ ~ 1.5 pc (~ $3 \times 10^7$ cm$^{-3}$) and the clumps would be stable against tidal disruption. The amount of gas in the CND is also quite uncertain. If the CND clumps (typical diameters ~ 0.2 − 0.3 pc) are virialized and stable, $H_2$ columns and masses must be very large (<$N(H_2)$>$_{clump}$ ~ $3 \times 10^{25}$ ($A_V$ ~ $10^4$), <$M_{clump}$> ~ 1 − $2 \times 10^4$ M$_\odot$) and the total CND mass may be a few $10^5$ to $10^6$ M$_\odot$ (Christopher et al. 2005, Montero-Castano, Herrnstein & Ho 2009). The far-IR to millimeter dust continuum emission from the CND is quite faint, however, suggesting a total CND gas mass of a few $10^4$ M$_\odot$ with standard (Galactic disk) submillimeter dust opacities and gas to dust ratios (Mezger et al. 1989, Davidson et al. 1992, Dent et al. 1993, Mezger et al. 1996). In that case the clumps and filaments, as well as the entire CND structure, may be transient features (Güsten et al. 1987). If the mass of the CND is near the upper range of the estimates discussed above, the CND has a significant impact on the stellar precession time scale in the central parsec (Figure 2.3.3). Subr, Schovancova & Kroupa (2009) note that if mass of the CND were larger than $10^6$ M$_\odot$, the resulting gravitational torque would have destroyed the disk of young massive stars within its age of 6 Myrs, supporting the idea of a light and/or transient CND.

There is presently no conclusive evidence confirming or excluding either the 'virial' or 'transient' scenarios. The line ratios of H$^{12}$CN (and other heavy top molecules) and its isotopologue H$^{13}$CN in different rotational transitions (J = 1-0, 3-2, 4-3) are sensitive to the local gas density. The available observations can be fit either by sub-Roche density ($10^6$ cm$^{-3}$), warmer (≥ 100 K) gas, or by super-Roche density (~ $10^8$ cm$^{-3}$) gas with a range of somewhat lower temperatures (50 − 100 K) (Wright, Marr & Backer 1989, Jackson et al. 1993, Christopher et al. 2005, Montero-Castano, Herrnstein & Ho 2009). The {HCN}/{H$_2$} abundance is a free parameter and mass estimates directly based on the HCN-line fluxes thus are very uncertain. While the faint submillimeter dust emission favors the lower column density, 'transient' solution, higher column densities may in principle be accommodated if dust emission properties in the CND deviate from the Galactic disk, or if the submillimeter emission is somewhat optically thick (as indeed is predicted in the 'virial' scenario). However, the HCN clumps should then be visible as 'black' spots ($A_K$ ~ 100 − 1000) on the near-infrared images. While the outline of the southern part of the CND is indeed recognizable as a clear drop in K-band surface brightness in the top left image of Figure 1.1, individual clumps are not (Mezger et al. 1996).

In the 'virial' scenario the clumps may then collapse and form stars locally. Near-infrared spectroscopic searches for early-type stars in or near the CND clumps have not been successful so far (Martins, priv. comm.) but, given the high extinctions implied, may not have been deep enough to exclude that some young stars have formed there. Latvakoski et al. (1999) find a few local dust temperature peaks in the mid-infrared dust emission distribution of the CND, which may be suggestive of intrinsic heating sources. Latvakoski et al. also conclude that the mid-infrared emission at a resolution of about 5" is optically thin. Yusef-Zadeh et al. (2008b) have found $H_2O$ and collisionally excited $CH_3OH$ masers toward nine locations in the CND clumps. These masers may be signposts for very recently formed, embedded massive stars, as elsewhere in the Galaxy, although some of the $H_2O$ masers may also originate in the envelopes of evolved, late-type stars.

The sharp inner edge of the CND and the large drop in average gas density within ~ 1.5 pc is undoubtedly a ***transient feature in the interplay between gas inflow from further out, angular momentum dissipation, and radiation and ram pressure*** from the central massive star cluster (Güsten et al. 1987). The current low gas density in the central cavity may be the result of the nuclear star cluster currently being in a post-main-sequence wind-dominated phase of the star formation episode 6 Myrs ago. Once the O/WR-stars have disappeared, the gas accretion into the central parsec may increase to a few times $10^{-2}$ $M_\odot$ $yr^{-1}$, as estimated from the gas mass and non-circular motions in the CND. Increased gas accretion may then lead to another star formation episode in the future, resulting in a continuous but episodic limit-cycle (Morris & Serabyn 1996).

## 3.3 Dust and interstellar extinction toward the Galactic Center

The seminal spectrophotometric studies of stars by Becklin & Neugebauer (1968), Becklin et al. (1978), Rieke & Lebofsky (1985) and Rieke, Rieke & Paul (1989) established that the Galactic Center is situated behind ~ 30 magnitudes of visible extinction, mainly due to diffuse Galactic dust along the line of sight. The near-infrared extinction curve falls of steeply with wavelength ($A(\lambda) \sim \lambda^{-1.7}$ at $\lambda \sim 1 - 2.5$ µm, Figure 3.3.1), with a $K_s$-band extinction toward Sgr A* of $3 \pm 0.2$ mag. This early work also indicated that this steep dropoff continues to $7 - 8$ µm but then the extinction increases again sharply in two spectral features near 10 and 20 µm. The empirical curve appears to agree broadly with the extinction curve derived in the Orion star forming region (Rosenthal et al. 2000) and with theoretical dust models of mixtures of silicate and graphite dust grains with a standard ratio of B-V reddening to total extinction ($R_V = A_V/E(B-V) = 3.1$, Draine 1989, Weingartner & Draine 2001, Draine 2003).

More recent work by Lutz et al. (1996), Lutz (1999) and Scoville et al. (2003) used near- and mid-infrared hydrogen recombination line ratios and calibrated with case B recombination theory either to an assumed $K_s$-extinction, or to the free-free radio continuum emission in the same aperture. These observations confirm ***the steep near-infrared slope and the high extinction near the silicate features but indicate a much shallower slope in between, perhaps with additional extinction maxima near 3 and 6 µm*** (Figure 3.3.1). Spitzer spectro-photometry on larger spatial scales by Nishiyama et al. (2009b) has since confirmed the $2 - 8$ µm extinction curve inferred by Lutz et al.

The shallow $2.5 - 8$ µm slope can be explained by a grayer dust model ($R_V = 5.5$, Weingartner & Draine 2001). Alternatively the dust of the line of sight toward the Galactic Center may have ice mantles with water, $CO_2$ and organic ices surrounding silicate and graphite cores. Such ice mantles would be naturally expected in cold molecular clouds that are known from molecular line spectroscopy to reside along the line of sight to the Galactic Center. Direct evidence for these ice features are also seen in the ISO-SWS spectrum of Sgr A* displayed in the right panel of Figure 3.3.1 (Lutz 1999).

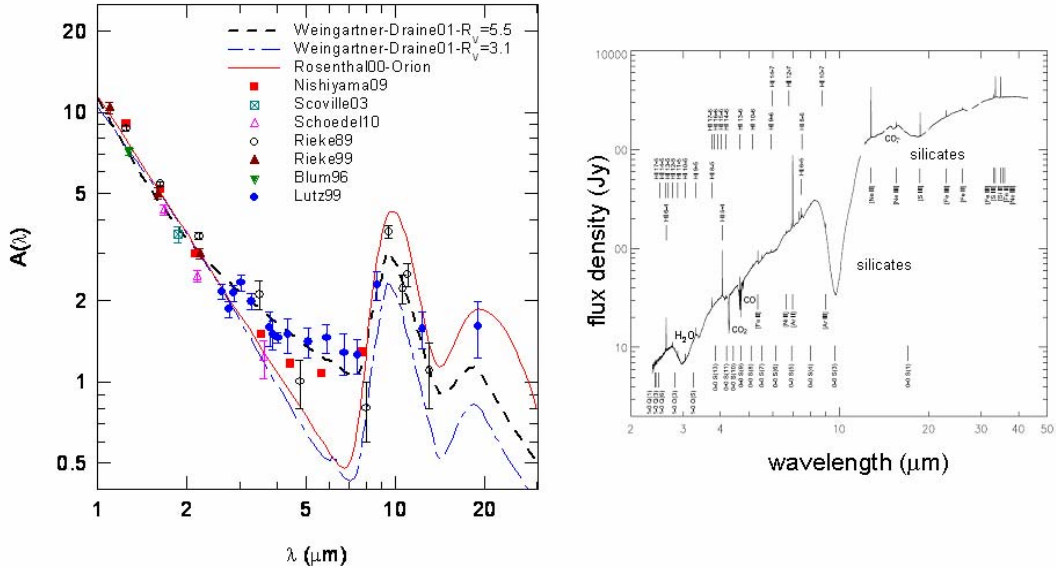

**Figure 3.3.1.** Left: compilation of near- and mid-infrared extinction measurements toward the central few arcseconds around Sgr A* from the literature, as well as two theoretical dust models from Weingartner & Draine (2001), for different choices of $R_V = A_V/E(B−V)$. The data points of Rieke, Rieke & Paul (1989, open black circles), Blum et al. 1996 (filled green triangles), Rieke (1999), Schödel et al. (2010, open magenta triangles) and Nishiyama et al. (2009b) rely mainly on the analysis of stellar spectrophotometry with known intrinsic spectral energy distributions. In contrast, the work of Scoville et al. (2003, open crossed square) and Lutz et al. (1996, 1999, filled blue circles) derived extinction from hydrogen recombination line ratios (and radio continuum) in comparison to case B recombination. For comparison, the magenta curve denotes the extinction curve derived for the Orion star-forming region by Rosenthal et al. (2000). Right: Near- to mid-infrared SED of Sgr A* from ~ 20" resolution SWS-spectroscopy with the Infrared Space Observatory (Lutz 1999), directly showing the silicate and ice dust absorption features ($H_2O$, $CO_2$, 6.5 μm etc.) that probably cause the flat infrared extinction curve on the left.

Figure 1.1 shows that there is also a component of spatially variable local extinction. The line of sight toward Sgr A* is a local minimum in extinction and the extinction rises on average by about 0.5 magnitude from Sgr A* to $p \sim 5$" (Schödel et al. 2007). In addition, the mini-spiral and even more strongly the CND contain dust that can be clearly seen by its imprint on extinction of the stellar light (Figure 1.1, Scoville et al. 2003). The dense +20 km/s cloud a few arcminutes south-east of Sgr A* has an even stronger impact and blocks out much of the stellar near-infrared light (Mezger et al. 1996).

## 3.4 Hot gas and high energy emission

The central parsec(s) also contain(s) several components of diffuse, hard X-ray line and continuum emission with equivalent temperatures of $10^7$ to $10^8$ K ($1 - 10$ keV, after subtraction of individual point sources, Muno et al. 2004, for a recent review see Goldwurm 2010). On scales of $10 - 30$ pc the 6.4 keV 'K$\alpha$'-line emission from 'neutral' Fe and some of the hard X-ray continuum shows a strong spatial correlation with large clouds of molecular line emission, such as Sgr B2 (Koyama et al. 1996, Sunyaev et al. 1993, Koyama et al. 2003). The most likely interpretation of this component is the ***Compton reflection/fluorescence of cold gas irradiated by (a) time variable hard X-ray continuum source***(s), which might be identified with Sgr A* (Koyama et al. 2003, Revnivtsev et al. 2004, Muno et al. 2007). Alternatively, the emission may be generated by the interaction of cold gas with energetic particles (Valinia et al. 2000, Predehl et al. 2003, Goldwurm 2010).

The more extended, 'Galactic ridge' continuum and the highly ionized Fe 6.7 keV line emission may either be diffuse gas or a superposition of many unresolved point sources (Muno et al. 2004). Suzaku observations support an interpretation of the diffuse X-ray emission in terms of distributed plasma with a range of temperatures (Yuasa et al. 2008, Koyama et al. 2009). The cooler plasma (~ 1 keV) exhibits a spatial structure on arcminute or smaller scales, and is plausibly the result of the interaction of stellar winds with each other and of stellar winds and expanding supernova remnants with cold and ionized gas (Muno et al. 2004, Baganoff et al. 2003). The hot component (~ 10 keV) is spatially more uniform (Yuasa et al. 2008) but broadly correlated with the softer component, perhaps suggesting a related physical origin (Muno et al. 2004). Because of its high sound speed ~ 10 keV diffuse gas would not be bound to the Galactic Center region. Its temperature is above that usually associated with stellar winds and supernove remnants, and its origin may be magnetic reconnection or acceleration in supernova shocks. To maintain such a component in dynamical equilibrium requires $10^{40..43}$ erg s$^{-1}$ (Muno et al. 2004, Revnivtsev, Molkov & Sazonov 2006, Revnivtsev et al. 2009). These energy requirements are very large. Alternatively, the hard X-ray continuum emission may be due to the superposition of many unidentified compact sources, such as cataclysmic variables or coronally active binary sources (Muno et al. 2004, Revnivtsev, Molkov & Sazonov 2006). Revnivtsev et al. (2009) have presented out very deep Chandra imaging of the Galactic ridge emission near the Center, but ~ 1.4° below the Galactic plane where the interstellar absorption and confusion is mimimized. This observation resolves > 80% of the ~ $6 - 7$ keV line and continuum emission into ***discrete sources*** and thus makes a very strong case for the compact stellar sources interpretation.

In the central parsec, the diffuse emission is consistent with thermal emission from ~ 1 keV plasma with a mean electron density of ~ $30 f^{-1/2}$ cm$^{-3}$, where $f$ is the filling factor of the emission (Figure 1.1, Baganoff et al. 2003). In addition there is a compact source associated with Sgr A* itself (§ 7, Baganoff et al. 2001, 2003, Xu et al. 2006), as well as a number of compact stellar sources (Muno et al. 2009) and several transients (§ 2.4, Muno et al. 2005), There are local emission maxima on IRS 13E (§ 4.7) and a north-south 'streak' of emission located ~ 7" arcseconds north-west of Sgr A* (Baganoff et al. 2003). The latter is not obviously related to any feature in the radio of infrared maps and may be a pulsar wind-nebula (Muno et al. 2008). The X-ray emission in Sgr A West is connected to more extended emission from the entire Sgr A complex and the supernova remnant Sgr A East (Baganoff et al. 2003).

Ground-based Cerenkov Telescopes have detected a counterpart of the Sgr A complex (Aharonian et al. 2004, Kosack et al. 2004, Albert et al. 2006). The currently most sensitive and highest angular resolution γ-ray observations are from the HESS observatory, reporting the detection of a point-like TeV source 7" ± 10" from Sgr A* (van Eldik 2008, Acero et al. 2009). The luminosity of this γ-ray source is about $10^{35}$ erg/s in the TeV range and its spectrum follows a power-law $\nu L_\nu \sim \nu^{-0.25}$ between 0.2 and 9 TeV (Aharonian et al. 2004, van Eldik 2008). The TeV emission might be created in the relativistic accretion zone, in shocks or in dark matter annihilation in the vicinity of Sgr A* (Bergström 2000, Atoyan & Dermer 2004, Aharonian & Neronov 2005), originate from accelerated relativistic particles in the supernova remnant Sgr A East (Aharonian et al. 2004), or could come from inverse Compton upscattering of infrared photons by relativistic electrons in the bright pulsar wind nebula 359.95-0.04 located ~ 9" north-west of Sgr A* (Figure 1.1, Wang 2006, Muno et al. 2007, van Eldik 2008, Acero et al. 2009). Given the lack of TeV time variability even during an X-ray flare (Aharonian et al. 2008) and its location in the central 10" the ***TeV source near SgrA\* may simply be due to the pulsar wind nebula*** (van Eldik 2008).

# 4. Testing the black hole paradigm: is Sgr A* a massive black hole?

It is now well established that the mass distribution in the central (few) parsec(s) can be well described by a ***combination of a central compact mass associated with Sgr A*, and a dense nuclear star cluster*** that is embedded in the larger scale bulge and disk of the Milky Way (c.f. Genzel & Townes 1987, Genzel, Hollenbach & Townes 1994, Mezger, Duschl & Zylka 1996, Melia & Falcke & 2001, Reid 2009). In this and the next chapter we discuss the very significant progress over the last two decades in our quantitative knowledge of the relative importance and masses of these two components, as well as the nature of the central mass.

## 4.1 Evidence for a central compact mass from gas motions

The first dynamical evidence for a central mass concentration emerged in the late 1970s and early 1980s, when the group of C.H. Townes at the University of California, Berkeley, discovered that the radial velocities of ionized gas (in the 12.8 μm line of [NeII]) increase to a few hundred km/s in the central parsec of our Milky Way (Wollman et al. 1977, Figure 4.1.1). Applying a virial analysis to these gas velocities suggested the presence of a central mass concentration of 2 − 4 million solar masses in the central parsec, in excess of what can be plausibly assigned to stars. The Berkeley group concluded that this mass concentration might be a massive black hole plausibly associated with the compact radio source Sgr A* (Lacy et al. 1980, Lacy, Townes & Hollenbach 1982).

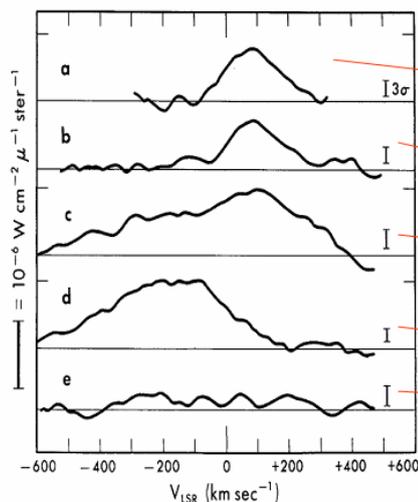
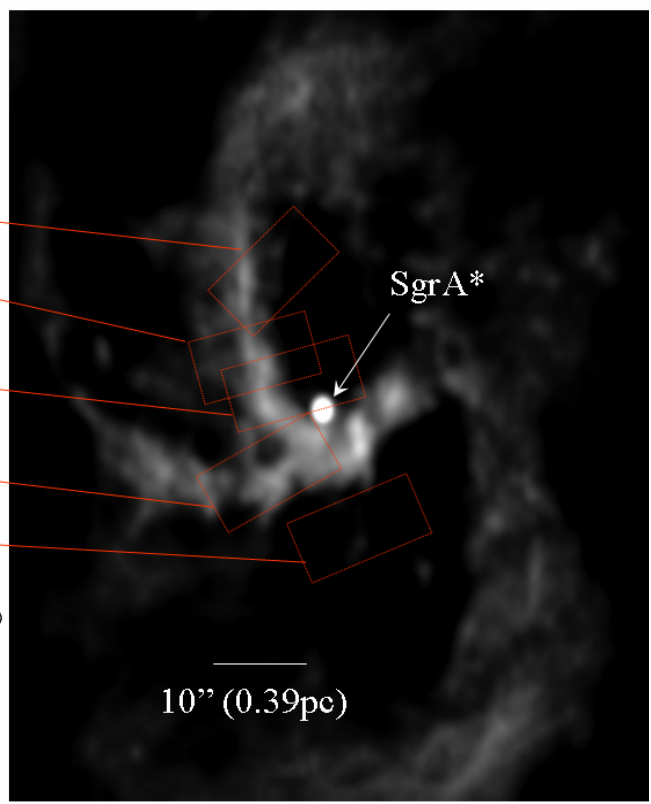

**Figure 4.1.1.** 3.6 cm VLA radio continuum map of the central parsec (right, Roberts & Goss 1993), and 12.8 μm [NeII] line profiles from Wollman et al. (1977, left) for the apertures indicated on the radio map. The radio emission delineates ionized gas streamers (the 'mini-spiral') orbiting the compact radio source Sgr A*. The Wollman et al. observations provided the first dynamic evidence from large gas velocities that there might be a hidden mass of 2 to $4 \times 10^6$ $M_\odot$ located near Sgr A*.

The case further improved throughout the 1980s with the advent of increasingly more detailed measurements of the ionized gas within the central ionized 'cavity', as well as of atomic and molecular gas outside the central parsec (Serabyn & Lacy 1985, Crawford et al. 1985, Mezger & Wink 1986, Serabyn et al. 1988, Güsten et al. 1987, Schwarz, Bregman & van Gorkom 1989, Lacy, Achtermann & Serabyn 1991, Jackson et al. 1993, Herbst et al. 1993, Roberts, Yusef-Zadeh & Goss 1996). However, *many considered the evidence for a central mass concentration based on gas dynamics unconvincing, mainly because gas is sensitive to forces other than gravity*, and also because there was no accompanying detection of a luminous infrared or X-ray source associated with Sgr A* (Rees 1982, Allen & Sanders 1986, Kormendy & Richstone 1995). Further progress required stellar dynamics.

## 4.2 Evidence from stellar motions

The first stellar velocity dispersion measurements came from spectroscopy of the 2 μm CO overtone absorption bands in late-type giants and supergiants (from $p \sim 10"$ to 3', Rieke & Rieke 1988, McGinn et al. 1989, Sellgren et al. 1990, Haller et al. 1996) and from 18 cm OH maser emission stars (Lindqvist, Habing & Winnberg 1992). These data broadly confirmed the gas measurements but sampled the mass distribution at too large a radius to give a conclusive constraint on the existence of a central non-stellar mass. At around the same time Forrest et al. (1987), Allen, Hyland & Hiller (1990) and Krabbe et al. (1991) detected the first blue supergiants in the central 10". Krabbe et al. (1995) used about a dozen of these stars for a virial estimate of the mass within 10" of Sgr A*, strengthening the evidence in favor of a few million solar mass central, non-stellar mass.

Genzel et al. (1996) and Haller et al. (1996) carried out a more quantitative analysis of the nuclear mass distribution from ~0.1 pc (2.5") to a few parsecs (~1'), based on radial velocities of ~ 200 late-type stars and two dozen early-type stars available at that time. Employing a combination of statistical projected mass estimators (virial and Bahcall-Tremaine (1981) estimators), as well as statistical modeling with the Jeans equation, separately for the early- and late-type stellar components, Genzel et al. (1996) inferred a combination of a $3.0 \times 10^6$ $M_\odot$ central mass and a $10^6$ $M_\odot$ star cluster with a core radius of ~ 0.4 pc (~ 10"). The resulting average stellar density in the core is ~ $10^6$ $M_\odot pc^{-3}$. Haller et al. (1996) employed Jeans modeling as well and arrived at a combination of a $1.5 - 1.8 \times 10^6$ $M_\odot$ central mass, plus a stellar cluster with a mass of $1 - 2.5 \times 10^6$ $M_\odot$ (all values are re-scaled from the published values to the $R_0$ = 8.3 kpc distance used in this paper).

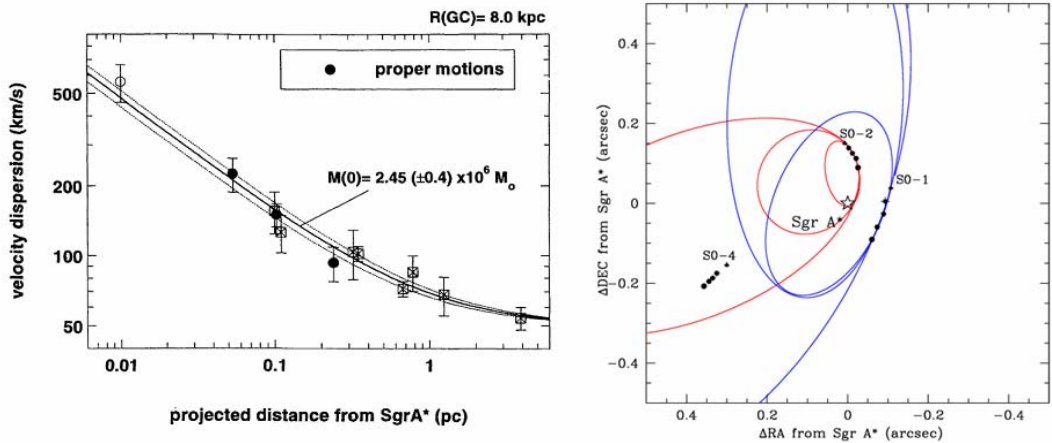

**Figure 4.2.1.** Stellar motions in the immediate vicinity of Sgr A*. Left: stellar velocity dispersion as a function of projected separation from Sgr A* (Eckart & Genzel 1996, 1997, Genzel et al. 1997). Circles are data derived from proper motions, crossed squares from line of sight velocities. The best fitting point mass model and its 1σ uncertainty are shown as continuous curves. Right: first detections of orbital accelerations for the stars S1 (S01), S2 (S02) and S8 (S04) and inferred possible orbits (from Ghez et al. 2000).

A major breakthrough occured with the first detections of stellar proper motions in the central few arcseconds and, in particular, with the determination of proper motions of the fast moving (up to ~ $10^3$ km/s) 'S'-stars within ≤ 1" of Sgr A*, based on near-infrared speckle imaging observations with the 3.5 m ESO NTT since 1992 (Eckart & Genzel 1996, 1997, Genzel et al.1997). Ghez et al. (1998) confirmed and improved these results by higher resolution speckle imaging with the 10m Keck telescope (since 1995). Both data sets showed that ***velocity dispersion of the stars follows a Kepler-law around a compact mass*** ($\sigma(v) \sim R^{-1/2}$) to a scale of about 0.01 pc (0.4 light months, Figure 4.2.1). The case for a compact central mass was then considered convincing by most practitioners in the field (c.f. Kormendy 2004). The number and quality of proper motions in the S-star cluster rapidly improved in the following years but the lack of z-coordinates initially dictated a statistical approach. This inevitably limits the innermost scale to the average radius of the S-star cluster (~ 0.5" or 0.02 pc) and the significance of the results because of the Poisson noise for a modest sample of stars.

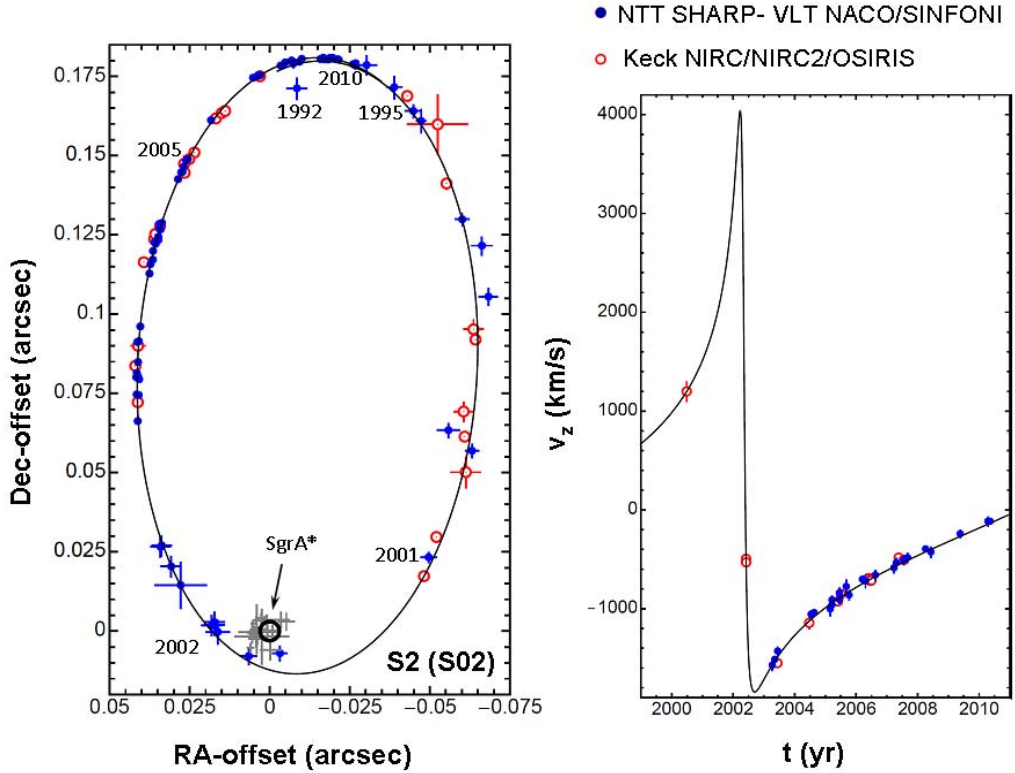

**Figure 4.3.1.** Orbit of the star S2 (S02) on the sky (left panel) and in radial velocity (right panel). Blue, filled circles denote the NTT/VLT points of Gillessen et al. (2009a,b, updated to 2010), and open and filled red circles are the Keck data of Ghez et al. (2008) corrected for the difference in coordinate system definition (Gillessen et al. 2009a). The positions are relative to the radio position of Sgr A* (black circle). The grey crosses are the positions of various Sgr A* IR-flares (§ 7). The center of mass as deduced from the orbit lies within the black circle. The orbit figure is not a closed ellipse since the best fitting model ascribes a small proper motion to the point mass, which is consistent with the uncertainties of the current IR-frame definition. Adapted from Gillessen et al. (2009a).

## 4.3 Constraints from stellar orbits

It was clear that the next big step would come from the determination of individual stellar orbits (Genzel & Eckart 1999, Fragile & Matthews 2000, Ghez et al. 2000, Rubilar & Eckart 2001, Eckart et al. 2002). The first success toward this goal was the detection of accelerations for three S-stars (Ghez et al. 2000, right panel of Figure 4.2.1, Eckart et al. 2002). The breakthrough came with the ***unambiguous determination of the first orbit of the star S2*** (S02, Schödel et al. 2002, Ghez et al. 2003) revolving with a period of 15.8 years. Since S2 is on a highly elliptical orbit with $e = 0.88$, its peri-center distance from Sgr A* in spring 2002 was a mere 17 light hours, or 1400 $R_S$ for a $4.4 \times 10^6$ $M_\odot$ black hole (Figure 4.3.1). The data from the NTT/VLT and Keck telescopes agreed very well: the first orbital analyses gave $4.1 \times 10^6$ (Schödel et al. 2002) and $4.6 \times 10^6$ $M_\odot$ (Ghez et al. 2003, both re-scaled to $R_0 = 8.3$ kpc), in agreement with each other to within the uncertainties, and with the statistical estimates at larger radii.

The most recent work on the S-star orbits (Schödel et al. 2003, Ghez et al. 2005b, Eisenhauer et al. 2005, Ghez et al. 2008, Gillessen et al. 2009a, b) has corroborated and further substantiated this evidence. Many more and higher quality data are now available, mainly due to the advent or high quality adaptive optics (AO) data, based either on near-infrared wavefront sensing, or on laser guide star AO. High quality radial velocities have been obtained for roughly two dozen of the S-stars (Eisenhauer et al. 2005, Gillessen et al. 2009a,b). Detailed analysis has significantly improved the astrometric precision by elimination of a number of systematic uncertainties, especially in the distortion of the infrared cameras and the long-term definition of the reference frame (to between 150 and 300 µas: Ghez et al. 2008, Gillessen et al. 2009b, Fritz et al. 2010a). At the time of writing the number of well-determined S-star orbits has grown to about 30 (Gillessen et al. 2009b). For the star S2 a complete orbit is now available (Figure 4.3.1, Gillessen et al. 2009a). Given uncertainties, *a near perfect agreement between the NTT/VLT and Keck data sets is reached* by allowing for a linear drift between the two reference frames (Gillessen et al. 2009a). Ghez et al. and Gillessen et al. report a mass of $M_\bullet = 4.4 \times 10^6$ $M_\odot$ and $4.28 \times 10^6$ $M_\odot$ respectively (for $R_0 = 8.3$ kpc), with a statistical uncertainty of about $\pm 0.07 \times 10^6$ $M_\odot$ (Gillessen et al.) at fixed distance $R_0$. The mass scales roughly as $M_\bullet \sim R_0^2$, which is the result of mixing astrometry ($M_\bullet \sim R_0^3$) and radial velocity information ($M_\bullet \sim R_0$). Including the dominating systematic uncertainty of the distance (see § 5.3) gives a total mass uncertainty of $\pm 0.4 \times 10^6$ $M_\odot$.

The combined data set, including also the stars S1, S8, S12, S13 and S14, yields a marginal improvement over the numbers cited in Gillessen et al. 2009b:

$$R_0 = 8.28 \pm 0.15 \pm 0.29 \text{ kpc, and}$$
$$M_\bullet = 4.30 \pm 0.20 \pm 0.30 \times 10^6 \text{ M}_\odot, \qquad (2)$$

where the first error is the statistical fit error and the second the systematic error (Gillessen et al. 2009a). *The extended mass component within the orbit of S2 (visible stars, stellar remnants and possible diffuse, dark matter) contributes less than 4 to 6.6% of this central mass* (2σ, Ghez et al. 2008, Gillessen et al. 2009b).

The analysis has two main sources of uncertainty. First, the S2 data during the peri-center passage is suspicious. The star was brighter than usual in 2002 and its position may have been confused (and offset) by another, weaker source on/near Sgr A*. Such 'astrometric' confusions by faint ($K_s > 16.5$) stars very close to Sgr A* have since been seen on other occasions, for instance in the astrometry of Sgr A* itself (e.g. Dodds-Eden et al. 2010b). For this reason, Gillessen et al. (2009b) assigned lower weights (larger uncertainties) for the 2002 data, and Ghez et al. (2008) ignored their 2002 points. This is unfortunate since the peri-center data are most constraining for the gravitational potential. Second, the motion of the reference frame cannot yet be determined from the orbital data itself to a high precision. Ghez et al. (2008) showed two orbital fits, one with the 3D reference system fixed (yielding $R_0 = 8.4 \pm 0.4$ kpc), and one with the reference system motion treated as free parameters ($R_0 = 8.0 \pm 0.6$ kpc), the mass scaling correspondingly. In particular the line-of-sight velocity of the massive black hole is degenerate with mass and distance. Gillessen et al. (2009a) used priors on the coordinate system derived from tests of the accuracy of the coordinate system. These authors presented also a table of fit results for various combinations of priors and selections of orbital data.

The position of the mass and that of the radio source coincide to ± 2 mas. The mapping between radio and infrared coordinates is hampered by the abscence of any extragalactic reference sources in the Galactic Center field. The measurement is nevertherless achieved by comparing the positions of SiO maser stars that are visible both in the near infrared and the radio (Menten et al. 1997, Reid et al. 2003, 2007). The level of agreement matches the expectation following from the formal uncertainties of the radio and near infrared positions of the SiO maser stars. It is also worth noting that positions of various near-infrared flares agree with the same position. Sgr A* is the only galactic nucleus for which the coincindence of mass and variable emission can be shown to the remarkable precision of 2 mas.

In summary, from the stellar orbits it is now established that the Galactic Center contains a highly concentrated mass of ~ 4 million solar masses within the peri-center of S2, i.e. within 125 AU. This requires a minimum density of $5 \times 10^{15}$ $M_\odot pc^{-3}$. The mass centroid lies within ± 2 mas at the position of the compact radio source Sgr A*, which itself has an apparent size of < 1 AU only (Shen et al. 2005, Bower et al. 2006, Doeleman et al. 2008). Taken together, this makes the ***Galactic Center Black Hole the currently best case for the existence of astrophysical black holes***. Further support for this conclusion comes from the fact that near-infrared and X-ray flares are observed from the same position, which naturally can be ascribed to variations in the accretion flow onto the massive black hole.

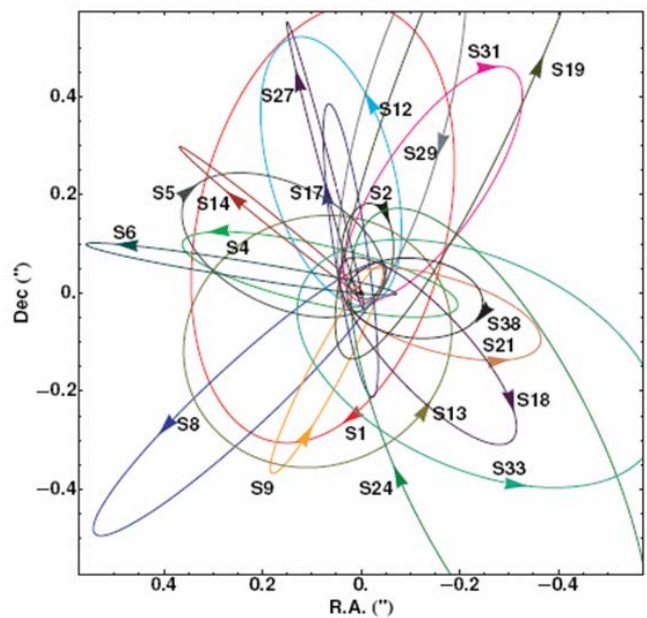

**Figure 4.3.2.** A summary of 20 of the ~ 30 S-star orbits delineated by the most recent orbital analysis of Gillessen et al. (2009b) [5].

---

[5] For movies of these orbits see:
http://www.mpe.mpg.de/ir/GC/index.php
http://www.eso.org/public/news/eso0846/
http://www.astro.ucla.edu/~ghezgroup/gc/pictures/orbitsMovie.shtml

Given this accurate determination of the central black hole mass one can ask how well the Galactic Center black hole matches the $M_\bullet$-$\sigma$-relation in nearby galaxies (Gebhardt et al. 2000, Ferrarese & Merritt 2000, Tremaine et al. 2002). These authors found that empirically central black hole mass $M_\bullet$ and velocity dispersion $\sigma$ for many galaxies correlate over several orders of magnitude. The importance of the relation comes from the fact that the velocity dispersion $\sigma$ is measured over a region much larger than the sphere of influence of the central black hole, thereby suggesting that the black hole and the bulge must have co-evolved. Given a bulge velocity dispersion of 105 km/s (Gültekin et al. 2009), the Milky Way falls below the best fitting relation, although its position is compatible with the general scatter in the relation (Gültekin et al. 2009) and the relatively large uncertainty in the Milky Way's bulge velocity dispersion (± 20 km/s) in principle allows it to fall on the relation if the velocity dispersion were as small as 85 km/s. While in Tremaine et al. (2002) the Milky Way was off by a factor of 5 (mainly due to the too low mass of Sgr A*), the current mass estimates from stellar orbits make the discrepancy slightly less than a factor of ~ 2 or ~ 0.3 dex. Graham (2008b) and Gültekin et al. (2009) redetermined the intrinsic scatter of the relation and found an increased value (Gültekin et al.: 0.44 ± 0.06 dex) compared to Tremaine et al. (2002, 0.30 dex). Even if the extreme outlier Circinius is excluded from the Gültekin et al. (2009) sample, the intrinsic scatter stays as high as 0.36 dex. This scatter is large enough to include the Milky Way point. Furthermore, the Milky Way has a pseudobulge (Binney 2009). Graham (2008a) and Hu (2008) noted that galaxies with pseudobulges tend to lie below the relation. Hence, the Milky Way actually might be an excellent example of the $M_\bullet$-$\sigma$-relation for pseudobulges.

## 4.4 Very Long Baseline Interferometry of Sgr A*

Very Long Baseline interferometry observations deliver the highest resolution images and sub-mas astrometry. This places strong additional constraints on the properties of Sgr A*. The left panel of Figure 4.4.1 shows the currently best determinations of the intrinsic size of the radio source as a function of wavelength across the cm- and mm-bands, as obtained by Bower et al. (2004, 2006), Shen et al. (2005) and Doeleman et al. (2008). The various data sets and analyses show consistently that – once the observed source diameters are corrected for wavelength dependent interstellar scattering – the intrinsic size of Sgr A* decreases with decreasing wavelength. At the shortest wavelength currently reached by these challenging measurements (1.3 mm, Krichbaum et al. 1998, Doeleman et al. 2008) the source size is 37 (+16, −10) (3$\sigma$) μas. This is a remarkable result as this diameter corresponds to a ***mere 3.7 times the size of the event horizon*** of a $4.3 \times 10^6 \, M_\odot$ black hole. Because of the foreground scattering and the limited u-v-coverage of the interferometry, the current data are not yet good enough to yield a proper two-dimensional image of the source and by necessity make the strongly simplifying assumption of a circular or elliptical Gaussian brightness distribution. Yuan, Shen & Huang (2006) show that a radiatively inefficient accretion flow model reproduces the sizes of the 7 mm and 3.5 mm emission reasonably well, and they predict a brightness distribution at 1.3 mm that cannot be described at all by a Gaussian. The increased angular resolution and spatial frequency coverage of VLBI at short mm-wavelengths should soon allow observations of more complex structures. The 1.3 mm VLBI data of Doeleman (2008) can also well be fit by a uniform thick ring of inner diameter 35 μas and outer diameter 80 μas when convolved with the interstellar scattering. Such a distribution is motivated by physical models of the Sgr A* accretion region with general relativistic ray tracing (Falcke, Melia & Algol 2000, Broderick & Loeb 2006) and magneto-hydrodynamic effects (Noble et al. 2007). These models predict a 'shadow' or null in emission in front of the black hole position, especially in the case of face-on accretion disks. As a result of the strong light bending and lensing near the event horizon the minimum FWHM diameter of even intrinsically very compact light distributions (centered on the black hole and azimuthally symmetric) would be at least ~ 5 $R_S$ ~ 50 μas for a Schwarzschild hole and ~ 4.5 $R_S$ ~ 45 μas for a maximally rotating Kerr hole (Broderick & Narayan 2006). This lower limit is just marginally consistent with the 3σ upper limit of the 1.3 mm VLBI observations of Doeleman et al. (2008). A plausible solution is that the mm-emission is offset from Sgr A* due to Doppler boosting and inhomogeneities ('hot spots') in the rapidly rotating accretion zone (Broderick & Loeb 2006), or because the emission comes from a jet (Falcke & Markoff 2000).

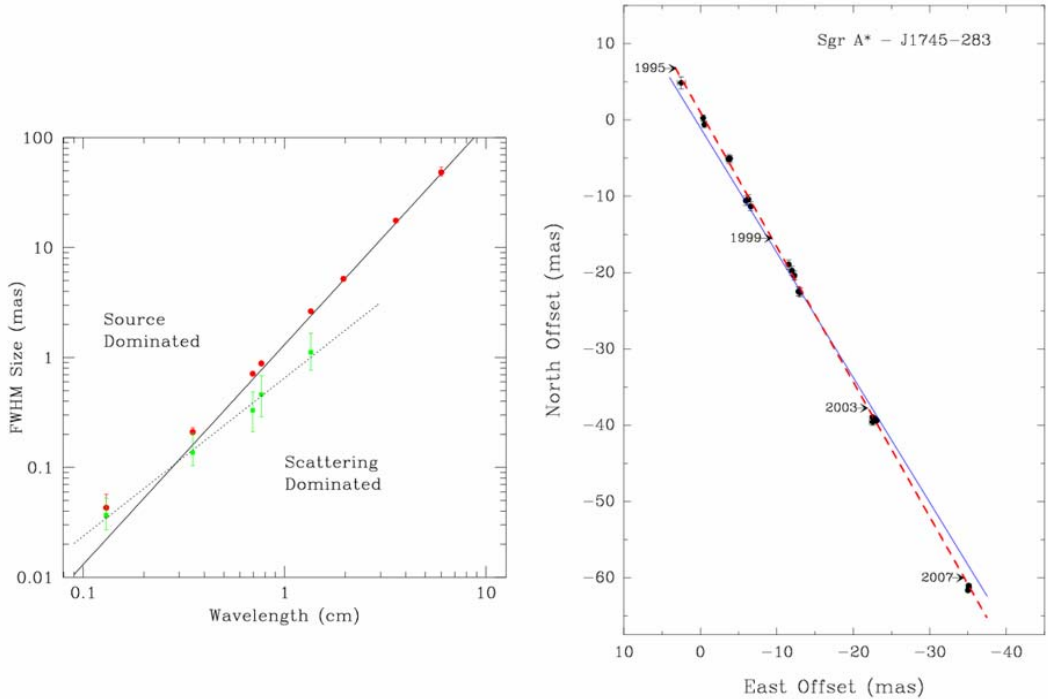

**Figure 4.4.1.** Current constraints on the intrinsic size (left) and motion (right) of Sgr A*, as obtained by Very Long Baseline Interferometry Observations. Left: Observed (red) and intrinsic (green) sizes of Sgr A* as a function of wavelength (adapted from Doeleman et al. 2008). Red circles show major-axis observed sizes of Sgr A* from VLBI observations (all errors 3σ). Data from wavelengths of 6 cm to 7 mm are from Bower et al. (2006), data at 3.5 mm are from Shen et al. (2005), and data at 1.3 mm are from Doeleman et al. (2008). The solid line is the best-fit $\lambda^2$ scattering law from Bower et al. (2006), and is derived from measurements made at 1.17 cm. Below this line, measurements of the intrinsic size of Sgr A* are dominated by scattering of foreground interstellar electrons, while measurements that fall above the line indicate intrinsic structures that are larger than the scattering size. Green points show derived major-axis intrinsic sizes (subtracting in squares the expected sizes for a point source from the measured sizes) from 2 cm > λ > 1.3 mm and are fitted with a power law (~ $\lambda^{1.44}$) shown as a dotted line. Right: The apparent motion on the sky of the compact radio

source Sgr A* relative to a distant quasar (J1745–283) (adapted from Reid 2009). The dashed line is the variance-weighted best-fit motion of 6.379 ± 0.024 mas/yr
(= 30.24 ± 0.12 km/s/kpc) for the data published by Reid & Brunthaler through 2004. Recent data from 2007 shown here confirms the published result. All of the apparent motion of Sgr A* can be accounted for by the ≈ 210 Myr period orbit of the Sun about the Galactic center. The solid line gives the orientation of the Galactic plane, and the difference in orientation of the two lines is caused by the 7.16 km s$^{-1}$ motion of the Sun perpendicular to the Galactic plane. The residual, intrinsic motion of Sgr A* perpendicular to the Galactic plane is extremely small: −0.4 ± 0.9 km s$^{-1}$. A massive black hole perturbed by stars orbiting within its gravitation sphere of influence is expected to move ≈ 0.2 km s$^{-1}$ in each coordinate (Merritt, Berczik & Laun 2007).

Monitoring Sgr A*'s position with respect to background QSOs since 1981, initially with the VLA (1981-1998, Backer & Sramek 1999) and more recently with the VLBA (1995-2007, Reid & Brunthaler 2004, Reid 2009), has established that the radio source is stationary, once the solar reflex motion is corrected for (Figure 4.4.1). The latter was independently measured from HIPPARCOS data. The residual motion of Sgr A* in this reference frame is very small. The motion in the Galactic plane is −7.2 ± 8.5 km/s (Reid et al. 2009a), but the precise value depends on the circular speed at the solar radius, for which values between 220 km/s and 255 km/s are reported. A more precise number can be given for the motion perpendicular to the Galactic plane: −0.4 ± 0.9 km/s (Reid & Brunthaler 2004). The ***extremely small intrinsic motion orthogonal to the plane is close to the expected Brownian motion*** of a massive black hole in the potential of the surrounding dense star cluster (Chatterjee et al. 2002, Dorband et al. 2003, Reid & Brunthaler 2004). The fact that Sgr A* moves ~ 100 to 3000 times slower than the surrounding S-stars (with masses of ~ 10 to 15 $M_\odot$) implies that the radio source must contain a significant fraction of the dynamical mass inferred from the stellar orbits. In equilibrium, the motions of stars and central black hole tend toward equipartition of kinetic energy. Analytical as well as numerical work indicates that a $4 \times 10^6\ M_\odot$ black hole would move at ~ 0.2 km s$^{-1}$ per coordinate (Chatterjee et al. 2002, Dorband et al. 2003, Reid & Brunthaler 2004, Merritt, Berczik & Laun 2007). The observed limits to the motion of Sgr A* then require a mass of at least $4 \times 10^5\ M_\odot$.

The 3σ upper limit of the intrinsic source size obtained at 1.3 mm (53 μas: Doeleman et al. 2008), combined with this lower limit of the mass of Sgr A* of $4 \times 10^5\ M_\odot$ then yields a conservative lower limit for the mass density of ~ $8 \times 10^{22}\ M_\odot \mathrm{pc}^{-3}$. This limit is only two orders of magnitude below the 'effective' density of a massive Schwarzschild black hole of $4.4 \times 10^6\ M_\odot$,
$\rho_{eff}(M_\bullet = 4.4 \times 10^6\ M_\odot) = M_\bullet/(4\pi/3 R_S^3) = 1.8 \times 10^{25}\ M_\odot \mathrm{pc}^{-3}$.

## 4.5 Does Sgr A* have an event horizon?

Broderick & Narayan (2006) and Broderick, Loeb & Narayan (2009a) have recast earlier considerations by Narayan et al. (1997, 1998) to argue that Sgr A* must have an event horizon (but see the arguments in Abramowicz, Kluzniak & Lasota 2002 and the MECO model mentioned below in § 4.7). The basic argument of Broderick et al. is as follows. Matter accreting onto a hypothetical hard surface lying outside the gravitational radius $R_g = GM_\bullet/c^2$, but within the upper limit ~ 10 $R_g$ set by the VLBI images, will emit on the way some of its gravitational energy as non-thermal radiation (and possibly particles). Once the matter hits the surface, it will shock, thermalize, and emit *all* its remaining energy as black-body radiation in the IR range. Such a component is not observed in the SED, and this sets an upper limit on the mass accretion rate. This limit is so low, that even the low level of observed quiescent non-thermal emission from the infalling gas requires an assumed radiative efficiency of nearly 100% (Figure 4.5.1). This can be ruled out, which then leads to the conclusion that the ***central object does not have a hard surface, but rather an event horizon***.

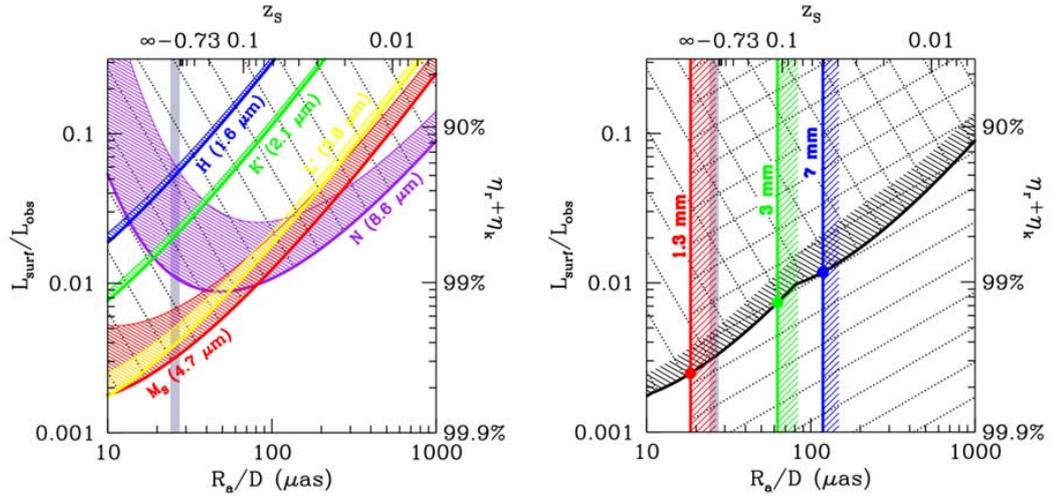

**Figure 4.5.1.** Left: limits of the ratio of the surface luminosity of a putative infrared photosphere to the observed luminosity of Sgr A*, $L_{surf}/L_{obs}$, as a function of the photosphere size as seen at infinity for the infrared measurements of Hornstein et al. (2007, 1.6, 2.1 and 4.7 µm), Ghez et al. (2005a, 3.8 µm) and Schödel et al. (2007b, 10 µm). The hatched bands denote the 3σ upper bounds. The peculiar behavior of the 10 µm-band constraint is a result of the transition from the Rayleigh-Jeans limit to the Wien limit around $R_a/D \approx 50$ µas as the surface becomes cooler. The region above any of the limits is necessarily excluded. The right-hand vertical axis shows the corresponding limits upon the accretion flow's radiative efficiencies. The top axis gives the redshift associated with a Schwarzschild space time given the apparent source radius and the thick grey line shows the apparent radii associated with the photon orbit for Kerr space times. Right: The limits upon $R_a/D$ implied by recent VLBI observations (Bower et al. 2004, 7 mm), Shen et al. (2005, 3 mm) and Doeleman et al. (2008, 1.3 mm), superposed on the combined constraint implied by IR flux measurements. Regions to the right of the leftmost (smallest) size constraint are excluded. When combined with the limits from the IR flux measurements, the permissible parameter space is reduced to a small corner in the $R_a - L_{surf}/L_{obs}$–plane. Adapted from Broderick et al. (2009a).

## 4.6 Could Sgr A* be a binary?

The available data place interesting limits on the mass of a possible second black hole in the Galactic Center (Figure 4.6.1). A second black hole of similar mass as the main one in an orbit within the peri-center distances of the S-stars (< 10 light hours separation) would coalesce by gravitational radiation within a few hundred years. Any second object would have to be of significantly smaller mass, and thus be an 'intermediate' mass black hole.

The first constraint comes from the lack of motion of Sgr A*. If Sgr A* and a nearby intermediate mass black hole were orbiting each other, the orbital reflex motion of Sgr A* might show up in the infrared astrometry. The upper limits on the velocity set by Ghez et al. (2008, ~ 30 km/s 1σ error per axis) and Gillessen et al. (2009, ~ 11 km/s 1σ error per axis, assuming that the nuclear star cluster rests with respect to Sgr A*) correspond to lines in a phase space plot of the mass of the intermediate mass black hole versus the black hole-black hole distance (Figure 4.6.1), separating configurations at smaller masses from systems with higher masses. However, it is not possible to detect such an orbital motion of the massive black hole if the orbital period $P$ is much shorter than that of S2 ($P < 5$ yr). Taken together, this excludes an area toward higher masses and larger distances. An even stronger constraint of the same type comes from limit on the motion of radio Sgr A* in galactic latitude ($-0.4 \pm 0.9$ km s$^{-1}$, Reid & Brunthaler 2004). Also for these data, it seems reasonable to assume that only systems with $P > 5$ yr would have been discovered. Similar arguments constraining the binarity of Sgr A* have been put forward by Hansen & Milosavljevic (2003), whose results we also reproduce in Figure 4.6.1.

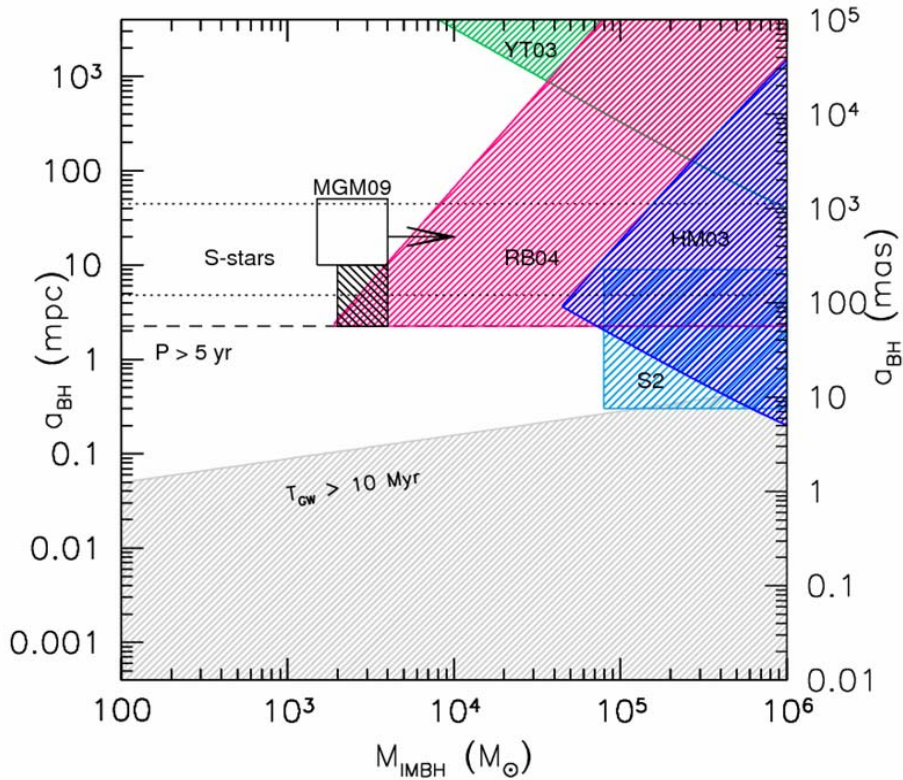

**Figure 4.6.1.** Constraints on the orbital parameters of a hypothetical second (intermediate mass) black hole in the Galactic region. The shaded areas represent regions of parameter space that can be excluded based on observational or theoretical arguments. The dotted lines mark the distances at which the S-stars are currently observed. The dashed line represents the 5 yr orbital period corresponding to discoverable systems. The parameters enclosed in the empty rectangular box are required for an efficient randomization of inclinations in the cluster infall scenario (Merritt, Gualandris & Mikkola 2009). The small rectangular region just below the empty box represents the parameter space excluded by numerical simulations (adapted from Gualandris & Merritt 2009, see also Yu & Tremaine 2003).

Secondly, black holes in close orbits will lose energy via gravitational waves and thus spiral in. Demanding a life-time of $> 10^7$ yr for the massive black hole – intermediate mass black hole system excludes configurations with smaller separations and higher masses. Dynamical stability can also be assumed for the S-star cluster as a whole. Mikkola & Merritt (2008) have shown that an intermediate mass black hole of $10^{-3} M_\bullet$ at a distance of 1 mpc would make the S-star cluster unstable. Based on simulations, Gualandris & Merritt (2009) conclude that an intermediate mass black hole will reach a stalling radius that is proportional to the mass of the intermediate mass black hole: $R_{stall} = 3.5$ μas $\times M_{IMBH}$ [$M_\odot$]. Since one does not expect an intermediate mass black hole to reside at a much smaller radius, this puts another constraint on the binary. Finally, the S2 orbit itself excludes part of the phase space. Motivated by the findings of Ghez et al. (2008) and Gillessen et al. (2009b), we assume that no mass larger than 0.02 $M_\bullet$ can be hidden inside 0.5 × the peri-center distance of S2.

Finally Gualandris & Merritt (2009) have studied with numerical integrations the impact of a black hole binary on the properties of the S-star orbits and derived some further constraints. The parameters enclosed in the empty rectangular box in Figure 2 are required for an efficient randomization of inclinations in the cluster infall scenario (Merritt, Gualandris & Mikkola 2009). The small rectangular region just below the empty box represents the parameter space excluded by numerical simulations.

In summary *there currently is no empirical evidence for a second massive black hole in the central parsec. To be consistent with the observations such a second hole anywhere in the central parsec has to be less massive than about ~$10^5 M_\odot$. The measurements do allow the presence of an intermediate mass black hole ($10^3…10^{4.5} M_\odot$) if it is either very close (≤ 1 mpc) or at > 100 mpc from SgrA\**.

## 4.7 Alternatives to a black hole configuration

This section briefly summarizes how the continuously improving constraints on the size and density of the central mass over time have eliminated ever more of the alternatives to the black hole hypothesis for Sgr A*.

With the 1996 – 1998 detections of the O($10^3$ km/s) proper motions of the S-stars on scales of ≤ 0.02 pc the case for a compact central mass was firmly established. Its implied density is ≥ $10^{12}$ $M_\odot$pc$^{-3}$, more than $10^6$ times denser than the visible nuclear star cluster and similar to the density of the dark mass in NGC 4258 from the H$_2$O maser work of Miyoshi et al. (1995). At that density any astrophysical cluster of faint stars (e.g. white or brown dwarfs), dark stellar remnants (stellar or hypothetical planetary mass black holes and neutron stars) or other collisional matter (e.g. rocks, asteroids etc.) must be short-lived because relaxation and collisions lead to core collapse and/or evaporation on a time scale of ~ $10^{7...8}$ yr (Maoz 1998). Such a short-lived dark cluster thus is a fairly implausible explanation of the central mass. The determination of stellar orbits since 2002 (especially S2) have increased the minimum density of the central object by another four orders of magnitude, to ~ $10^{16}$ $M_{\odot}pc^{-3}$. This eliminates the dark cluster scenario on the basis of the Copernican principle. At that density a dark astrophysical cluster would have a lifetime less than a few $10^5$ years (Maoz 1998), just a small fraction of the lifetime of the stars in the central cusp.

Viollier, Trautmann & Tupper (1993) proposed that the dark matter concentrations in galactic nuclei, including QSOs, may be due to compact 'fermion balls' (for example made of hypothetical, massive neutrinos) supported by degeneracy pressure. The size of a degenerate object of a given mass increases the lighter the fermion, and the maximum stable 'Chandrasekhar' mass of a fermion ball (or its relativistic analog, the 'Oppenheimer-Volkoff' mass) also increases the lighter the fermion. In this scenario the largest observed central masses in elliptical galaxies, a few $10^9$ $M_{\odot}$, would approach the Oppenheimer-Volkoff mass, resulting in an upper limit to the mass of the constituent fermions to about 17 $keV/c^2$, with larger black hole masses requiring smaller masses. At that mass a fermion ball in the Galactic Center would have a radius of about 15 light days (Munyaneza, Tsiklauri & Viollier 1998), 36 times larger than the peri-center distances of the stars S2 and S14. This is obviously excluded by the orbital data. To still 'fit' within these peri-center distances the fermions would have to have a mass of

$$m_f \sim 70 \text{ keV c}^{-2} \left(\frac{R_{peri}}{10 \text{ light hours}}\right)^{-3/8} \left(\frac{g}{2}\right)^{-1/4} \left(\frac{M_\bullet}{4.4 \times 10^6 M_\odot}\right)^{-1/8} \quad (3),$$

where $g$ is the spin degeneracy factor. Smaller peri-center distances would hence require even larger fermion masses. The two limits are incompatible, hence massive fermions cannot explain all massive black holes. Such a hypothetical fermion also cannot be the cosmological dark matter particle, since it would be $2 \times 10^5$ times more massive than allowed by the upper limit to the mass of stable neutrinos in the standard Big Bang model with the input of CMB constraints in WMAP 3 (Goobar et al. 2006). The fermion ball hypothesis thus is probably excluded.

Another proposed non-black hole configuration is the 'boson star' scenario advanced by Torres et al. (2000). A wide range of boson star masses can be realized, including ones with massive black hole masses, depending on the assumptions about the specific boson particle masses and their self-interactions. Since such an object consists of weakly interacting particles it is unclear how it may have formed. A boson star does not extend much beyond its event horizon and is highly relativistic. It does not possess a singularity, a horizon or a hard surface. A boson star is unstable to collapse to a black hole if it experiences baryonic accretion, as observed in the Galactic Center to occur frequently (§ 7). Furthermore, accreted baryons will amass in the interior, providing a surface satisfying the requirements for the surface-emission arguments from § 4.5. We conclude that also the boson star scenario is excluded by the present evidence. Since a boson star is a factor of a few larger than the event horizon of a massive black of the same mass, it may in addition be possible to test this proposal with future interferometric observations (Eisenhauer et al. 2008).

In summary then, ***the evidence that Sgr A\* is a massive black hole is compelling and beyond any reasonable doubt, as long as General Relativity holds***. Some authors have argued, however, that adding quantum phenomena may prevent the formation of black holes due to early evaporation by Hawking radiation (Vachaspati, Stojkovic & Krauss 2007). There also are proposals for alternative, non-black hole configurations of very high density (but without an event horizon or a central singularity) motivated by quantum phase transitions. One such configuration is the 'grava-star' (Mazur & Mattola 2001) or 'dark energy' star (Chapline 2005) where inside $R_{grava} \sim R_S + O(\lambda_P = 1.6 \times 10^{-33}$ cm) the volume is filled with a de Sitter vacuum solution of General Relativity, with $\rho = -p$ ('dark energy'). As this configuration has practically the same size as the event horizon of a black hole (and hence a very large gravitational redshift from its surface to a distant observer) it will be very difficult (or impossible: Abramowicz, Kluzniak & Lasota 2002, but see also Broderick & Narayan 2006) to verify or falsify such a hypothetical configuration through present or future empirical measurements.

# 5. Mass distribution in the nuclear star cluster

In this chapter we summarize our current knowledge about the distribution of the extended mass in the nuclear star cluster within the sphere of influence of the massive black hole (~ 3 pc). In this region the extended mass is dominated by stars and stellar remnants (stellar black holes and neutron stars). Gas and dark matter probably do not contribute significantly. We also discuss the status of the determnations of the distance to Galactic Center.

Figure 5.1.1 displays the best direct mass estimates to $R \sim 6$ pc. The most recent measurements from the young stars based on the S-star orbits (mainly S2) at the smallest scales (Ghez et al. 2008, Gillessen et al. 2009a,b) and the 'orbital roulette' estimate at $\sim 0.1 - 0.2$ pc from the clockwise rotating O/WR disk (Beloborodov et al. 2006, Bartko priv.comm.) show that the massive black hole completely dominates the mass distribution out to $\sim 0.5$ pc. No significant extended mass is (as yet) detected from the S-star orbits on scales of $< 0.02$ pc (Ghez et al. 2008, Gillessen et al. 2009b), or from the clockwise disk determination at $\sim 0.1 - 0.2$ pc.

Trippe et al. (2008) and Schödel et al. (2009) have derived high quality proper motions ($\delta v \sim 3 - 10$ km/s) for a large number (~ 6000) of late-type stars across the entire central ~ 1 pc. Trippe et al. (2008) also obtained radial velocities for about 660 of these stars. After elimination of the (unrelaxed) early-type stars in the sample both papers proceed to derive the mass distribution of the nuclear cluster, including earlier measurements as needed. Trippe et al. first decompose the late-type star motions into a random component and a rotation in the plane of the Milky Way, and then use Jeans equation modeling to feed the rotational and random motions back into the mass modeling. Schödel et al. do not factor out the rotational component and apply both isotropic and anisotropic modeling, as well as projected mass estimators to the observed velocity dispersion distribution. The results of the two studies are in good agreement with each other, which perhaps is not surprising, as both analyses use partially overlapping NACO/VLT data sets. Both papers find that the stellar velocities at $R \sim 0.5 - 1$ pc require a statistically very significant extended mass, in addition to the central black hole, consistent with but much improved over the earlier work of Genzel et al. (1996) and Haller et al. (1996). The inferred dynamical mass within the central parsec is $5.4 \pm 0.4 \times 10^6$ M$_\odot$. The cluster mass is thus $M_{cluster} \sim 1 \pm 0.4 \times 10^6$ M$_\odot$ within $R \leq 1$ pc. This mass is also consistent with the stellar light distribution, combined with a 2 μm light to mass ratio calibrated at larger radii (e.g. Genzel et al. 1996). The quoted uncertainty is systematic and is dominated by assumptions on the power law slope of the stellar density distribution.

Further constraints between 0.1 pc and 10 pc come from ionized and molecular gas (Serabyn & Lacy 1985, Serabyn et al. 1988, Lacy, Achtermann & Serabyn 1991, Herbst et al. 1993, Güsten et al. 1987, Jackson et al. 1993, Roberts, Yusef-Zadeh & Goss 1996, Christopher et al. 2005, Zhao et al. 2009). Based on the assumption that the various gas streamers indeed follow unique Keplerian orbits and that the gas velocities are not strongly affected by forces other than gravity, the derived dynamical masses are in reasonable agreement with those derived from the stellar dynamics (e.g. Zhao et al. 2009).

The extended mass at larger radii obviously depends on the slope $\gamma$ of the stellar density distribution ($M(\leq R) \sim R^{3-\gamma}$). Figure 5.1.1 uses $\gamma = 1.8$ motivated from the light distribution of the star cluster (and a constant $M/L_{Ks}$, Genzel et al. 1996, Trippe et al. 2008). Stellar dynamical mass determinations at $R >> 1$ pc suggest a somewhat flatter slope ($\gamma \sim 1.5$, Launhardt, Zylka & Mezger 2002, Trippe et al. 2008) but by themselves are not precise enough to constrain the power law slope accurately. If the circular velocity of the CND at $R = 1.4 - 4$ pc ($v_{rot} = 110 \pm 10$ km/s, Serabyn & Lacy 1985, Güsten et al. 1987, Jackson et al. 1993, Christopher et al. 2005) is used as an additional constraint, a steeper slope $\gamma \sim 1.8 - 2$, and smaller stellar mass is favored, giving then a significantly better fit.

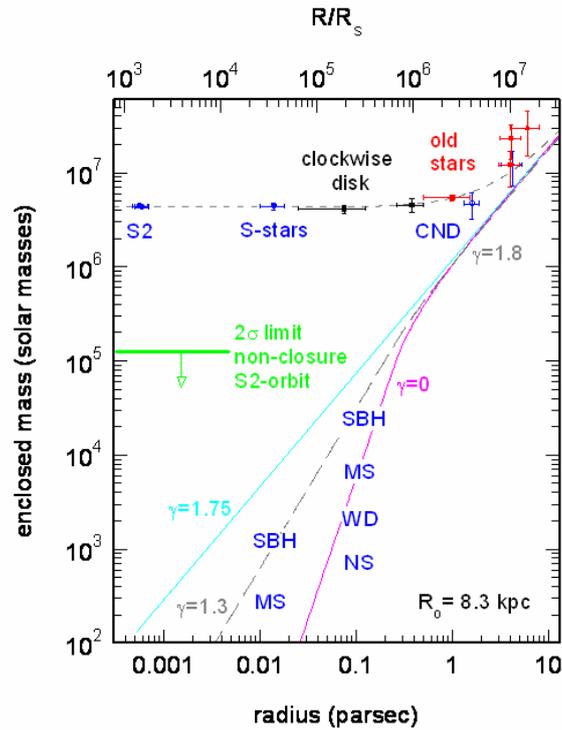

**Figure 5.1.1.** Mass distribution in the central few parsec of the Galaxy for $R_0 = 8.3$ kpc. The distribution is based on the most recent analyses of the orbit of S2 (blue, Ghez 2008, Gillessen 2009b), of 'roulette estimations' of the clockwise rotating disk of O/WR stars (filled black squares, Beloborodov et al. 2006, Bartko, priv. comm.), of modeling of the proper motions (Schödel et al. 2009) and proper motions and radial velocities (Trippe et al. 2008, Genzel et al. 1996) of late-type stars (filled red squares), and from the light of the late-type stars (Launhardt et al. 2002 (open red triangles), and the rotation of the molecular gas:in the CND (Güsten et al. 1987, Jackson et al. 1993, Christopher et al. 2005 (open blue circles)). In addition to a central mass ($4.35 \times 10^6$ M$_\odot$), the grey dashed model curve includes a star cluster with a broken power law density distribution ($\rho(R) \sim R^{-\gamma}$ with $\gamma = 1.3$ ($\gamma = 0$) for $R \leq R_0 = 6$", or 0.25 pc and $\gamma = 1.8$ for $R > R_0 = 6$") for the long-dashed grey (and continuous magenta) curves. This is the best fit ($\chi^2 \sim 6.4$ for 12 data points) to the S-Star-cluster, orbital roulette and late-type cluster data, plus the gas measurements of the $R \geq 1.5$ pc circum-nuclear disk, for fixed power law exponents and $R_{break} = 0.25$ pc. The best fitting stellar cluster model has a density at $R_0$ of $\sim 1.35 \times 10^6$ M$_\odot$ pc$^{-3}$. We also give the 2$\sigma$ upper limit of any extended mass within the S2 orbit from Gillessen et al. (2009a,b) and the estimates of masses in main-sequence stars (MS), stellar black holes (SBH), white dwarfs (WD) and neutron stars (NS) at 0.01 and 0.1 pc in the 'standard' simulation of Freitag et al. (2006, blue symbols).

Motivated by the observations of the stellar surface density distribution discussed in § 2.4, we adopt a broken power law, fitting function for the intrinsic density distribution of the overall stellar (plus remnant) mass, with a break radius at ~ 6" (0.25 pc, Schödel et al. 2007a). Depending on whether the slope within the break radius follows that of the early-type stars ($\gamma \sim 1.3$), or that of the late-type stars ($\gamma \sim 0$), the best fitting density and mass distributions of the stellar dynamical constraints, plus the dynamical constraints from the CND are

$$\rho_*(R) = 1.35(\pm 0.5) \times 10^6 \left(\frac{R}{0.25 pc}\right)^{-\gamma} \text{ M}_\odot \text{pc}^{-3} \quad (4)$$

with $\gamma = 1.3$ ( or $\gamma = 0$) for R < 0.25 pc
and $\gamma = 1.8$ for R ≥ 0.25 pc, and

$$M_*(R) = 1.54 \times 10^5 \left(\frac{R}{0.25 pc}\right)^{1.7} \text{ M}_\odot \text{ for R < 0.25pc and } \gamma=1.3$$

$$M_*(R) = 9.0 \times 10^4 \left(\frac{R}{0.25 pc}\right)^{3} \text{ M}_\odot \text{ for R < 0.25pc and } \gamma=0$$

$$M_*(R) = 2.2 \times 10^5 \left\{\left(\frac{R}{0.25 pc}\right)^{1.2} - 1\right\} + M(0.25 pc) \text{ M}_\odot \text{ for R } \geq 0.25\text{pc} \quad (5).$$

The quoted mass distribution has a systematic uncertainty of about 50% and is in good agreement with previous estimates (Genzel et al. 2003a, Mouawad et al. 2005, Schödel et al. 2007a). For comparison we show in Figure 5.1.1 the masses of different stellar components and remnants within 0.01 and 0.1 pc in the simulations of Freitag et al. (2006), which are based on the overall $\gamma \sim 1.3$ power law slope in the mass distribution. In this model and the work by Hopman & Alexander (2006a), the mass of the stellar remnants starts to dominate over the stellar mass within the central ~ 0.5 pc and the stellar black holes attain a slope close to that of the equilibrium Bahcall-Wolf solution ($\gamma \sim 1.75$). According to these simulations, 50 to 100 stellar black holes may reside inside the orbit of S2, resulting in a mass 200 to 300 times below the current limit on an extended mass around Sgr A* (Ghez et al. 2008, Gillessen et al. 2009b). Merritt (2010) has pointed out that the remnant densities and masses in the central 0.1 pc would be significantly less than estimated by Freitag et al. (2006) and shown in Figure 5.1.1 if the overall stellar mass were to follow the distribution of the late-type stars ($\gamma \sim 0$) and mass segregation were tied to the non-resonant two-body rate.

## 5.1 Outlier 'high velocity' stars

If the dynamical mass in the central region is dominated by a compact central mass and all stars are on bound orbits around this mass a conservative lower limit of this mass can be obtained from the observed 3D space velocity $v_{tot}$ and its 1σ uncertainty $\Delta v_{tot}$ and the projected distance from Sgr A*, $p$, from

$$M_{min} = \frac{(v_{tot} - 3\Delta v_{tot})^2 p}{2G} \qquad (6).$$

The true mass is always greater than $M_{min}$ and only equals $M_{min}$ for parabolic orbits ($e = 1$) at the peri-center distance ($p = R_{peri}$) and for $\Delta v \sim 0$. Reid et al. (2007) have used this estimator to show that for 14 out of 15 SiO maser stars in the nuclear star cluster this minimum mass is below the best estimate of the mass of Sgr A*, while for the bright maser star IRS 9 ($p = 0.34$ pc) this minimum mass is at least $5.1 \times 10^6$ M$_\odot$. Reid et al. consider a number of explanations, including orbital motion in a binary and an offset of the reference frame relative to the local standard of rest. They conclude that the most plausible explanations are either that IRS 9 is not bound to the central parsec and is on a near parabolic orbit with an apo-center radius ~10 pc, or that the Galactic Center distance must exceed 9 kpc. IRS 9 is not the only bright AGB star with these properties. The late-type star sample of Trippe et al. (2008: ~ 6000 stars in the central parsec) contains 12 stars for which $M_{min}$ is 3σ larger than the mass of Sgr A* as estimated from the S-star orbits. The overall space velocity distribution is very well fit by a Maxwellian distribution to the highest velocities of 450 km/s. Most of these outlier stars give even more extreme lower mass limits than IRS 9. All but IRS 9 would still not be bound to the central parsec even if the distance were $R_0 = 9.2$ kpc, 3σ above the current best estimate (§ 5.5). This means that of the various explanations discussed by Reid et al. (2007) the scenario of highly parabolic orbits remains as the most plausible one. Remarkably, of these 12 outlier stars 8 have $K_s < 13$ and 5 have $K_s < 12$. If the selection of outlier stars is relaxed by subtracting only 2 times the velocity uncertainty, the number of outliers increases (only) to 18, with 10 stars that have $K_s < 13$ and 6 with $K_s < 12$. The identification of outliers thus is fairly stable. This indicates that at least half of the outlier stars are bright and cool, TP-AGB stars of age < 1 Gyr. While the 12 (18) stars only make up ~ 0.2 (0.3)% of the entire sample of Trippe et al. (2008), these 5 (6) bright stars make up ~ 5 (6)% of all K < 12 AGB stars in the central parsec. Since their lifetime is very probably less than the relaxation time in the nuclear cluster (~ 2 Gyr or even longer), these highly elliptical/parabolic orbits probably trace their initial orbital parameters.

## 5.2 Dark Matter in the central parsec?

One obvious question is whether dark matter might contribute in a measurable way to the extended mass in the Galactic Center. Given the standard dark matter profile obtained by Navarro, Frenk & White (1997) from N-body simulations, the amount of dark matter inside a region of the size of the sphere of influence of a massive black hole actually is negligible compared to the mass of astrophysical sources. Still, the presence of the black hole can create a spike in the cold dark matter distribution, as pointed out by Gondolo & Silk (1999) and Gnedin & Primack (2004). The argument is the same as for the stellar cusp, in which the lightest component would settle in a power-law density profile with a slope of −3/2 (Alexander 2005). For a Navarro-Frenk-White profile, the actual slope is −7/3 (Quinlan et al. 1995). The dynamics of the dark matter cusp has been recently discussed by Vasiliev & Zelnikov (2008). The evolution of dark matter in the Galactic Center is driven by the scattering of dark matter particles by bulge stars, their accretion into the massive black hole, and self-annihilation. If the Milky Way is older than the two-body relaxation time, a universal dark matter profile is reached, regardless of the initial conditions. Roughly 40% of the original reservoir has evaporated from the sphere of influence of the massive black hole in these calculations. Only 10%-15% of the reservoir is accreted into the hole. Hence this is not the dominant mechanism to grow it, and is lower compared to what previous studies (Zhao, Haehnelt & Rees 2002, Zelnikov & Vasiliev 2005) had found neglecting either heating or assuming that the loss cone would always be filled. This also renders the dark matter accretion driven explanation of the $M_\bullet$-$\sigma$-relation less probable.

Even with the enhancement of a spike, the mass of the dark matter in the Galactic Center is orders of magnitude too small to be dynamically relevant (Vasiliev & Zelnikov 2008), when compared with other forms of unseen, extended mass, such as stellar remnants (Morris 1993). Also, dynamical friction with WIMPs is irrelevant for the stellar orbits (Salati & Silk 1998). Regardless of this, the effect of dark matter on stellar dynamics has been discussed in the literature. Hall & Gondolo (2006) find that from the infrared-based dynamical measurements at most 10% of the mass of the black hole could be dark matter, implicitly assuming that unseen mass would be dark matter. The same confusion is found in Zakharov et al. (2007), who claim to derive constraints on the dark matter distribution from the absence of an observed apo-center shift for the S-stars. To be clear: While formally the limits are correct, a detection of an extended mass component would not necessarily imply a dark matter spike.

Another route to detecting dark matter in the Galactic Center is the search for a radiation signal. In many models, the dark matter particles can annihilate with themselves, creating high-energy photons. This process leads to a flux proportional to the square of the density. Any such signal from a spike would thus be even more peaked than the spike; but also a standard NFW-halo would have a pronounced concentration of the emission. The actual flux values expected (and thus the perspectives for detecting the signal with current or future instruments) are quite uncertain, depending on the assumptions and models used. For example, Amin & Wizansky (2008) argue that for a certain class of supersymmetric dark matter models, the cross section for annihilation is dependent on the relative velocities of the particles, which leads in the potential well of the massive black hole to a flatter halo profile and an enhanced signal. Also astrophysical processes might be important. Merritt et al. (2002) investigate the effect of galaxy mergers and find that these events transfer energy to the dark matter particles. This lowers their density and decreases the fluxes by an order of magnitude.

Essentially all high-energy signals from the Galactic Center have been suggested to originate from dark matter annihilation. Boehm et al. (2004) claim that the INTEGRAL-based observation of the 511 keV line from the galactic bulge (Knödlseder et al. 2003) is consistent with annihilation of dark matter particles with masses of ~ 1 to 100 MeV. Similarly, Cesarini et al. (2004) attribute the EGRET signal from the Galactic Center region (Belanger et al. 2004) to the same sort of process. Hooper et al. (2004) interpret the detections of TeV-$\gamma$-rays at the time available (Tsuchiya et al. 2004, Kosack et al. 2004) from the central parsec as a dark matter signal. While the detection from Tsuchiya et al. (2004) is now considered erroneous, the MAGIC and HESS telescopes have also observed a TeV-source in the Galactic Center. The properties of the Sgr A TeV source, however, are similar to other Galactic, astrophysical TeV-sources. Horns (2005) has analyzed the TeV-signal in the light of a WIMP origin and concluded that an unexpectedly high mass and interaction cross-section for the WIMP particles would be needed in order to explain the TeV-data. The problem apparently is that the WIMP-predicted TeV-spectrum would be peaked, while the TeV-data are perfectly consistent with a simple power law (see Horns 2005, Hall & Gondolo 2006).

Interestingly, not only high-energy data constrain the dark matter content of the Galactic Center. Gondolo (2000) showed that the radio flux upper limit from Sgr A* at 408 MHz excludes a spike of neutralinos. Hence, if the neutralino in the minimal supersymmetric standard model is the dark matter particle, there cannot be a dark matter spike in the Galactic Center. Conversely, if a dark matter cusp extends to the central parsec, the neutralino cannot be the dark matter in our galaxy. Aloisio, Blasi & Olinto (2004) revisited the problem with more realistic assumptions for the accretion flow and the magnetic field. They reaffirmed the previous conclusions.

Taking all wavebands together, Regis & Ullio (2008) conclude: "none of the components, which have been associated to Sgr A*, nor the diffuse emission components from the Galactic Center region, have spectral or angular features typical of a dark matter source. Still, data-sets at all energy bands, namely the radio, near infrared, X-ray and gamma-ray bands, contribute to place significant constraints on the WIMP parameter space."

Another aspect of a possible dark matter peak in the Galactic Center is its influence on stellar evolution. Salati & Silk (1989) have suggested that stars could catalyze the annihilation of WIMPs. The additional energy source would alter the evolution of the stars. Salati & Silk have shown that for massive stars, the effect is small. For low-mass stars, the luminosity is increased by a factor $10 - 100$ and the effective temperature of the star is lowered, i.e. these stars change their appearance to be more giant-like. Moskalenko & Wai (2007) show that stars can capture WIMPs, which would make them dark matter burners since the energy released by the annihilation can exceed the thermonuclear reactions. The most efficient burners should be white dwarfs. Scott, Fairbairn & Edsjö (2008) presented the results of studies with a stellar evolution code that includes energy injection by WIMP annihilation, applied to the nuclear stars and assuming an isothermal dark matter halo. They conclude that for the OB stars residing in two disk-like systems with small orbital eccentricities, the WIMP luminosity would not exceed 1% of the nuclear power. The situation is different for solar-mass stars on highly elliptical orbits (similar to the S-stars). Then the additional energy source can even exceed the nuclear fusion energy release. Detecting such low-mass stars would thus provide a strong test for the WIMP properties. The authors also mention a scenario in which the S-stars have extended main-sequence life times due to the WIMP burning, thus maybe resolving the paradox of youth. But they also point out, that the stars would need to be formed in a region where the dark matter density is not enhanced. Formation and transport of the S-stars to the central arcsecond would still take longer than their lifetimes and thus the explanation is incomplete.

### 5.3 Comparison to earlier statistical mass estimates

A possible concern are the differences in derived (central) dynamical masses in different publications (once all are re-scaled to 8.3 kpc), exceeding in some cases the quoted uncertainties. This is especially relevant when comparing the more recent values obtained from the S-star orbits (mainly S2: $M_\bullet = 4.32 \times 10^6$ $M_\odot$) with those obtained in the initial statistical proper motion and radial velocity determinations prior to the availability of the orbits (e.g. Genzel & Townes 1987, Genzel et al. 1996, 1997, Eckart & Genzel 1996, 1997, Ghez et al. 1998, Chakrabarty & Saha 2001: $M_\bullet \sim 2 - 3.2 \times 10^6 \, M_\odot$). A majority of these earlier statistical estimates used the Bahcall-Tremaine (1981, for radial velocities) and/or Leonard & Merritt (1989, for proper motions) projected mass estimators. For a more detailed discussion of these estimators we refer the reader to the original papers and the summary in Alexander (2005). The projected mass estimators are sensitive to the orbital structure (isotropic, radially or tangentially anisotropic) and the radial surface density distributions assumed. They also are predicated on the assumption that the motions are determined by a central mass and that the entire cluster volume is sampled. The impact of these various effects on the mass estimators were explored in Genzel et al. (2000) by simulating and 're-observing' model clusters. These simulations revealed the need for correction factors to the mass estimators and/or systematic uncertainties of $O(10 - 20\%)$. When such corrections are included the mass estimators result in a central mass of $\sim 3.5 - 4.1 \times 10^6 \, M_\odot$, consistent with (to within the $1\sigma$ uncertainties) but still slightly below the orbital estimates (e.g. Genzel et al. 2000, Schödel et al. 2003, 2009). Modeling the surface density and velocity dispersion data of the late-type stars with isotropic, anisotropic or Jeans equation techniques (Genzel et al. 1996, Trippe et al. 2008, Schödel et al. 2009) also resulted in low central mass values $(2.5 - 3.6 \times 10^6 \, M_\odot)$. Here the most plausible cause is the radial distribution of the late-type star population in the Galactic Center. As discussed in § 2.4 these stars are less frequent in the central few arcseconds (Sellgren et al. 1990, Genzel et al. 1996, Haller et al. 1996, Genzel et al. 2003, Schödel et al. 2009, Buchholz et al. 2009, Do et al. 2009a), thus strongly decreasing the sensitivity of projected velocity dispersion measurements to a central point mass. Some modeling also combined the early-type stars together with the late-type stars in a single estimate (Chakrabarty & Saha 2001). The early-type stars have a different spatial distribution than the late-type stars and are now known to be unrelaxed and dominated by moderately elliptical orbits, such that a combination of the data is not appropriate.

Jeans modeling of the O/WR stars gave $3.8 \pm 0.8 \times 10^6 \, M_\odot$ in Genzel et al. (2000). Applying the 'orbital roulette' technique proposed by Beloborodov & Levin (2004) to the clockwise disk O/WR stars gives a mass of $4.3 \pm 0.4 \times 10^6 \, M_\odot$ (Beloborodov et al. 2006, Bartko et al. 2009, Bartko, priv. comm.), both in excellent agreement with the orbital estimates. The dynamical estimates obtained from fitting orbits to the ionized gas in the 'mini-spiral' (western arc, northern arm and central regions: Serabyn & Lacy 1985, Serabyn et al. 1988, Lacy, Achtermann & Serabyn 1991) and from the circular velocities estimated in the CND under the assumption of pure rotation (Güsten et al. 1987, Jackson et al. 1993, Christopher et al. 2005 and references therein) give $2 - 4 \times 10^6 \, M_\odot$. These values are lower than but, given their uncertainties, probably still consistent within the more recent data. We note here that several of the mini-spiral estimates are quoted as lower limits in the original papers.

We conclude that ***most (or all) of the differences between different mass estimates arise as the result of now plausibly understood systematics in the analyses, and differences in distribution and kinematics of different stellar components*** used for these. It is clear, however, that the earlier work underestimated the systematic uncertainties present in the data.

## 5.4 Does IRS 13E contain an intermediate mass black hole?

IRS 13E (Figure 1.1, 5.4.1) is a remarkable grouping of three bright blue supergiants (two Wolf-Rayet stars and one O/B supergiant, E1, E2 and E4), all within a compact region of diameter ~ 0.6". IRS 13E is also associated with X-ray emission and a prominent local enhancement in the western ridge of the radio 'mini-cavity' (Figure 1.1). In addition there are a number of stars and emission maxima on $H/K_s/L'$ maps in the immediate vicinity of the O/WR-stars (including the bright central knot E3, right inset of Figure 5.4.1, Maillard et al. 2004, Paumard et al. 2006, Schödel et al. 2005). The O/WR stars have similar 3D space velocities, which are counter-clockwise on the sky.

For these reasons Maillard et al. (2004) proposed that IRS 13E might be a compact, self-gravitating star cluster with an embedded intermediate mass black hole. The basic argument is that such an overdensity of stars (as well as of dust and gas) cannot be a statistical fluctuation and thus must be a true 3D concentration. If so IRS 13E must then be self-gravitating, otherwise it would have been long destroyed by the strong tidal forces. The residual proper motion and radial velocity dispersion of the brightest four or five objects in IRS 13E then implies a mass of at least $10^3$ $M_\odot$ (Maillard et al. 2004), and possibly as large as several $10^4$ $M_\odot$ (Schödel et al. 2005). The observed stellar mass associated with IRS 13E is no larger than a few $10^2$ $M_\odot$ (~ 350 $M_\odot$: Paumard et al. 2006), depending on the adopted stellar mass function and on how many of the fainter stars seen in projection toward the core of IRS 13E are actually physically associated with the concentration of O/WR-stars (Schödel et al. 2005, Paumard et al. 2006, Trippe et al. 2008). The excess mass thus might be an intermediate mass black hole at the center of the stars. Such an intermediate mass black hole might be the natural result of the core collapse of a massive, dense star cluster spiraling into the central parsec (Portegies Zwart & McMillan 2002). There is also a second group of compact, dusty (L') sources just north of IRS 13E, IRS 13N (left panel of Figure 5.4.1). The IRS 13N knots share a common proper motion. Muzic et al. (2008) concluded that IRS 13N is a concentration of embedded, luminous and perhaps very young stars. The mean proper motion of IRS 13N, however, is significantly different from IRS 13E (left panel of Figure 5.4.1) and it is thus not obvious at all that IRS 13E and N are physically related.

The evidence for an intermediate black hole associated with IRS 13E is potentially very exciting. Since the case rests on the determination of the (residual) velocity dispersion of only a few stars in a highly crowded region, it is obviously desirable to test and improve the robustness of the result. Trippe et al. (2008) found that most of the somewhat brighter stars outside the core containing the O/WR-stars E1-4 do not share the average proper motion of these stars and are thus almost certainly unrelated. If these unrelated stars are removed from the surface density counts, the maximum of stellar density found by Paumard et al. (2006) and Schödel et al. (2007a) (see also Figure 3.2.2) becomes much less significant (Trippe et al. 2008). Fritz et al. (2010b) deconvolved the best $H/K_s/L'$ NACO images of IRS 13E between 2003 and 2008. The results are shown in the right panel of Figure 5.4.1. By adding in the fainter components in the estimate of the stellar density Fritz et al. find that IRS 13E represents a ~ 3.5σ overdensity above the local background and also derive proper motions for half a dozen additional, fainter components, as shown in Figure 5.4.1. The overall 1d velocity dispersion of all IRS 13E members identified by Fritz et al.

(2010b) is 85 ± 15 km/s, in good agreement with the dispersion of the four O/WR-stars (75 ± 21 km/s), and suggesting perhaps a central mass of a few $10^4$ $M_\odot$.

The main concern with this interpretation is the nature of the faint components in the core of IRS 13E. Are they really stars? Some of these components are closely associated with emission peaks in the ionized gas (radio continuum) and the dust emission (L' map), which are co-moving with the H/$K_s$-components (left panel of Figure 5.4.1). An analysis of the spectral energy distributions indicates that most of these knots, including IRS 13E3, are peaks of hot dust and ionized gas, with no evidence for an embedded star (Fritz et al. 2010b). The X-ray emission from IRS 13E (bottom right panel of Figure 1.1, Baganoff et al. 2003) suggests as well that wind-gas interactions play an important role. If IRS 13E3 and the fainter knots are interstellar in nature they cannot be used as independent 'test particles' for determining the underlying velocity dispersion and gravitational potential.

This leaves the remarkable clustering of two or three massive stars in a very small volume, which is statistically unlikely to be due to a random fluctuation and would thus favor the presence of a central mass (Fritz et al. 2010b). However, if such a mass in the form of an intermediate mass black hole is present, accelerations of the O/WR-stars should be observable. They are not detected, making the presence of a $10^4$ $M_\odot$ black hole unlikely at a similar level as a random fluctuation hypothesis. We conclude that ***the case for an intermediate mass black hole in IRS 13E remains tantalizing but presently unconvincing***.

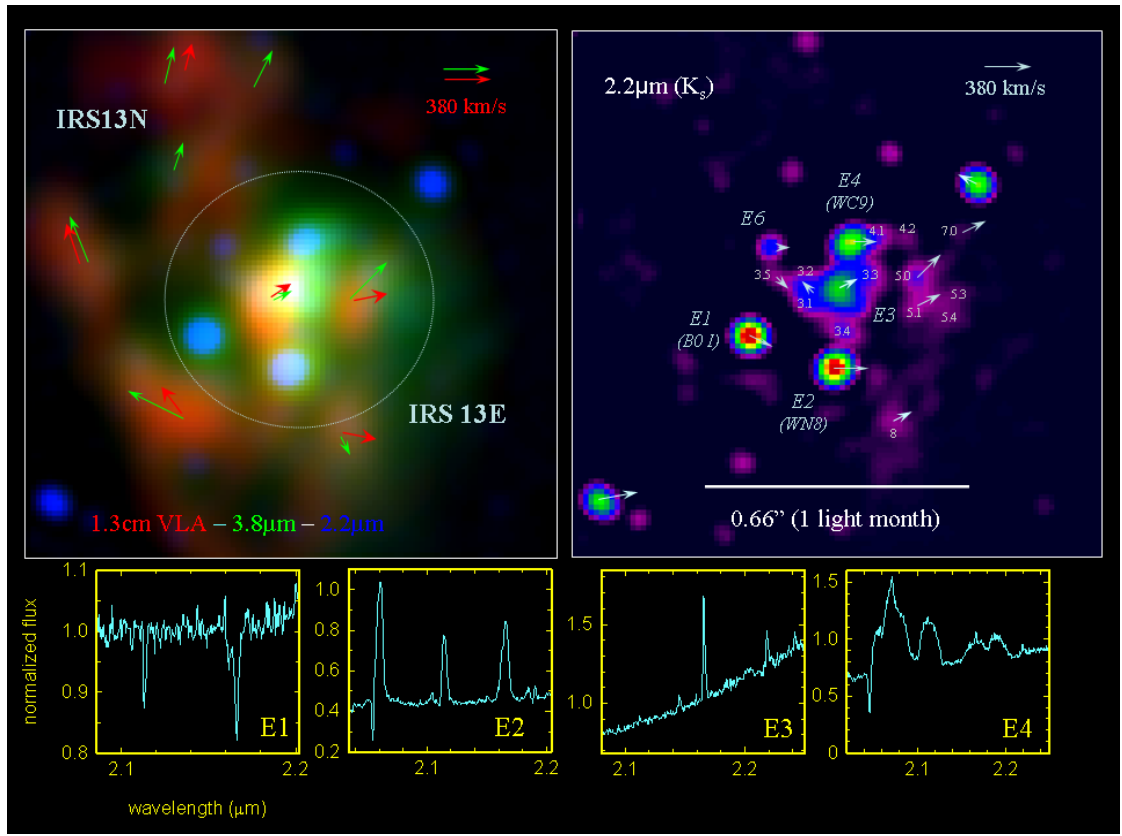

**Figure 5.4.1.** Right: High quality 'Lucy-deconvolved' NACO-AO image in the $K_s$-band of the central cusp around the WR-'cluster' IRS 13E (right, co-add of four $K_s$ images between 2002 and 2007). The restoring beam of this image has a FWHM of 40 mas and the dynamic range is about 6 magnitudes. The faintest trustworthy stars are around $K_s \sim 16.5$. Derived proper motions (Fritz et al. 2010b) are indicated by arrows. The left panel shows a three color composite of the $K_s$ image on the right (blue), an L' (3.8 µm) image (green, also NACO) and a 1.3 cm VLA image of the radio continuum (Zhao et al. 2009). Proper motions of the radio knots (red) are from Zhao et al. (2009) and of the L' knots (green) from Muzic et al. (2008). The vector lengths are on the same scale as those of the stars in the right panel. The bottom panel shows SINFONI K-band spectra of the four brightest stars in IRS 13E.

## 5.5 The distance to the Galactic Center

The distance to the Galactic Center ($R_0$) is one of the fundamental parameters for all models of the Milky Way. It sets a length scale to the models and its value has impact on distance and mass estimates of objects throughout the Milky Way and extragalactic distance estimators such as period-luminosity-relationships. Various reviews on this subject have been written (de Vaucouleurs 1983, Kerr & Lynden-Bell 1986, Reid 1989, 1993, Nikiforov 2004). The currently recommended IAU value is 8.5 kpc.

The methods available to estimate $R_0$ can be divided into 1) direct estimates, 2) indirect estimates, 3) model-based estimates and 4) others. Beyond a classical parallax, a direct estimate compares an angular dimension, such as an apparent size or a proper motion with an absolute scale, for example a radial velocity or a light travel time. Indirect measurements rely on some secondary calibration, for example a period-luminosity relationship. A model-based estimate is a more global approach, where some model of the Milky Way is fit to data and $R_0$ is one of the model parameters. Finally, there are other estimates that usually involve strong assumptions. We do not discuss such estimates here.

*5.5.1 Direct Estimates*

The most direct way to obtain a distance is to measure the **_trigonometric orbital parallax_** induced by Earth's motion around the Sun. The expected parallax of Sgr A* in front of the extragalactic background is $\approx 100$ µas. The astrometric accuracy of VLBI observations has reached the 10 µas regime (Pradel, Charlot & Lestrade 2006) and in principle one could hope for a determination of $R_0$ from radio observations. Unfortunately, Sgr A* is strongly scatter broadened (Bower et al. 2004, Shen et al. 2005), limiting the longest VLBA baseline to 1500 km. Observing a less scattered source located close enough in space to Sgr A* can lower these difficulties. The complex Sgr B2 almost certainly lies within 300 pc from the Galactic Center (Reid et al. 1988) and contains bright $H_2O$ masers. Reid et al. (2009b) have determined the parallax with VLBI measurements of these masers. Their result is $R_0 = 7.9 \pm 0.8$ kpc.

The determination of 3D **_stellar orbits_** around Sgr A* (Schödel et al. 2002, Eisenhauer et al. 2003, Ghez et al. 2005, Eisenhauer et al. 2005, Ghez et al. 2008, Gillessen et al. 2009a,b) yields another direct estimate of $R_0$. This measurement relies on the knowledge of radial velocities and proper motions for individual stars orbiting the massive black hole (Salim & Gould 1999). The difficulty is the large number of parameters that need to be determined from the same data, namely the orbital elements of each star as well as the coordinate system parameters. The first such estimate was presented by Eisenhauer et al. (2003), who obtained
$R_0 = 7.94 \pm 0.38|_{stat} \pm 0.16|_{sys}$ kpc. This work used the orbital data of one star only,

namely S2, which is in a 16-year orbit around Sgr A*. Ghez et al. (2008) and Gillessen et al. (2009a,b) have published updated geometric distance estimates. Ghez et al. (2008) find $R_0 = 8.4 \pm 0.4$ kpc from the S2-orbit, where the error is of statistical. The systematic uncertainty is of similar size. Gillessen et al. (2009b) derive $R_0 = 8.33 \pm 0.17 \pm 0.31$ kpc from a combined fit of six stars (S2, S1, S8, S12, S13, S14). These papers demonstrate that the earlier papers underestimated the systematic uncertainties dominating the current accuracy. The main problem is source confusion in the dense stellar cluster in the immediate vicinity of Sgr A*. Gillessen et al. (2009a) presented a combined fit, taking both VLT- and Keck-data sets and obtained $R_0 = 8.28 \pm 0.15 \pm 0.29$ kpc. Zucker et al. (2006) noted that fitting Keplerian instead of relativistic models might systematically underestimate $R_0$. Using the data from Eisenhauer et al. (2005), the effect amounted to $\Delta R_0 = +0.11$ kpc. Gillessen et al. (2009b) found for their data $\Delta R_0 = +0.18$ kpc.

Assuming axisymmetry of the star cluster around the rotation axis of the Galaxy, 3D velocity measurements of nuclear stars in the Galactic Center yield a ***statistical parallax*** estimate of $R_0$ from a comparison of the radial velocity and proper motion velocity dispersions. Taking into account the corotation of the cluster with the Galaxy and the finite field of view results in a small correction factor (Trippe et al. 2008). The first statistical distance estimate by Huterer, Sasselov & Schechter (1995) compared the velocity dispersion for two samples of ≈ 50 giants and yielded $R_0 = 8.21 \pm 0.98$ kpc. The most recent update is based on 664 full 3D velocities of late-type giants in the central parsec and yields $R_0 = 8.07 \pm 0.35$ kpc (Trippe et al. 2008). A statistical parallax distance to a star forming region has also been obtained for $H_2O$ masers in Sgr B2 and gives $R_0 = 7.1 \pm 1.5$ kpc (Reid et al. 1988).

*5.5.2 Indirect Estimates*

The spherical halo of ***globular clusters*** around the dynamical center of the Milky Way allows one to determine the position of the Galactic Center by measuring the three-dimensional distribution of globular clusters. In total, 154 galactic globular clusters are currently known[6]. The method is indirect since getting the distances to the individual clusters requires an additional step of cross calibration. Different results are mainly due to the subset of clusters considered and due to the choice of secondary calibration. A review of the latter is beyond the scope of this article. The most recent work by Bica et al. (2006) yields a low value of $R_0 = 7.2 \pm 0.3$ kpc (statistical error), similar to earlier works. Speculating why this is a bit shorter than other recent estimates, one might ask how reliable the apparent magnitudes obtained in the crowded stellar fields actually are. Stellar confusion would mimic too high apparent magnitudes, and would place the cluster closer than it is. Also, probably some clusters behind the Galactic Center are missing from the sample.

***RR Lyrae stars*** are variable stars with a characteristic absolute magnitude of $M(RR) \approx 0.75$. They are abundant throughout the Galaxy and trace the stellar density. The position of the Galactic Center can thus be found as the center of the population of galactic RR Lyrae stars. The main obstacle for this approach is extinction, at least in the optical. There exist a few low extinction "windows" through which RR Lyrae stars can be found out to 15 kpc. The first such work from Baade (1951) used

---

[6] A catalog of globular clusters can be found at http://physwww.mcmaster.ca/~harris/Databases.html

M(RR) = 0 and yielded $R_0$ = 8.7 kpc. A major uncertainty is the value of M(RR) and its dependence on metallicity (Reid 1993). Blanco (1985) and Carney et al. (1995) show that using either a galactic or LMC calibration, $R_0$ changes by as much as 1 kpc. Fernley et al. (1987) were the first to use infrared observations of RR Lyrae stars in the Galactic Bulge to overcome the problem of extinction. Their result was $R_0$ = 8.0 ± 0.65 kpc. Dambis (2009) argue that six different subpopulations of RR Lyrae stars with slightly different absolute magnitude calibrations can be defined and derive $R_0$ = 7.58 ± 0.40 kpc. Majaess (2010) conclude from OGLE RR Lyrae variables observed in the direction of the bulge $R_0$ = 8.1 ± 0.6 kpc.

Using stars brighter than RR Lyraes for deriving distance moduli allows one to observe stars further out and in more extincted regions. Groenewegen, Udalski and Bono (2008) used 49 *Cepheids* (bright variables with a well-defined period-luminosity relation) in the Galactic bulge. As a 'by-product', they also obtained data for 37 RR Lyrae stars. Their combined estimate is $R_0$ = 7.94 ± 0.37 ± 0.26 kpc. The apparent magnitude uncertainty due to extinction is reduced to 0.1 mag for *red giants* (and long period variables, such as Miras). The downside is that the calibration of the absolute magnitudes is more complicated for cooler stars. Glass & Feast (1982) obtained in this way $R_0$ = 8.8 kpc using a (probably too large) distance modulus of 18.69 mag for the LMC. Groenewegen & Blommaert (2005) found from a sample of 2691 Mira star candidates in the Galactic Bulge OGLE fields that $R_0$ = 8.8 ± 0.7 kpc. Matsunaga et al. (2009) use 143 Miras from a region more concentrated on the Galactic Center and obtained $R_0$ = 8.24 ± 0.08 ± 0.42 kpc.

In many cases, a stellar population as a whole has some *characteristic absolute magnitudes in the Hertzsprung-Russell-Diagram*, which can be used to determine a distance modulus. Van den Berg & Herbst (1974) for example used the position of the main-sequence turn off and reported $R_0$ = 9.2 ± 2.2 kpc. More frequently the so-called *red clump* feature was used. It appears roughly at absolute magnitude M = 0 with a typical width of ± 0.5. Paczynski & Stanek (1998) used the feature in Baade's low extinction window and found $R_0$ = 8.4 ± 0.4 kpc. Recently, Babusiaux & Gilmore (2005) conducted a deep, wide-angle NIR photometric analysis of the central regions of the Galaxy and found $R_0$ = 7.7 ± 0.15 kpc (statistical error). Nishiyama et al. (2006) gave $R_0$ = 7.52 ± 0.10 ± 0.35 kpc from an independent bulge red clump data set, obtained at a 1.4 m telescope in the NIR.

*5.5.3 Model-based Estimates*

One can determine geometric distances to star forming regions by computing a *cluster parallax distance* from the 3D-motions of $H_2O$ masers in the regions with VLBI. Combined with a kinematic model of the Milky Way, such an approach puts tight constraints on $R_0$. Genzel et al. (1981) and Schneps et al. (1981) presented the first attempts of this type and determined the distance to W51. Unfortunately that source is found in an unfavorable direction to constrain $R_0$ (Reid et al. 1988). Better suited is W49, for which Gwinn, Moran & Reid (1992) obtained $R_0$ = 8.1 ± 1.1 kpc. Recently, Reid et al. (2009a) measured *trigonometric parallaxes* of a total of 18 masers ($CH_3OH$, $H_2O$ and SiO), along with proper motions and radial velocities. Their parallaxes have errors as small as 10 μas and the proper motion errors are a few km/s. They obtained $R_0$ = 8.4 ± 0.36 ± 0.5 kpc, noting that the model parameter that actually is best constrained from the measurements is $\Theta_0/R_0$ = 30.3 ± 0.9 km/s/kpc.

Many objects in the Galaxy have been used to constrain the distance to the Galactic Center by tying a model to the spatial distribution (and sometimes radial velocities). Joy (1939) inferred $R_0$ = 10 kpc from a Cepheid-based Galaxy model. Feast (1967) extended the sample and added B-stars and open clusters to deduce
$R_0$ = 8.9 ± 1.4 kpc. Metzger, Caldwell & Schechter (1998) found
$R_0$ = 7.66 ± 0.32 kpc. Recently, Vanhollebeke, Groenewegen & Girardi (2009) used a population synthesis code to model the Galactic bulge with several classes of objects. Their best fit yielded $R_0$ = 8.7 ± 0.5 kpc. Van de Hulst, Muller & Oort (1954) mapped the 21 cm neutral hydrogen emission through a substantial part of the Milky Way. From a kinematic model they obtained $R_0$ = 8.26 kpc. Rybicki, Lecar & Schaefer (1974) sampled the rotation curve in several directions, yielding $R_0$ = 9 kpc. Honma & Sofue (1996) concluded from the geometry of the HI disk $v_0$ = 200 km/s and thus quoted a lower value of $R_0$ = 7.6 kpc.

### *5.5.4 Combined best Estimate*

Reid (1993) determined a weighted 'best average' of $R_0$ for the data available up to 1992 and concluded $R_0$ = 8.0 ± 0.5 kpc. Similarly, Nikiforov (2004) calculated an average and proposed $R_0$ = 7.9 ± 0.2 kpc. The error of the latter is however inconsistent with the scatter in the data up to 2004. Probably the sample variance was not taken into account. Figure 5.5.1 shows a comprehensive compilation of published values for $R_0$. Many of these are not independent and using a simple, weighted average is not correct. Instead, we average the most recent measurements from the main methods, with two modifications: The most recent Globular Cluster analysis by Bica et al. (2006) cites a statistical error only. The value of 7.2 ± 0.3 kpc is much lower than the Surdin (1999) value of 8.6 ± 1.0 kpc. Hence, we assign it a larger error of 1.0 kpc. Secondly, from Groenewegen, Udalski & Bono (2008) we only use the Cepheid-based value ($R_0$ = 7.98 ± 0.51 kpc) since for the RR Lyrae stars, the approach by Dambis (2009) seems to be more complete. In this way, we obtain
$R_0$ = 8.15 ± 0.14 ± 0.35 kpc, where the first error is the variance of the weighted mean and the second error the unbiased, weighted sample variance. ***Using only direct estimates yields $R_0$ = 8.23 ± 0.20 ± 0.19 kpc and additionally using the maser data from Reid et al. (2009a) gives $R_0$ = 8.25 ± 0.19 ± 0.19 kpc.***

### *5.5.5 Discussion*

The indirect methods have a long history, yet the various calibrations continue to be a serious error source. On the other hand, the direct methods will continue to improve gradually, as they have done over the last years. As a consequence, probably in a few years from now, it will become pointless to estimate $R_0$ in an indirect way.

The value of $R_0$ has little direct influence on estimates of the halo mass of the Milky Way. Smith et al. (2007) used high velocity stars from the RAVE survey to estimate the escape velocity of the Milky Way and deduced a (low) halo mass of
$1.4 \times 10^{12}$ $M_\odot$. They found that their results change by less than 1% if $R_0$ is varied by 0.5 kpc. However, there is an indirect coupling of $R_0$ and halo mass. The value of $\Theta_0$ is strongly coupled to the mass of the halo. Since $\Theta_0/R_0$ is well determined by the work of Reid et al. (2009a), an independent measurement of $R_0$ will fix $\Theta_0$ and thus influence the estimates for the halo mass.

On the other hand, the value of $R_0$ might have direct influence on the shape of the dark matter halo. Olling & Merrifield (2001) claim that if $R_0$ > 7 kpc, the halo would have to be close to spherical. This conclusion is weakened, however, by the fact that these authors also require the circular speed at the Sun to be < 190 km/s, a value which is strongly disfavored and would be inconsistent with Sgr A* being at rest in the Galactic Center. A spherical halo excludes that decaying massive neutrinos or a disk of cold molecular hydrogen are responsible for the dark matter halo. Independently, Koposov, Rix & Hogg (2010) conclude from the observation of a tidal stream that the potential flattening is $q = 0.87 \pm 0.6$ ($q = 1$ corresponds to a spherical halo).

| First Author & Year | Method | $R_0$ [kpc] | Comment |
|---|---|---|---|
| Bica 2006 | Globular Clusters | 7.2 ± 0.3 (1.0) | Error seems to be statistical only |
| Nishiyama 2006 | Red clump | 7.52 ± 0.45 | |
| Groenewegen 2008 | Bulge Cepheids | 7.98 ± 0.51 | infrared |
| Ghez 2008 | Stellar Orbit S2 | 8.4 ± 0.4 | Error is statistical only |
| Trippe 2008 | Statistical parallax | 8.07 ± 0.35 | corrects for cluster rotation |
| Vanhollebeke 2009 | Galaxy Model | 8.7 ± 0.50 | Model of Bulge |
| Gillessen 2009b | Stellar Orbits S1, S2, S8, S12, S13, S14 | 8.33 ± 0.35 | |
| Dambis 2009 | RR Lyrae | 7.58 ± 0.40 | 6 subsamples with different calib. |
| Reid 2009a | H$_2$0 masers and Galaxy Model | 8.4 ± 0.6 | |
| Matsunaga 2009 | Miras | 8.24 ± 0.43 | DM LMC: 18.45 |
| Reid 2009b | Direct parallax to SgrB2 | 7.9 ± 0.8 | |

**Table 5.5.1:** Values for $R_0$ used to calculate a weighted average. The table indicates first author, year of publication, method used and the values for $R_0$. The errors quoted are the square sum of statistical and systematic errors (if given in the original work).

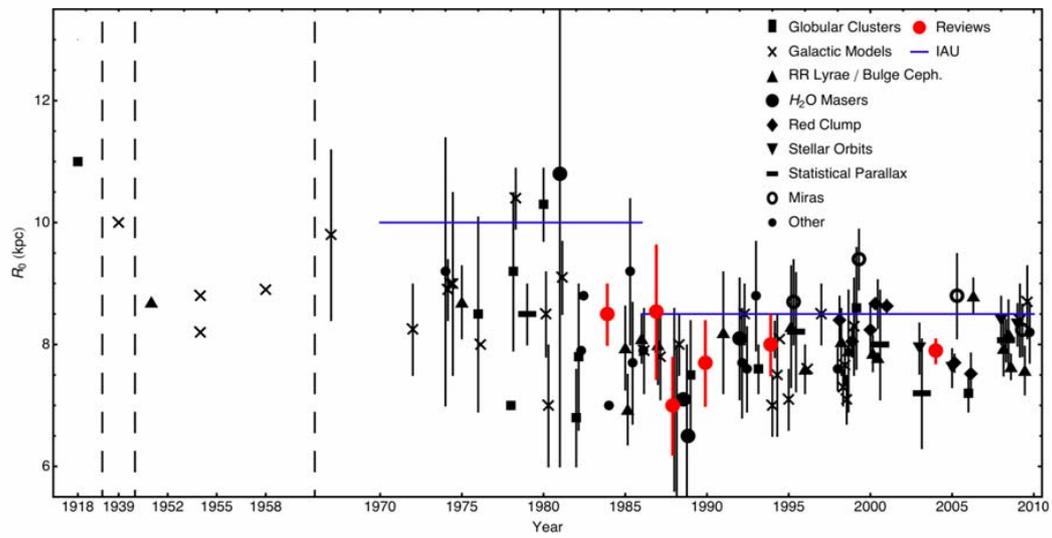

**Figure 5.5.1:** Graphical compilation of published values for $R_0$. The publication dates have been adjusted slightly such that the overlap between the various error bars is minimized. The time axis before 1970 has three gaps from 1920 to 1938, 1940 to 1950 and 1960 to 1970, denoted by vertical, dashed lines.

# 6. Paradox of youth: how did the young stars get into the central parsec?

We have discussed in section 2 that the central parsec contains ~ 200 young, massive stars, and is one of the richest 'massive star formation regions' in the entire Milky Way. This is highly surprising. If there is indeed a central black hole associated with Sgr A* the presence of so many young stars in its immediate vicinity is unexpected (Allen & Sanders 1986, Morris 1993, Ghez et al. 2003, Alexander 2005). For gravitational collapse to occur in the presence of the tidal shear from the central mass, gas clouds have to be denser than the critical '*Roche*' density,

$$n_{Roche}(R) \sim 6 \times 10^{10} \left(\frac{R}{0.1 pc}\right)^{-3} cm^{-3} \qquad (7),$$

which *exceeds by several orders of magnitude the density of any gas currently observed in the central region* (~ $10^3$ to $10^8$ cm$^{-3}$, Jackson et al. 1993, Christopher et al. 2005, Montero-Castano et al. 2009). In situ star formation in the central parsec thus *requires substantial compression* of the gas (Morris 1993). Furthermore, the near-diffraction limited AO spectroscopy with both the Keck and VLT shows that most of the cusp stars brighter than K ~ 16 mag and within 0.5" of Sgr A* appear to be normal, main-sequence B stars (Ghez et al. 2003, Eisenhauer et al. 2005, Martins et al. 2008b, Gillessen et al. 2009b). If these stars are conjectured to have formed in situ as well, the required cloud densities approach conditions in outer stellar atmospheres, which is fairly implausible. Transport of massive stars from more benign formation regions into the central core by two-body relaxation also is not straightforward. The two-body relaxation time at ~ 1 pc is equivalent to the main-sequence lifetime of stars more massive than ~ 1.5 − 2 M$_\odot$ (Figure 2.1.1). Several scenarios have been proposed to account for this '*paradox of youth*' (Ghez et al. 2003), which will be discussed in more detailed in this section. The most promising ones are in situ formation in a dense gas accretion disk that can overcome the tidal forces (Levin & Beloborodov 2003), rapid in-spiral of a compact, massive star cluster that formed outside the central region (Gerhard 2001), and re-juvenation of older stars by collisions or stripping (Lee 1987, Genzel et al. 2003a, Davies & King 2005). Another possibility is efficient transport mechanisms, such as relaxation by massive perturbers and three-body collisions (Hills 1988, Hopman & Alexander 2006b, Perets, Hopman & Alexander 2007, Alexander 2007) that act on time scales faster than the classical two-body relaxation time (Figure 2.3.3). We begin with a summary of what is known about the star formation history in the Galactic Center.

## 6.1 Star formation history in the central parsec

There are a number of complementary estimates of the Galactic Center star formation rate as a function of look-back time t. We have already discussed in § 2.2 and § 2.5 that the O- and Wolf-Rayet stars in the star disk(s) were formed in a well defined star-formation episode ($t = 6 \pm 2$ Myrs) with a top-heavy IMF (Paumard et al. 2006, Bartko et al. 2010). The results confirm and put on solid footing earlier estimates by Tamblyn & Rieke (1993: $t = 7 - 8$ Myrs), Krabbe et al. (1995:

$t = 3 - 7$ Myrs) and Tamblyn et al. (1996). Depending on star formation histories chosen (single burst vs. a somewhat extended star formation episode), and on the input metallicities and stellar tracks, any finite duration burst could not have lasted longer than about 2 Myrs. Given the total number of observed O/WR-stars ($N_{O+WR} \sim 100$) and with the best parameters of the burst ($t = 6$ Myrs, an e-folding decay time of $\tau = 2$ Myr, for $\gamma = 0.45$ to $0.85$) the resulting peak star formation rate of an exponentially decaying model is $\sim 5.5 \times 10^{-3}$ $M_\odot \mathrm{yr}^{-1}$.

The star formation rate just prior to this burst is constrained by the number of red supergiants. There are 2 (respectively 15) spectroscopically confirmed, red supergiants with $M(K) = -7.6\ldots-11.6$ (the brightest is IRS 7) within 0.5 pc (respectively 2.5 pc, Blum et al. 2003). An analysis of the stellar tracks suggests that at most one of these red supergiants (IRS 7) may have formed during the 6 Myr burst; all others had to have formed between $t \sim 10$ to 40 Myrs (Blum et al. 2003). The fraction of red supergiants in a cluster of given star formation history depends sensitively on the metallicity and the stellar tracks chosen (Maeder & Meynet 2000). Stellar rotation, mass loss, convection and metallicity all strongly affect the time an $m_\square \leq 40$ $M_\odot$ star spends on the red supergiant branch, and how many 'blue loops' occur during that time. As a result the uncertainty in converting the observed number of supergiants into a star formation rate is substantial. As an indication, Figure 6.1.1 shows an estimate of the time averaged star formation rate between $t = 1$ and $4 \times 10^7$ yrs within $R = 0.5$ pc (and 2.5 pc) if the IMF is flat as for the 6 Myr burst and with solar metallicity tracks with rotation from the work by Maeder & Meynet (2000 and references therein). This estimate would indicate that the star formation activity in this time frame on average was less by a factor of several relative to the 6 Myr burst. This is a statement about an average over a time much longer than the burst duration of the 6 Myr event, so one can obviously not exclude that the star formation rate during that period varied, with a few short epochs of active star formation and little star formation in between.

Estimates of the star formation prior to the red supergiant phase rely on quantitative modeling of the distribution in the HR-diagram of older, cool red giants, AGB stars and red clump stars. Blum et al. (2003) carried out the first detailed stellar population analysis of $\sim 80$ bright ($M(K) \leq -7.1$, $K_s \leq 10.5$) red supergiants and bright AGB stars, including very cool ($T \sim 2750$ K), dusty long period variables (LPVs) at the tip of the AGB phase, during which the stars experience a series of thermal pulses with very high mass loss rates and dust formation. Treating the complex evolution of the LPVs (20 of 78) in an approximate manner, Blum et al. (2003) compare the data and model star formation histories in four time bins, using an input IMF (Salpeter or somewhat flatter) with a variable lower mass cutoff $m_l$. The cutoff mass is then constrained to be $\sim 0.7$ $M_\odot$ from matching the integrated current stellar mass (including remnants, but subtracting mass lost during stellar evolution) to the dynamical mass of the star cluster within 2.5 pc ($\sim 1 - 3 \times 10^6$ $M_\odot$ in Blum et al.). The resulting star formation history is shown as green triangles in Figure 6.1.1. Its salient characteristic is a higher star formation rate recently ($t < 0.1$ Gyr) and perhaps again in the distant past ($8 - 12$ Gyrs), with a minimum in between. Given the fairly high flux limit, the relatively small number of stars and the uncertainties of the TP-AGB analysis mentioned above, however, it is clear that the earlier star formation history is comparatively poorly constrained. The estimate in the most recent bin ($t = 10^7 - 10^8$ yr) agrees reasonably well with the simple estimate from the number of red supergiants discussed above, despite the strong differences in assumed IMFs.

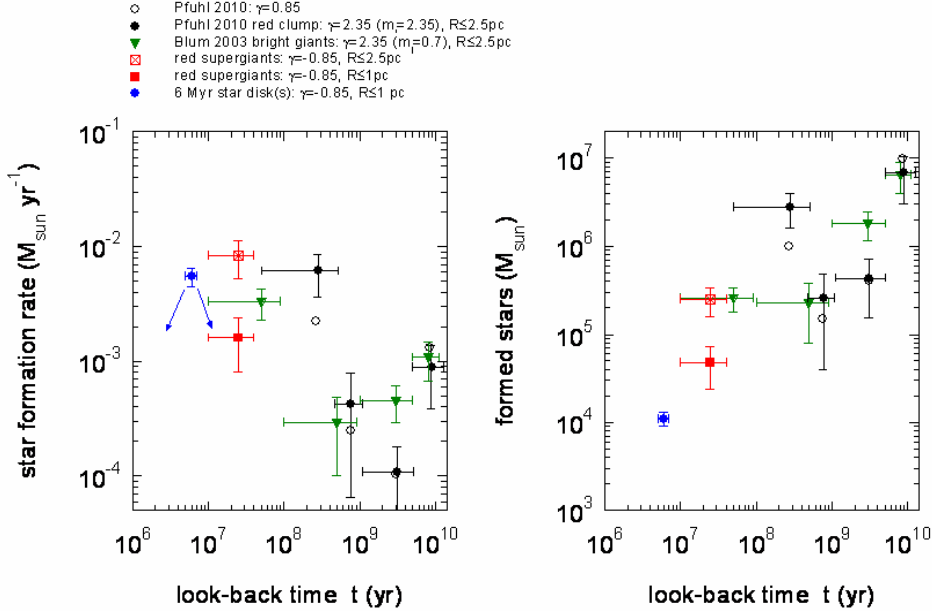

**Figure 6.1.1.** Star formation history of the central $1 - 2.5$ pc as a function of look-back time. The left panel shows the derived star formation rates in various epochs, while the right panel refers to the mass of stars formed during that epoch, thus taking into account the widely different time bins. Different types of squares denote estimates based on the assumption that the Galactic Center IMF was always flat ($\gamma = 0.85$), as proposed by Paumard et al. (2006) and Maness et al. (2007, § 6.1). Triangles denote models based on a truncated Salpeter IMF ($\gamma = -2.35$) with a lower mass cutoff of $m_l = 0.7$ M$_\odot$, favored by Blum et al. (2003). For all data points the bar in the time direction denotes the time-width of each bin. The blue filled square fits the number of O+WR stars ($N_{O+WR} = 100$) in the $t = 6$ Myr star disk(s). Red squares mark the constraints derived from the number of red supergiants inferred by Blum et al. (2003) for $R \leq 1$ pc ($N_{RSG} = 3$, filled) and $R \leq 2.5$ pc ($N_{RSG} = 15$, open-crossed). The filled green triangles denote the best fit star formation history from observations of bright red giants/supergiants (M($K_s$) $\leq -7.5$) derived by Blum et al. (2003). The filled and open black circles denote the star formation rates and masses derived from observations of red-clump stars with a $\gamma = 2.35$ and $\gamma = 0.85$ IMF, respectively (Pfuhl et al. in prep., see also Maness et al. 2007). The models assume solar metallicity. The total stellar mass formed in these models is consistent with the dynamical mass constraint of $4$ ($+5$, $-2$) $\times 10^6$ M$_\odot$ in the central 2.5 pc (Figure 5.1.1).

Maness et al. (2007) and Pfuhl et al. (in prep.) classify 300, respectively 800, late-type giants and red clump stars within 1 pc of Sgr A* with deep, adaptive optics assisted, $K_s$-band integral field spectroscopy. These observations represent the deepest spectroscopic data set so far obtained for the Galactic Center (~ 6 magnitudes deeper than Blum et al.), reaching a 50% completeness threshold at the approximate magnitude of the 'red clump' ($K_s = 15.5$ mag, M(K) = $-1.5$). Maness et al. (2007) and Pfuhl et al. (in prep.) then proceed as in Blum et al., combining the spectroscopic indices with extinction corrected photometry to construct an HR-diagram and then comparing with Padua tracks and various star formation histories through a Monte Carlo likelihood analysis. This method has the obvious advantage that it samples a well populated part of the HR-diagram (the helium-burning red clump/horizontal branch), in which stellar evolutionary models are fairly well understood. The disadvantage is that it attempts to construct the overall star formation history of the central parsec by sampling only a relatively small fraction of the total volume. Maness et al. (2007) find that the average red clump star is hotter and more luminous than predicted in the Blum et al. (2003) model, or in most other obvious input scenarios based on a standard IMF. Maness et al. then conclude that either the IMF is top-heavy or that there was little star formation prior to a few Gyrs ago. Using an improved calibration of the equivalent width of the CO-bands Pfuhl et al. (in prep.) find that this temperature excess disappears and that the red-clump data are fully commensurate with a standard IMF. The filled and open black circles in Figure 6.1.1 give the derived star formation history from the work of Pfuhl et al. To within the uncertainties the Pfuhl et al. results are in excellent qualitative and quantitative agreement with the Blum et al. (2003) model.

It thus appears that *most of the stars in the Galactic Center formed about 9±2 Gyrs ago, plausibly at the same time as the Galactic Bulge. The star formation rate then dropped to a minimum a few Gyrs ago*. This minimum is also consistent with the low abundance of A-stars in the central parsec found in the Pfuhl et al.(2010) study. *For the past few hundred million years the star formation rate in the central few parsecs has been approximately constant at a level of a few $10^{-3}$ $M_\odot yr^{-1}$* and adding no more than $10^6$ $M_\odot$ of stellar mass during that epoch. As witnessed by the star formation episode ~ 6 Myrs ago, this roughly constant rate of star formation may have consisted of a number of *star formation events*. This scenario may be consistent with the picture of a recurring 'limit cycle' developed by Morris & Serabyn (1996).

## 6.2 In situ star formation or in-spiral of a star cluster?

In § 2.2 and § 2.3 we discussed the empirical evidence for the 'paradox of youth'. In the following we review the recent theoretical work on the two most plausible scenarios for explaining this remarkable population of massive stars. *One is that dense gas fell into the nucleus about 6 Myrs ago and formed a disk around the black hole*. Shock compression and cooling drove the disk into global gravitational instability, resulting in an intense star formation episode (Morris 1993, Genzel et al. 1996, Levin & Beloborodov 2003). *The alternative is that a dense and massive star cluster formed outside the hostile central parsecs, subsequently spiraled into the nuclear region by dynamical friction and then finally was disrupted tidally there* (Gerhard 2001). The following section then discusses fast relaxation, scattering and migration processes that could bring young stars into the immediate vicinity of the hole.

*6.2.1 Star formation near the black hole*

Theoretical considerations lead to the generic prediction that parsec-scale gas disks around a central black hole can become self-gravitating and form stars (Paczynski 1978, Kolykhalov & Sunyaev 1980, Lin & Pringle 1987, Shlosman & Begelman 1989, Collin & Zahn 1999, Goodman 2003). This theoretical work predated by more than a decade the discovery of the stellar disks in the Galactic Center (§ 2.2) and M31 (Bender et al. 2005).

A cooling, self-gravitating gas disk of mass $M_{disk}$ around a black hole of mass $M_\bullet$ becomes unstable to fragmentation on a dynamical time scale if the Toomre (1964) $Q$-parameter drops below unity,

$$Q = \frac{c_s \Omega}{\pi G \Sigma_{disk}} \sim \left(\frac{h_z}{R}\right)\left(\frac{M_\bullet}{M_{disk}}\right) \leq 1 \qquad (8).$$

Here $c_s$ is the effective sound speed (or turbulent speed), $h_z$ the z-scale height, $\Omega = v/R$ the orbital frequency of the disk, and $\Sigma_{disk}$ its surface density. Goodman (2003) and Nayakshin & Cuadra (2005) use standard accretion disk theory and energy arguments to show that radiative feedback from star formation or the central black hole are probably insufficient to prevent a disk with an accretion rate above ~ 0.01 times the Eddington rate from becoming strongly self-gravitating at a distance of $10^4$ - $10^5$ Schwarzschild radii from the central black hole (0.1" − 1" for the Galactic Center).

Larson (2006) argued that at the high temperatures likely prevalent near a black hole newly-formed stars should have large average masses, simply because of the temperature dependence of the Jeans mass, $M_{Jeans} \sim T^{3/2} n^{-1/2}$. However, the high densities required for overcoming the large tidal shear might also drive the Jeans mass downward, rather than upward. For this reason the initial fragmentation mass in a $Q \sim 1$ disk around a $10^6$ $M_\odot$ black hole may be quite small, $M_{frag} \sim 0.01$ $M_\odot$ (Collin & Hure 1999). Levin (2007, see also Levin & Beloborodov 2003) applies results from protostellar disk theory and numerical results from Gammie (2001) to argue that the initially fragmenting gas clumps can grow rapidly by additional accretion and clump merging in the disk to a high final mass. This mass could possibly approach the maximum 'isolation mass' at the radius of the orbiting initial clump (Lissauer 1987, Milosavljevic & Loeb 2004). A lower limit to this isolation mass is given by the disk mass contained with an annulus of width $2h_z$,

$$M_{isolation-1} \sim 4\pi R h_z \Sigma_{disk} \underset{Q=1}{=} 4\left(\frac{M_{disk}}{M_\bullet}\right) M_{disk} \sim 100 \left(\frac{M_{disk}}{10^4 M_\odot}\right)^{3/2} M_\odot \qquad (9).$$

An upper limit to the isolation mass is the maximum accreted mass in an annulus that is stable to tidal disruption as given by Lissauer (1987),

$$M_{isolation-2} \sim \beta \left(\frac{M_{disk}}{M_\bullet}\right)^{1/2} M_{disk} \sim 500 \beta \left(\frac{M_{disk}}{10^4 M_\odot}\right)^{3/2} M_\odot \qquad (10),$$

where $\beta \sim O(10)$ is a constant. The isolation mass probably cannot be reached because of the opening of gaps in the disk. Levin (2007) estimates a reasonable upper mass range between 10 and $10^2$ $M_\odot$.

An isothermal $Q \sim 1$ disk has a surface density that scales with radius as $\Sigma_{disk} \propto R^{-1.5}$. If the mass of stars scales as the (Lissauer 1987) isolation mass, these stellar masses in turn scale as $R^{0.75}$ and the surface number density of stars scales as $R^{-2.25}$ (Levin 2007). These simple estimates are remarkably, and perhaps fortuitously, close to the observed surface density of stars in the O/WR-stars in the Galactic Center (§ 2.2). In the Galactic Center the current star disk(s) have $M_\bullet/M_d$ of a few $10^2$ and $h_z/R \sim 0.1$ so that $Q >> 1$. For star formation to have occurred in these disk(s) 6 Myrs ago, the initial gas disk would have to have been much thinner, or the star formation efficiency would have to have been low (<< 10%), or a combination of both. Since stellar disks always increase their velocity dispersions with time and the current disk(s) are well defined, it is quite likely that the initial gas disk(s) indeed was (were) quite thin.

Unavoidably these analytical considerations are too simplistic. A number of numerical simulations have explored the physical processes in more detail. Sanders (1998) was the first to consider the fate of a low angular momentum molecular cloud entering the high tidal shear region in the central parsec. With a sticky particle scheme he calculated the evolution of clouds on different orbits. The generic feature of these simulations is the formation of a '***dispersion ring***' (top left panel of Figure 6.2.1). The infalling cloud is tidally stretched into a long filament. When the upstream part of this filament reaches peri-center it starts losing angular momentum through shocks. As a result the gas orbits in a near-circular or elliptical orbit near that radius. Downstream gas in the filament on the first passage crashes into the upstream gas on its second passage, increasing dissipation and resulting in circularization on an orbital time scale. The result is the formation of a rapidly cooling, dense inner gas disk or ring that is prone to fragmentation and star formation.

Nayakshin, Cuadra & Springel (2007) report the first in a series of recent SPH simulations of a cooling disk with a simple cooling time recipe, $t_{cool} = b\, t_{dyn}$, motivated by earlier work of Gammie (2001). They treat star formation through a sink particle technique when the local density significantly exceeds the tidal (Roche) critical density and include growth of the initial stellar cores by accretion through a Bondi-Hoyle formalism with an upper limit given by the Eddington rate. As predicted from the analytical work, stellar cores in this relatively 'benign' simulation grow substantially through accretion and mergers. Nayakshin et al. find that the accretion disk forms stars vigorously inside out. The rate of fragmentation into cores and the resulting stellar mass function are strongly dependent on the parameter b. Fragmentation sets in for $b \leq 3$ (Gammie 2001). Just below that critical limit the IMF is strongly top-heavy and dominated by very massive (~ 100 $M_\odot$) stars. For smaller b the disk fragments more rapidly and the IMF becomes more 'normal'. Alexander et al. (2008a) arrive at similar conclusions and, in addition, find that for smaller b there appears to be a high abundance of captured binaries. Alexander et al. (2008a) conclude that the fragmentation processes are not significantly different for eccentric gas disks. However, the average stellar mass in eccentric gas disks is higher than in circular disks since the stronger and variable tidal forces near peri-center suppresses the fragmentation of weakly bound clumps.

Bonnell & Rice (2008) go beyond these 'static' simulations and study the dynamical result of the plunging of a low angular momentum giant molecular cloud ($10^4$ and $10^5$ $M_\odot$) into the central parsec around a $10^6$ $M_\odot$ black hole (top middle panel of Figure 6.2.1). Their SPH scheme also includes radiative transport. The basic sequence of events resembles the findings of Sanders (1998): tidal disruption and stretching of the cloud, compressional heating at the impact parameter radius of ~ 0.1 pc, dissipation and capture of ~ 10% of the cloud's initial mass in an eccentric, clumpy and filamentary disk near this circularization radius, rapid fragmentation and star formation. The resulting stellar masses depend on the balance between compressional heating, radiative cooling and optical depth effects. Near the inner radius the heating is largest, increasing the gas temperature to several thousand K, and the resulting typical stellar mass to $10 - 50\ M_\odot$ for the $10^5\ M_\odot$ case, matching the prediction by Larson (2006). The resulting IMF in this simulation is very top-heavy (top right panel of Figure 6.2.1) but much less so for the $10^4\ M_\odot$ case. Another SPH simulation for a $4 \times 10^4\ M_\odot$ infalling cloud by Mapelli et al. (2008) yields similar results. Molecular line mapping shows that there are several smaller molecular clouds in the central 10 pc that probably are on highly eccentric orbits as required by this scenario (Stark et al. 1991, Jackson et al.1993, Mezger, Duschl & Zylka 1996). Wardle & Yusef-Zadeh (2008) have proposed the alternative scenario that a small central gas disk forms by partial capture of a large, massive cloud, such as the '+50 km/s' cloud. As such a cloud engulfs the Sgr A region on part of its orbit, low angular momentum gas could be created from cancellation of the angular momentum on either side of Sgr A*. Such an 'engulfing' cloud may perhaps also be capable of simultaneously forming stars with opposite angular momenta.

Hobbs & Nayakshin (2009) explore the more complex situation of what happens when two clouds fall into the central parsec simultaneously and collide there, as motivated by several of the observational papers to account for the simultaneous presence of clockwise and counterclockwise young stars (bottom left in Figure 6.2.1). The numerical approach is the same as in Nayakshin, Cuadra & Springel (2007) but in this case two clouds of different mass are injected at $R \sim 1$ pc with somewhat elliptical orbits and at large angles with respect to each other. Low angular momentum gas created in the first collision settles at $R \sim 0.04$ pc and forms a dense small disk that forms many (high mass) stars. Higher angular momentum gas on different trajectories is initially tidally unstable and first has to cool before it can form (somewhat lower mass) stars in several outer filaments. Lower angular momentum gas falls back onto the central disk and forms additional stars in an overall strongly warped inner disk. The distribution and angular momentum distribution of gas and stars of different masses near the end of the simulations at $t \sim 10^5$ yrs are shown in the bottom right panel of Figure 6.2.1. The angular momentum orientations of inner disk and outer filament stars still reflect the initial conditions of the two clouds. The inner disk is close to circularization while the filaments contain stars with elliptical orbits. Overall the salient results of these simulations match the data reasonably well (not so the radial changes in the IMF). This includes the strong warping and, in particular, the steep overall stellar surface density distribution ($\Sigma_* \sim p^{-2}$). The details of the 3D-spatial and velocity structure depend sensitively on the initial conditions of the colliding clouds. For instance, fairly massive star clusters form in some but not in other simulations. In all simulations the amount of gas accreted through the inner disk radius is large, enough to provide Sgr A* with gas to radiate near its Eddington throughout the star formation episode. This is in contrast to Nayakshin, Cuadra & Springel (2007) who found little gas to accrete onto the central black hole. The work of Hobbs & Nayakshin (2009) shows that to create a highly warped main clockwise disk with a significant population of surrounding counterclockwise stars at the time of formation probably does require the collision of two separate gas clouds.

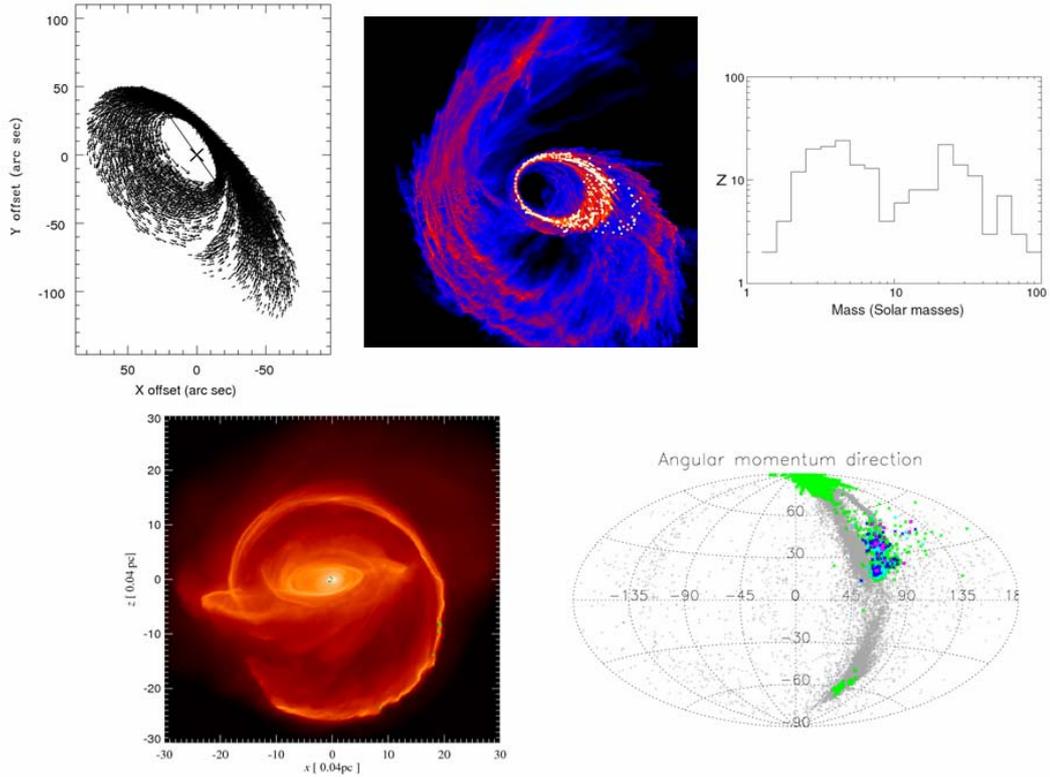

*Figure 6.2.1.* Simulations of star formation near a massive black hole. Top left: snapshot of the velocity vectors of the gas of a tidally stretched, infalling gas cloud that is beginning to fall back on itself near the circularization radius and that then forms a 'dispersion ring' (from Sanders 1998). Top middle: The final state of the simulation of a $10^5$ M$_\odot$ molecular cloud falling toward a $3 \times 10^6$ M$_\odot$ supermassive black hole. The image shows the region within 0.25 pc of the black hole located at the center; colors denote column densities from 0.75 g cm$^{-2}$ to 7500 g cm$^{-2}$. A portion of the cloud has formed a disk around the black hole, while—at the stage shown here—most of the mass is still outside the region shown. The disk fragments very quickly, producing 198 stars with semi-major axes between a = 0.04 pc and a = 0.13 pc and eccentricities between e = 0 and e = 0.53. Top right: The resulting IMF of the stars formed in the top middle simulation. Because of the high temperatures of the compression shock in this simulation, the resulting IMF is top-heavy. Both panels adopted from Bonnell & Rice (2008). Bottom left: Gas density (red colors) and newly formed stars $t \sim 10^5$ yrs after the collision at $R \sim 1$ pc of two somewhat unequal mass molecular clouds of mass $\sim 3 \times 10^4$ M$_\odot$ on slightly elliptical orbits. Stars with mass $\leq 1$ M$_\odot$ are marked green, $1 - 10$ M$_\odot$, $> 10$ M$_\odot$ stars blue and $> 150$ M$_\odot$ stars magenta. The line of sight is along the z-direction. This is simulation 1 of Hobbs & Nayakshin (2009). Bottom right: Location on the sky (same coordinate system as in § 2.4) of stars (colors) and gas particles (grey). This is simulation 2 of Hobbs & Nayakshin (2009.

In summary these recent simulations suggest that *rapid dissipation and cooling in a molecular cloud plunging into the vicinity of a massive black hole can plausibly overcome the demanding critical Roche density at ~ 0.1 to 1 pc*. *The resulting clumpy, filamentary disk fragments and may form stellar clumps efficiently*. Initially eccentric cloud orbits may result in eccentric stellar orbits. More complex scenarios including several cloud systems may create more than one stellar system and also include warps. All these results are in good qualitative agreement with the observations in the Galactic Center. Nevertheless caution is warranted given that star formation is theoretically not fully understood even in more benign environments. Top-heavy IMFs seem to be the natural outcome of several of these simulations, although for different reasons. In the 'dynamic' models of infalling clouds the top-heavy IMF appears to be mainly the consequence of the high Jeans mass and rapid accretion in the post-compression gas, while in the more 'static' models the top-heavy IMF is a result of more prolonged accretion and merging of the initially small stellar cores. These differences can be easily understood from the basic underlying physics but do not (yet) answer convincingly the question what causes the top-heavy IMF in the Galactic Center.

### *6.2.2 In-spiral of a massive star cluster*

As mentioned in the introduction of § 6 the relaxation time scale $T_{NR}$ is too long for any star more massive than about 1.5 M$_\odot$ to drift within its main-sequence lifetime into the central region from significantly outside the central parsec (Morris 1993). However, the mass segregation time scale of an object of mass $M_{cl}$ is $<m*>/M_{cl}$ times smaller than $T_{NR}$. Gerhard (2001) proposed that a massive star cluster might have formed outside the hostile central few parsecs, then spiraled in by dynamical friction, getting tidally disrupted on its way in and finally depositing its massive stars as a disk near the center. To explain the Galactic Center disk(s) the in-spiral process cannot take longer than the current age of the massive stars (~ 6 Myrs). The dynamical friction in-spiral time from an initial radius $R_{i,10}$ (in units of 10 pc) to the central parsec is given by (Binney & Tremaine 2008, Gerhard 2001)

$$t_{df} \sim 3x10^6 \varepsilon R_{i,10}^2 v_{130} M_{cl,5}^{-1} \quad \text{yr} \qquad (11),$$

where the cluster mass $M_{cl,5}$ is in units of $10^5$ M$_\odot$ and the initial velocity of the cluster is in units of the circular speed (~ 130 km/s) at 10 pc. The number $\varepsilon \sim 2$ takes into account that if the cluster is tidally disrupted en route only a fraction of the mass arrives at the Center and the in-spiral time is correspondingly longer. This equation shows that young clusters similar to the Arches or Quintuplet cluster (mass ~ $10^4$ M$_\odot$, Figer 2008) can in-spiral in 6 Myrs from $R_i$~a few parsecs, while $R_i$~ 10 pc requires much more massive clusters than those observed currently in the central tens of parsecs (Figer 2008, Ivanov et al. 2005).

A second requirement is that the radius at which the cluster is tidally disrupted ($R_{dis}$) must be comparable to the final radius at which the massive stars in the core of the cluster are deposited. This places a lower limit on the mean density of the star cluster, which is (Gerhard 2001)

$$\langle\rho\rangle_{dis} \sim 10^{7.5} M_{\bullet,4.4} \left(\frac{R_{dis}}{0.4pc}\right)^{-3} \quad \text{M}_\odot pc^{-3} \qquad (12).$$

This is obviously a tough criterion even at $R_{dis}$ ~ 0.4 pc (10") and still more difficult at smaller radii. For comparison, the Arches and Quintuplet cluster have mean densities of $10^{5.6}$ and $10^{3.2}$ M$_\odot$pc$^{-3}$ (Figer 2008), far below this limit. The in-spiral mechanism thus faces the combined challenges of requiring an unusually massive and dense cluster, when compared to currently observed, circum-nuclear star clusters.

Numerical simulations with SPH or direct N-body codes have confirmed these simple analytical estimates and show that the in-spiral hypothesis can only be met with fairly extreme assumptions (e.g. Kim & Morris 2003). Portegies Zwart,

McMillan & Gerhard (2003) make the important point that the high-density criterion may be achievable in the post core-collapse state of a stellar cluster. They find that with suitable combinations of initial virial radius, initial Galacto-centric radius and mass, the cluster undergoes core-collapse during the in-spiral phase and thus can survive to a smaller final radius. Core collapse clusters are mass segregated and deposit their most massive stars preferentially at small Galacto-centric radii, thus matching the observed top-heavy in-situ mass function in the Galactic Center disk(s). However, in all cases where core collapse occurred (for an initial cluster mass of $6 \times 10^4$ M$_\odot$) the final radius reached in the simulations is still ~ 1 pc, far outside the current location of the density maximum of the O/WR-stars ($R_{peak}$ ~ 0.1 pc). An initial elliptical orbit helps some but does not solve the problem that *even massive clusters do not appear to penetrate to a small enough final radius* (Kim & Morris 2003).

An *intermediate mass black hole* of mass $10^3 - 10^4$ M$_\odot$ would *stabilize the in-spiraling cluster core against tidal disruption and thus permit deeper penetration* (Hansen & Milosavljević 2003). As the intermediate mass black hole sinks further into the central region it drags massive stars in the cluster's core along with it. Such an intermediate mass black hole with mass fraction ~ $1 - 2 \times 10^{-3}$ of the cluster mass can form plausibly during core collapse of a dense star cluster (Portegies Zwart & McMillan 2002, Gürkan, Freitag & Rasio 2004, Fuji et al. 2009). Portegies Zwart et al. (2006) even conclude that 10% of all massive star clusters born in the central 100 pc form intermediate mass black holes and that 50 such intermediate mass black holes might be presently located within 10 pc of Sgr A*. In their simulations of $10^5$ to $2 \times 10^6$ M$_\odot$ clusters with a range of central densities and masses of embedded intermediate mass black holes, Kim, Figer & Morris (2004) and Gürkan & Rasio (2005) show that the inclusion of an intermediate mass black hole indeed helps bringing in more massive stars deeper into the potential of the Galactic Center. The most successful models, however, have uncomfortably large cluster and intermediate mass black hole masses. In one of the $10^5$ M$_\odot$ simulations of Kim et al. (2004) the central cluster density is low enough to come close to the central density for the early phase of the Arches cluster and deposits about 10% of the original cluster mass ($8 \times 10^3$ M$_\odot$) in the central parsec, reasonably close to the observed stellar content of the clockwise star disk (Paumard et al. 2006). The intermediate mass black hole in this simulation has a mass of $2 \times 10^4$ M$_\odot$, which would seem implausibly high in terms of the formation scenario of Portegies Zwart & McMillan (2002) and Gürkan, Freitag & Rasio (2004). Kim et al. (2004) express doubt that intermediate mass black holes can realistically be of help in solving the deep penetration problem.

The cluster in-spiral scenario also faces some problems explaining the observed properties of the star disk(s) in the Galactic Center. While the simulations match qualitatively the top-heavy mass function, the sharp inner edge and the thickness of the observed population, the smearing in radius of the tidally disrupted cluster core results in a fairly shallow surface density distribution, $\Sigma_\Box \sim p^{-0.75}$, while the observed distribution is much steeper (Berukoff & Hansen 2006). Further, if the IMF of the initial cluster is close to a 'normal' Chabrier/Kroupa distribution there should be 3 times as many $K_s \leq 16$ B-stars as O/WR-stars, which would be expected to be deposited just outside the surface density peak of the more massive stars. We have discussed in § 2.4 that this 'sea' of B-stars is not seen in the current data. There is also no direct evidence for the presence of intermediate mass black holes in the central parsec. A ~ $10^4$ M$_\odot$ black hole outside the central few arcseconds is consistent with the current data (Figure 4.6.1) but the evidence for an intermediate mass black hole in IRS 13E, for instance, is currently not convincing (§ 5.4).

In summary of this section it is clear that Gerhard's (2001) very promising cluster in-spiral proposal faces a number of theoretical and observational challenges. To complete the in-spiral from a more benign formation region in a dense molecular cloud more than a few parsecs from Sgr A* within ≤ 6 Myrs requires cluster masses much above the masses of young clusters currently observed in the central Bulge. Even if one assumes that an unusually massive, $10^5 - 10^6$ M$_\odot$ cluster did form and spiral into the central region, the cluster had to have been just dense enough to undergo core collapse during its travel and form an unusually large intermediate mass black hole to have a chance to deposit its O-stars at $R_{dis} \ll 1$ pc. The surface density profile of the observed O/WR-stars is probably too steep and the outer onion shell of somewhat lower mass B-stars is missing. Application of ***Occam's razor or a score card suggests to us that the cluster-inspiral scenario is not favored by the current observations*** (Paumard et al. 2006, Lu et al. 2009, Bartko et al. 2009, 2010).

## 6.3 Origin of B-stars in the central cusp: scattering or migration?

What is the explanation for the remarkable central S-star cluster containing ~ 80 $K_s \leq 17.5$ stars within 0.04 pc? It is now clear from the orbits and radial distributions that these S-stars are not directly part of the central O/WR-disk(s). An additional mechanism is required to account for their presence so close to the central black hole, above and beyond the options we have discussed in the last section. Two contenders have emerged as the most likely explanations. The first is the interaction of a massive binary star with the massive black hole, in which the binary is disrupted, one member captured and the other expelled at high velocity (Hills 1988). A variant of this mechanism is the capture or ejection of an incoming single star by a three-body interaction between star, massive black hole and an intermediate massive black hole in its vicinity (Yu & Tremaine 2003). To explain the observed number of S-stars successfully, Hill's proposal requires an efficient re-supply of massive binaries on highly elliptical orbits. One possible such mechanism is the interaction of binaries with giant molecular clouds at parsec distances from Sgr A* (Perets, Hopman & Alexander 2007). The second contender is the migration of massive stars from the disk(s) at 0.1 − 1 pc to the center of the cusp, perhaps aided by an intermediate mass black hole (Levin 2007, Fuji et al. 2010). Other possible explanations, such as the collision and merging of lower-mass stars (Genzel et al. 2003a), the stripping of large intermediate mass, late-type stars (Davies & King 2005, Dray, King & Davies 2006), or three-body exchange interactions of single, near-parabolic B-stars with a stellar black hole in the potential of the MBH, where the similarity in mass of the B-stars and stellar black holes enhances the exchange cross-section (Alexander & Livio 2004), now have lost favor because of low rates or other incompatibilities with the data.

### *6.3.1 The Hills capture mechanism*

Hills (1988) considered the fate of a binary of total mass $M_{12}$ and semi-major axis $a_{12}$ in a near-loss cone orbit around a massive black hole. The simulations of Hills (1991), Yu & Tremaine (2003) and Gualandris, Portegies Zwart & Sipior (2005) suggest that in > 50% of the encounters the binary is destroyed, with one member captured in a tight and highly elliptical orbit and one ejected as a hyper-velocity star with a velocity up to 4000 km/s (but see Sari, Kobayashi & Rossi 2010). Hills (1988) proposed that the detection of such hyper-velocity stars (HVS) emanating from the Galactic Center would be strong evidence for the presence of a central black hole. Gould & Quillen (2003) were the first to propose the 'Hills' capture process for the origin of the S2 orbit. In the Hills mechanism initial binary semi-major axis and final capture semi-major axis $a_{capture}$ are proportional to each other,

$$a_{capture} \sim a_{12} \left(\frac{M_\bullet}{M_{12}}\right)^{2/3} \sim 3.6 \times 10^3 \left(M_{12}/20 M_\odot\right)^{-2/3} a_{12} \quad (13),$$

so that orbits in the S-star cluster ($a_{S\text{-}stars} \sim 0.1" - 1" \sim 800 - 8000$ AU) would have to originate from binaries with $a_{12} \sim 0.1 -$ a few AU. Such orbital semi-major axes are in fact consistent with the average massive binary separation in the disk of the Milky Way ($<a_{12}> \sim 0.2$ (+0.6, −0.15) AU, Kobulnicky & Fryer 2007). The predicted initial ellipticity of the captured binary star member is very high (Gould & Quillen 2003, Ginsburg & Loeb 2006, Alexander 2007),

$$e_{capture} \sim 1 - \left(\frac{M_\bullet}{M_{12}}\right)^{-1/3} \sim 0.98 \quad (14).$$

The second member of the binary is ejected at a velocity

$$<v_\infty> \sim \left(\frac{1.4 G M_\bullet}{a_{capture}}\right)^{1/2} \sim 2400 \left(\frac{a_{capture}}{0.12"}\right)^{-1/2} \left(\frac{M_\bullet}{4.3 \times 10^6 M_\odot}\right)^{1/2} \text{ km/s} \quad (15).$$

A star ejected with a radial velocity of $\sim 10^3$ km/s from the Galactic Center will escape the Galaxy with $400 - 500$ km/s at R $\sim 100$ kpc (Kenyon et al. 2008).

Are the properties of the S-star cluster compatible with the 'Hills' proposal? First, we have shown in § 2.3 that the brightest S-stars are normal, main-sequence B-stars with a low rotational velocity (Eisenhauer et al. 2005, Martins et al. 2008a). Second, their orbital angular momentum distribution appears to be random on the sky, both inside and outside the central arcsecond (Gillessen et al. 2009b, Bartko et al. 2010). We have thirdly shown in § 2.5 that the KLF of the S-star cluster is steep and plausibly consistent with a normal (i.e. Galactic disk) IMF. All three facts are broadly consistent with a 'field' origin of the B-stars. The angular momentum orientations in the central arcsecond could have been randomized by a combination of vector resonant relaxation for all S-stars (Figure 2.3.3) and Lense-Thirring precession for S2 and perhaps S14 (Figure 2.3.3, Levin & Beloborodov 2003). The random orientation of the B-stars at $R > 1"$, however, may give a hint that these stars did not form in a planar structure, such as the O/WR-star disk(s).

Because of the simple scaling relationship between initial binary semi-major axes and orbital semi-major axis of the captured star in the Hills mechanism, the initial binary orbit distribution function should map directly into the semi-major axis distribution of captured stars since the time scale for energy relaxation is longer than the lifetime of B-stars. Taking into account the mechanism that puts binaries onto near loss-cone orbits, the semi-major axis distribution of the S-stars also reflects the periapse dependence of this process. This dependence is weak when the binaries gradually diffuse to near-radial orbits, the so-called 'empty loss-cone regime', and the semi-major axis distribution of the S-stars mirrors the one of the original binaries. In the 'full loss-cone regime' that distribution actually is a product of the intrinsic semi-major axis distribution and the periapse dependence of the scattering rate (Perets & Gualandris 2010).

The observed semi-major axis distribution derived from the orbital parameters of Gillessen et al. (2009b) fits a constant number distribution in $log(a_{12})$ space, similar to 'Öpik's' distribution (Öpik 1924), which plausibly approximates the major axis distribution of massive binaries in the solar neighborhood (e.g. Kobulnicky & Fryer 2007). The 'Öpik' distribution corresponds to a surface density distribution scaling as $\Sigma_{\Box} \sim p^{-2}$. This means that also the distribution of the B-stars outside of the S-star cluster in principle may originate from an 'Öpik' distribution and may be related to the original binary distribution. Given the substantial uncertainty of the input binary semi-major axis distribution (F. Martins priv.comm.) this agreement cannot be more than a plausibility check, however.

The highly eccentric orbital distribution of the captured binary members in the Hills process is obviously different from the only slightly super-thermal eccentricity observed distribution of the S-stars. Does this eliminate the Hills mechanism? Perets et al. (2009) carried out simulations of the interaction of a population of highly elliptical 'Hills' orbits, immediately in the aftermath of the capture, with the central cusp, including $10^3$ stellar black holes with 10 $M_\odot$ distributed isothermally around Sgr A*. Perets et al. find that within the central arcsecond (but not outside) the combination of the dense stellar black hole cusp and resonant relaxation can thermalize and isotropize an initial highly elliptical input distribution over a time of about 20 Myrs. This is consistent with the radial dependence of the relaxation times shown in Figure 2.3.3. Perets et al. infer a probability of 93% for the final model distribution of B-stars and the empirical distribution of S-stars to be drawn from the same distribution (bottom left panel of Figure 6.3.2). While this explanation is applicable to most of the fainter S-stars, the brighter and more massive star S2 has a main-sequence lifetime of ~ 6 Myr. Figure 6.3.2 shows that over this shorter time span the model distribution would still be somewhat more elliptical than the observed one. This would indicate that at least S2 must have been deposited within the last 6 Myrs and its current ellipticity (0.88) may still reflect more closely its orginal ellipticity. Levin & Beloborodov (2003) pointed out that the Lense-Thirring precession time scale for S2 is probably short enough to erase memory of its original angular momentum orientation. Bar-Or et al. (in prep.) show that also the existence of a dense, dark cusp would be make the Hills mechanism plus post-capture resonant relaxation evolution picture fully consistent with observed eccentricities of the S-stars, including the short-lived S2.

A variant of the Hills (1988) process is an exchange interaction between an incoming single star with a central binary consisting of the massive black hole and a stellar or intermediate mass black hole in its vicinity (Yu & Tremaine 2003, Hansen & Milosavljević 2003, Alexander & Livio 2004, Levin 2006, Yu, Lu & Lin 2007). In the three-body interaction energy exchange can result in capture of the B-star and ejection of the stellar black hole, or the star can also get ejected as a hyper-velocity star, which can then escape the Galaxy (Yu & Tremaine 2003). In the latter case, no star is captured of course and this process cannot account for the S-stars, but for the HVS. An interaction with an intermediate mass black hole in close orbit around the massive black hole is ten times more efficient in creating the most extreme HVS than the classical stellar binary-massive black hole process (Yu & Tremaine 2003, Yu, Lu & Lin 2007). Somewhat lower velocity HVS can be created in interactions with stellar black holes (Miralda-Escudé & Gould 2000, O'Leary & Loeb 2008). Quantitative estimates of the combination of the capture reaction with the rate of injection into the loss cone by standard relaxation processes in the central few parsec require a large density of B-stars at $R > 0.5$ pc (Alexander & Livio 2004), which are in excess of the observed surface densities found by Paumard et al. (2006) and Bartko et al. (2010). However, this process remains attractive for both capture and HVS ejection when combined with the more rapid relaxation processes discussed in § 6.3.3.

*6.3.2 Hypervelocity stars*

As discussed in the last section, one of the most straightforward consequences of a massive black hole residing in the center of a dense star cluster is that from time to time stars get ejected at sufficiently high velocity ($> 10^3$ km/s) that the star can escape the Milky Way.

From a spectroscopic survey at the Multiple-Mirror Telescope (MMT) Brown et al. (2005, 2006a, b, 2008, 2009a, b) have found more than a dozen high velocity B-stars (upper left panel in Figure 6.3.1). Their velocities and radii make them unbound to the Galaxy given the escape velocity model of Kenyon et al. (2008). These stars are excellent candidates for the Hills process discussed in the last section. Their observed properties are broadly consistent with a continuous ejection model (Brown 2008). Perets (2009) argues that the observed number of HVS B-stars and the number of S-stars near Sgr A* are best consistent with the classical stellar binary-massive black hole Hills interaction scenario.

The most important question thus is whether the HVS have indeed been ejected from the Galactic Center. Abadi, Navarro & Steinmetz (2009) have pointed out several of the HVS may be on a constant travel time track (dotted in the right panel of Figure 6.3.1). They thus could also be remnants of a dwarf accretion/disruption event, although the statistical significance of the HVS clustering around constant travel time tracks is quite marginal. Heber et al. (2008) proposed that some of the HVS instead may have been ejected from the Large Magallanic Cloud (LMC), or from the outer Galactic disk when a massive binary companion star exploded as a supernova.

From HST imaging Brown et al. (2010) have recently reported the first detection of an absolute proper motion of an HVS. HVS HE 0437−5439 is a short-lived B star (mass 9 $M_\odot$) located in the direction of the LMC. The measured velocity vector indeed points directly away from the Center of the Milky Way, in strong support of the Hills ejection scenario (bottom left panel of Figure 6.3.1). An origin from the center of the LMC is ruled out at the 3σ level. The flight time of the HVS from the Milky Way exceeds the star's main-sequence lifetime, suggesting that HE 0437−5439 is a rejuvenated 'blue straggler' that was originally in a binary system. The large space velocity rules out a Galactic-disk ejection. Combining the HVS's observed trajectory, stellar nature, and required initial velocity, Brown et al. (2010) conclude that **HE 0437−5439 was most likely a compact binary ejected by the Milky Way's central black hole**.

The velocity, spatial, temporal and rotation velocity distributions of the HVS can also in principle distinguish between the different variants of the Hills process (see Brown 2008 for references). The proposed binary nature of HE 0437−5439 at ejection may be evidence for the presence of an intermediate mass black hole in the Galactic Center (Lu, Yu & Lin 2007).

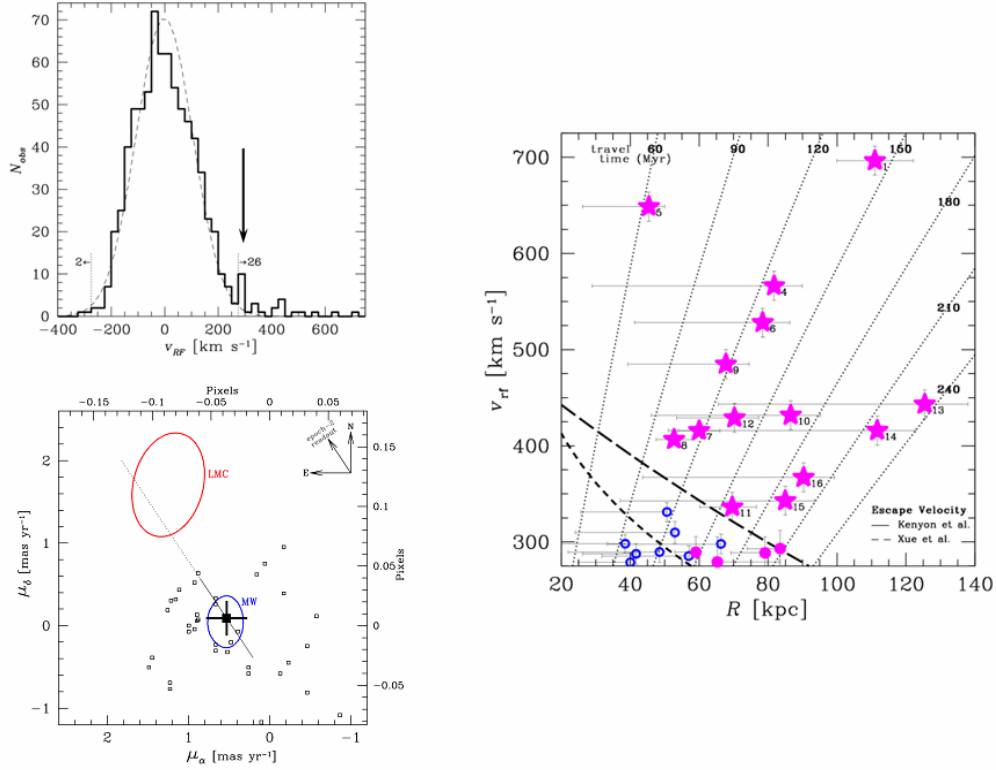

*Figure 6.3.1.* Hypervelocity stars. Upper left: Observed Galactic radial velocity distribution of the 759 B-stars in the CfA-MMT survey. There are 26 stars with $v_r > 275$ km/s and two stars with $v_r < -275$ km/s. Thus, 24 of the 26 cannot be explained by stellar interactions and 14 stars cannot be bound to the Galaxy. Right: Velocity-radius distribution of the HVSs. Magenta asterisks denote the 14 unbound stars, given the Kenyon et al. (2008) escape velocity model (long dashed curve, adopted from Brown 2008). Bottom left: Mean proper motion of HE 0437−5439 (solid black square) and the distribution of proper motions (open squares) measured from the individual epoch-1 and epoch-2 HST images. The 1σ statistical uncertainty (cross) and systematic uncertainty (solid line) are indicated, as well as the full correction for charge transfer inefficiency on the HST detector (dotted line). The blue ellipse shows the locus of proper motions with trajectories passing within 8 kpc of the Milky Way center, and the red ellipse denotes the locus of propert motions originating within 3 kpc of the LMC center at the time of pericenter passage (adopted from Brown et al. 2010).

### 6.3.3 Injection into the loss cone

Next, one needs to understand the source of the binaries and how stars get injected at a high enough rate into the highly elliptical orbits required for tidal disruption and capture. Perets, Hopman & Alexander (2007) consider the possibility that stars are injected into loss cone orbits by interactions with ***massive perturbers***, at parsec distances from Sgr A\*. Proposed first by Spitzer & Schwarzschild (1951) for molecular clouds in the solar neighborhood, massive clouds or clusters can dominate the overall relaxation rate when the product of their mass squared and their density is much greater than the corresponding product for the average star in the cluster,

$$N_p m_p^2 >> N_* <m_*>^2 \qquad (16),$$

The upper left panel of Figure 6.3.2 shows the effective relaxation time scales when considering different massive perturber candidates near the Galactic Center. From this comparison Perets, Hopman & Alexander (2007) conclude that the population of observed, dense and massive ($10^5 \ldots 10^7$ M$_\odot$) giant molecular clouds outside the central few parsecs are numerous and massive enough to inject field stars on highly elliptical orbits into the central cusp.

    Perets, Hopman & Alexander (2007) calculate the efficiency of the massive perturber induced re-filling of the loss cone, given a 75% binary fraction of equal mass binaries, a log-normal semi-major axis distribution around $<a_{binary}>$ ~ 0.2 AU, motivated by the observed distribution of massive binaries in the solar neighborhood (Kobulnicky & Fryer 2007). With these inputs and an assumed capture probability of 75%, Perets, Hopman & Alexander (2007) find that the massive perturber induced injection deposits several tens of young B-stars in the central arcsecond at any given time (upper right panel of Figure 6.3.2). At the same time, there are several hundreds of HVS ejected into the outer Galaxy. The massive perturber + Hills model thus appears to be successful in accounting for the puzzling B-star cusp around Sgr A*. It also accounts for the observed HVS if these indeed turn out to originate from the Galactic Center. Its philosophical shortcoming is that it appeals to fast relaxation processes twice, once to increase the injection rate into loss-cone orbits (by 2 − 3 orders of magnitude through the massive perturber scenario) and once to increase the thermalization rate of the captured B-stars (through resonant relaxation with stellar black holes). Both processes are difficult to test observationally.

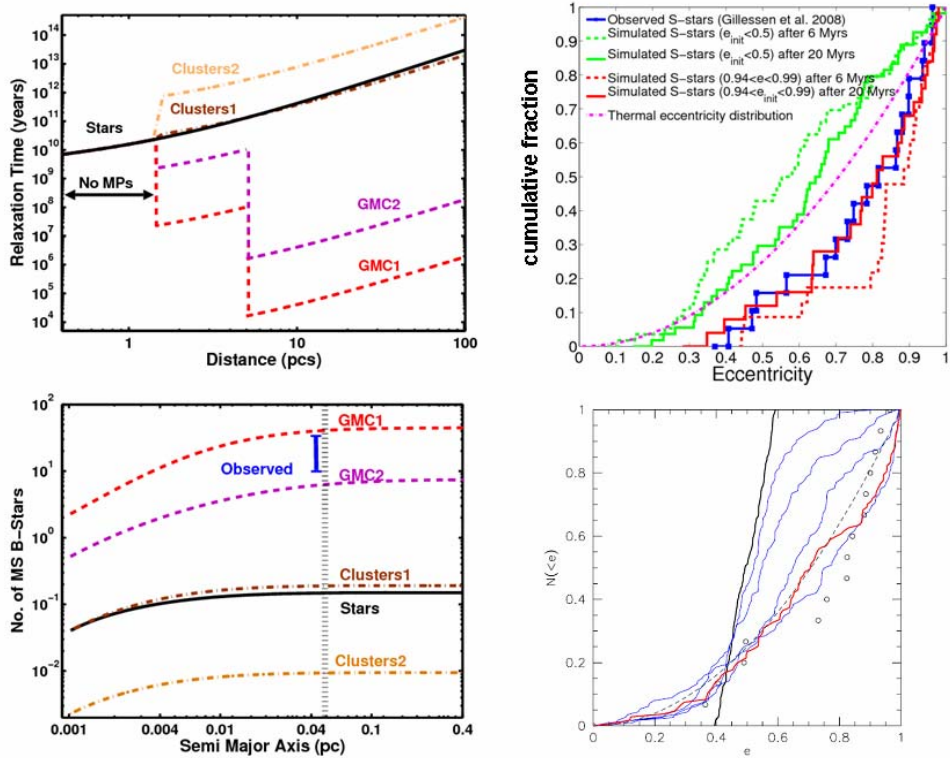

**Figure 6.3.2**. The massive perturber model for creating the S-star cluster. Top left: the impact of massive perturbers (giant molecular clouds, star clusters) on the relaxation time in the Galactic Center, as a function of distance from Sgr A* due to stars alone, the upper (GMC1) and lower (GMC2) mass estimates of the observed molecular clumps and giant molecular clouds (GMCs) and due to upper (Clusters1) and lower (Clusters2) estimates on the number and masses of stellar clusters. The sharp transitions at $R = 1.5$ and 5 pc are artifacts of the non-continuous mass perturber distribution assumed here. GMCs dominate the relaxation in the Galactic Center. Bottom left: Cumulative number of young B stars as predicted by the massive perturber model and by stellar two-body relaxation. The vertical bar represents the total number of observed young stars inside 0.04 pc (Eisenhauer et al. 2005; Gillessen et al. 2009). The hatched vertical line marks the approximate maximal distance within which captured B-stars are expected to exist (adopted from Perets, Hopman & Alexander 2007). Top right: The graph shows the cumulative distribution of eccentricities after 6 and 20 Myrs of stars migrating from the disk to the central cusp (green) and stars captured by the Hills process (red), compared to the data from Gillessen et al. (2009b). This shows that a migration process cannot explain the super-thermal distribution of eccentricities in the S-star cusp, but that the Hills mechanism can (adapted from Perets et al. 2009). Bottom right: Evolution of the distribution of stellar orbital eccentricities in a set of simulations with an intermediate mass black hole with a mass $M_{IMBH}/M_\bullet = q = 5 \times 10^{-4}$ and a semimajor axis $a_{IMBH} = 15$ mpc, $e_{IMBH} = 0.5$. The results of averaged simulations are shown for six times are shown: $t = 0$ (thick black curve), $t = (0.01, 0.02, 0.04, 0.2)$ Myr (thin blue lines), and $t = 1$ Myr (thick red line). The distribution is essentially unchanged at times greater than 1 Myr. Dashed line shows a thermal distribution, $N \propto e^2$, and open circles are the eccentricity distribution observed for the S-stars (Gillessen et al. 2009b). Adopted from Merritt, Gualandris & Mikkola (2009).

*6.3.4 Migration*

The second possibility is that the S-stars were formed relatively near their current location, in the O/WR-disk(s), and then migrated to the central cusp. The interaction of newly formed stars with their parent (gas) disk leads to inward migration that may be fast enough to bring stars from the disk(s) to the cusp in a fraction of their main-sequence lifetime (Levin 2007). Levin (2007) then appeals to resonant relaxation (Rauch & Tremaine 1996) to redistribute angular momenta in the central cusp. Löckmann, Baumgardt, & Kroupa (2009) consider the interaction of two disks with a top-heavy IMF and appropriate mass ratios and ages with the central massive black hole. As discussed in § 2.2 (Nayakshin et al. 2006), the disks perturb each other and cause precession and warping. This warping is stronger for counter-clockwise stars because of the dominant mass in the clockwise disk. In principle any star orbiting in the Galactic Center and on a highly inclined orbit relative to the orbit of a large mass, for instance the stars in the other disk, or an intermediate mass black hole, can undergo a Kozai resonance (Kozai 1962). Here the highly inclined, near circular orbit converts into a low inclination but highly elliptical one. Löckmann, Baumgardt, & Kroupa (2009) find from their numerical simulations that the stars in the clockwise and counter-clockwise systems can drive such a Kozai mechanism and perhaps inject enough B-stars from disk orbits into the central cusp where they are then captured by the Hills process. However, these simulations do not include a stellar cusp. From analytical calculations Chang (2009, see also Ivanov et al. 2005) finds that the Kozai resonance is suppressed when the mass in the cusp is larger than the mass in the perturbing disk, since then the apsidal precession induced by a (spherical) cusp is sufficient to destroy the resonance. Chang concludes that the Kozai process cannot work in the Galactic Center, as long as the cusp is spherical, an assumption that seems empirically well justified (Gillessen et al. 2009b, Trippe et al. 2008). Perets et al. (2008b) and Löckmann, Baumgardt & Kroupa (2009) have included spherical cusps in their numerical simulations and confirm the assessment but comment that the Kozai mechanism may still be helpful for bringing a binary already on a moderately eccentric orbit (e.g. through resonant relaxation) to the very high eccentricity needed to drive the Hills capture/ejection process. Numerically they find one to two pre-Hills binaries for a top-heavy IMF but seven such binaries for a Kroupa IMF.

The simulations of Perets et al. (2009) discussed above in the context of the massive perturber scenario also consider the alternative migration scenario from within the O/WR-disk(s). The green curves in the bottom left panel of Figure 6.3.2 show the results. The relaxation processes considered appear to be too slow to heat an initial 'cold' disk population to the observed properties of the S-star cluster within the lifetime of the current O/WR-stars. Even for a potentially older (~ 20 Myr) star disk component the resulting ellipticity distribution is still too cool in comparison with the data. Perets et al. conclude that migration and relaxation of an initially cold (disk) population cannot account for the S-star properties. This conclusion may not hold if an intermediate mass black hole (of mass $> 10^3$ $M_\odot$) is present. Merritt, Gualandris & Mikkola (2009) show that an intermediate mass black hole in a tight elliptical orbit around the massive hole can thermalize and randomize the orbits of stars in a plane (in their case due to an in-spiral of a star cluster) within a few Myrs (bottom right panel of Figure 6.3.2). Fuji et al. (2010) present N-body simulations of an infalling star cluster creating an intermediate mass black hole, producing a disk of young stars and accounting for the population of the S-stars. The surface density distribution of the young massive stars found by Fuji et al. is $\Sigma_{O/WR} \sim p^{-1.5}$ in the central 10", consistent with the data. As expected from the Perets et al. (2009) work, this migration simulation produces a thermal eccentricity distribution, but not hotter than that (figure 6.3.2 top right). Fuji et al. present only one simulation. Hence, it only shows that the scenario is possible, but does not shed light on how likely it is. Also, the authors used an unrealistically large mass for the IMBH excluded by the radio proper motion limits (Reid et al. 2009a). Finally, the resulting S-star population on average resides at a larger distance than what is observed.

The Newtonian precession time scale increases with ellipticity of the orbit of a given star (Madigan et al. 2009). Hence if the stars in the O/WR-disk(s) originally had non-zero eccentricity with a modest dispersion the somewhat higher eccentricity stars in this distribution precess slower and thus experience a coherent torque from the rest of the disk, increasing their ellipticity and angular momentum further (Madigan et al. 2009). Madigan et al. (2009) have shown form numerical simulations that this instability over time creates a spread of eccentricities with a tail of $e \sim 1$ orbits, which could then again be captured via the Hills (1988) process.

Perets & Gualandris (2010) show that many models over-predict the number of B-stars at R ≈ 0.5 pc, given the observed number of S-stars. Among these is the intermediate-mass black hole assisted cluster in-spiral scenario. On the other hand, the massive perturber induced binary disruption picture is consistent with both the S-stars and the extended population of B-stars further away. Another basic problem of the migration scenario is that the migration mechanism would have to selectively push the lower mass B-stars to the center, while leaving the more massive O-stars in the disk. Otherwise one would expect to see some O-stars from the present disk among the S-stars (Alexander 2010). The only loophole might be that the expected number of O-stars in the central arcsecond is small, such that not observing any such star is not a significant finding.

In summary of the last two sections then, **the properties of the S-stars around Sgr A\* as well as the hyper-velocity star in the halo of the Milky Way appear to be well accounted for by the field-binary-massive-perturber-Hills- scenario.** This explanation, however, is **very complex**. Especially if there were one or several intermediate massive black holes not yet discovered by the present observations (§ 4.6, Figure 4.6.1) it may be **premature to already dismiss the alternative of a migration process from the disks in the central parsec**. More precise measurements of the eccentricity distribution of the S-stars may offer a tool for distinguishing the migration and Hills scenarios. In the former the thermal distribution is reached from above, in the latter from below. A significant deviation from a thermal distribution therefore could discriminate between the two scenarios.

## 6.4 What powers the central parsec?

The question of the dominant luminosity source in the central parsec was a puzzle until a few years ago. We show in this section that the puzzle now appears to be resolved. The ultra-violet luminosity of the central parsec is $L_{UV} = 10^{7.5}\ (0.2/f_{FUV})\ L_\odot$ where $f_{FUV}$ is the fraction of the intrinsic far-ultraviolet luminosity that is absorbed in the circum-nuclear environment and converted to the observed far-infrared luminosity ($L_{FIR} \sim 5 \times 10^6\ L_\odot$, Davidson et al. 1992, Mezger, Duschl & Zylka 1996). The intrinsic number of Lyman-continuum photons produced in the central parsec is

$Q_{Lyc} = 10^{50.7}$ $(0.5/f_{ion})$ s$^{-1}$ ($L_{Lyc} \sim 10^{6.5}$ L$_\odot$, $L_{UV}/L_{Lyc} \sim 10$), where $f_{ion}$ is the fraction of hydrogen ionizing photons absorbed (Ekers et al. 1983, Mezger, Duschl & Zylka 1996). The UV radiation field is relatively soft. The ratio of helium to hydrogen ionizing photons in the HII region Sgr A West is ~ 0.06 (Krabbe et al. 1991) and the effective temperature of the ionizing UV field is 35,000 ± 2000 K, as estimated from fine structure line ratios (Lutz et al. 1996, Lacy et al. 1980, 1982, Shields & Ferland 1994). Najarro et al. (1994, 1997) and Krabbe et al. (1995) showed that the HeI emission line stars can account plausibly for the majority of the far-UV and hydrogen ionizing luminosity. The basic conclusion is that the central parsec is currently powered by an ageing cluster of hot stars, with a vanishing contribution from Sgr A* itself (Rieke & Lebofsky 1982, Allen & Sanders 1986, Genzel, Hollenbach & Townes 1994, Krabbe et al. 1995). The nuclear region of the Milky Way would be classified as 'HII region nucleus' by an extragalactic observer (Shields & Ferland 1994).

The $T_{eff} \sim 20{,}000 - 30{,}000$ K HeI stars detected in the 1990s are too cool, however, to be able to account for the nebular excitation, including the helium ionizing photons. Lutz (1999) pointed out that another problem is the small number of cool blue supergiants predicted by theoretical models at that time for an evolving cluster at $t_{burst} \sim 6$ Myrs. In addition, such models in conjunction with the stellar atmosphere models available at that time and photoionization modeling predict a ratio of the [NeIII] to [NeII] nebular mid-infrared fine structure lines of ~ 2 at $t_{burst}$. The observations with ISO SWS find a ratio of 0.05, however (Thornley et al. 2000).

The likely solution of this riddle is presented in Martins et al. (2007) and involves on the one hand the detection in the last few years of many hotter Wolf-Rayet and dwarf and giant main-sequence O-stars (mainly in Paumard et al. 2006), and on the other hand the availability of updated stellar atmosphere modeling. Martins et al. (2007) analyze with the CMFGEN stellar atmosphere code the H/K-band SINFONI spectra of 28 luminous massive stars covering the entire sequence of blue supergiants/WR-stars. CMFGEN includes updated treatments of non-LTE effects, winds, micro-turbulence and line blanketing due to metals. Figure 6.4.1 shows examples of the spectra and the best fits achieved. The analysis of Martins et al. represents a distinct improvement in matching the observed near-infrared spectra of the Galactic Center stars with remarkable detail, compared to the earlier work in Najarro et al. (1997). With some notable exceptions (e.g. 2.058 μm HeI) details of the line profiles, including those of carbon lines, are now matched very well by the theoretical atmosphere modeling and instill confidence in the extracted basic stellar parameters. The most important difference between the analyses of Martins et al. and Najarro et al. are the larger stellar temperatures in the new work, because of more recently discovered, earlier type WR-stars (e.g. IRS 16SE2 (WN 5/6) in Figure 6.4.1 with $T_{eff} \sim 41{,}000$ K, and a number of WC9 stars with $T_{eff} \sim 37{,}000 - 39{,}000$ K), but also because higher inferred temperatures in some of the previously analyzed WN 7-9 stars. As a result, the modeling now indicates that the massive stars can comfortably account for both the hydrogen and helium ionization rate required to explain the nebular excitation. With some extrapolation to the stars not specifically analyzed Martins et al. infer for the total output of the early-type cluster $Q_{Lyc} = 10^{50.8}$ s$^{-1}$ and $Q_{He}/Q_{Lyc} = 0.04$, in very good agreement with the observations. The new data and analysis also remove the discrepancy between stellar evolution models and observed stellar population noted by Lutz (1999). The empirical location of the WR-stars of all types are in excellent agreement with evolution models with an age of 6 ± 1 Myrs and tracks with metallicity 1 - 2 $Z_\odot$ (bottom right panel in Figure 6.4.1, Paumard et al. 2006). The low [NeIII]/[NeII] flux ratio can also be plausibly be accounted for by a combination of enhanced line-blanketing in the extreme ultraviolet and high gas densities in Sgr A West ($<n_e^2>^{1/2} > 10^5$ cm$^{-3}$ instead of $10^4$ cm$^{-3}$).

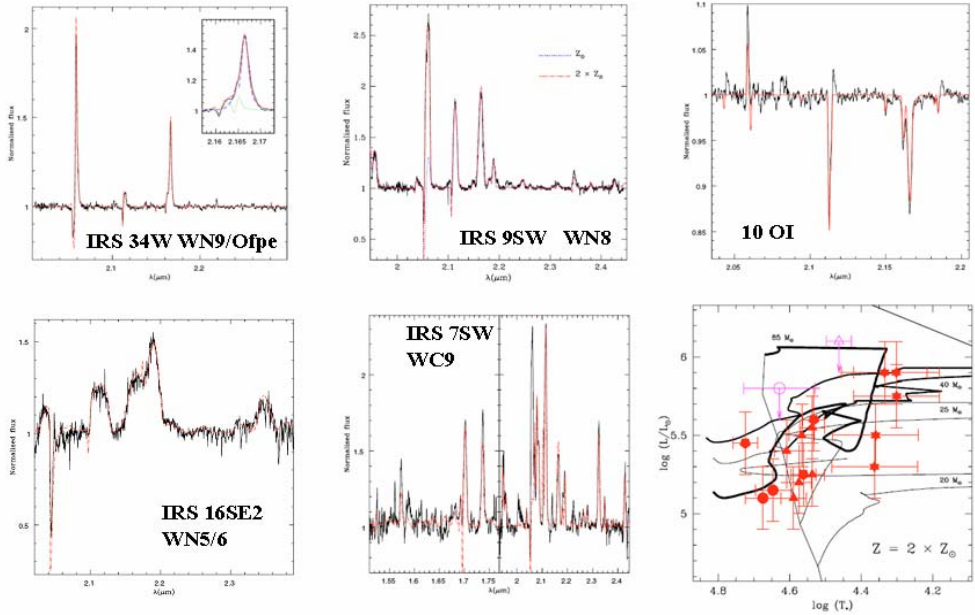

**Figure 6.4.1.** Five representative cases of O/WR-star H/K-band spectra (black) covering the entire range of blue supergiant types in the central parsec, along with the best stellar atmosphere models using CMFGEN (red dashed). Bottom right: HR diagram of WR stars. The assumed twice solar metallicity tracks include rotation. Star symbols are Ofpe/WN9 stars, triangles WN8 stars, pentagons WN 5-7 stars and circles WC9 stars. Different line thickness indicates different evolutionary tracks (the thicker the line, the higher the mass). ZAMS masses are marked for each track. The Humphreys-Davidson limit is also shown in the right upper part of each diagram. Adopted from Martins et al. (2007).

The new models also yield ~ 4 times lower mass loss rates ($dM_{tot}/dt > 5 \times 10^{-4}$ M$_\odot$ yr$^{-1}$ as compared to $> 2 \times 10^{-3}$ M$_\odot$ yr$^{-1}$) and lower helium abundances in the star cluster. The latter result suggests that the classical HeI stars (mainly WN9/Ofpe), which are thought to be closely related to Luminous Blue Variable (LBV) stars like η Car, are less evolved than previously thought and may be precursors to WN8 stars (which are found to be more helium-rich).

In summary of this section, *the observed bolometric, hydrogen and helium ionizing luminosity of the central parsec, as well as the excitation of the HII region Sgr A West can now be quantitatively explained by the combined radiation of the 100+ O/WR-stars in the $t_{burst}$ = 6 Myr star disk(s), without requiring any peculiar stellar evolution*.

# 7. Accretion and emission close to the central black hole

The Galactic Center black hole is most of the time in a *'steady' state*, emitting $\sim 10^{36}$ erg/s predominately at radio to submillimeter wavelengths (§ 7.1). On top of this quasi-steady component there is *'variable emission'* in the X-ray and infrared bands (§ 7.2). Some of this variable emission, especially at X-rays, appears as *'flares'*, typically a few times per day and lasting for about $10^2$ minutes. Flares release an additional luminosity of up to another $10^{36}$ erg/s (§ 7.3). The steady emission is well described by synchrotron radiation from a thermal distribution of relativistic electrons; the variable emission and the flares originate from transiently heated electrons, but details are still under debate. Compared to QSOs, which accrete and radiate at a significant fraction of the maximum rate given by the Eddington limit ($L_{Edd} \sim 1.5 \times 10^{11}$ L$_\odot$ for Sgr A*), the Galactic Center is *remarkably under-luminous by more than eight orders of magnitude with respect to $L_{Edd}$* ($<L_{Sgr\,A*}> \sim 10^{-8.5}\, L_{Edd}$), and by more than four orders of magnitude compared to the luminosity estimated from the accretion rate at the Bondi radius ($<L_{Sgr\,A*}> \sim 10^{-4.2}\, (0.1 \dot{M}(R_{Bondi} \sim 1") c^2)$) (§ 7.4). The luminosity of Sgr A* is currently dominated by accretion from stellar winds in its vicinity (§ 7.4). In addition, larger accretion rates probably occur from time to time when large, low angular momentum gas clouds fall into the center (§ 6.2), or when a star is scattered to within the tidal disruption radius ($R_t \sim 10\, R_S$) at a rate of once every 40,000 years (Freitag et al. 2006).

## 7.1 Steady emission from Sgr A* across the electromagnetic spectrum

The compact radio source Sgr A* is clearly detectable at all times and shows only moderate flux and spectral variations (Falcke et al. 1998, Zhao, Bower & Goss 2001, Herrnstein et al. 2004, Yusef-Zadeh et al. 2006a, 2009, Miyazaki et al. 2004, Marrone et al. 2008, Li et al. 2009). Steady faint X-ray emission from Sgr A* is also detected. It is spatially extended with an intrinsic size of about FWHM 1.4", consistent with the size of the Bondi accretion radius of the black hole (bottom central panel in Figure 1.1, Baganoff et al. 2003, Xu et al. 2006). Because of confusion with stars along the line-of-sight to Sgr A*, it is less clear whether there is a steady near-infrared counterpart to Sgr A* (Genzel et al. 2003b, Do et al. 2009b, Dodds-Eden et al. in prep.). We call the state in which Sgr A* emission is dominated by the radio flux the 'steady', or 'quiescent', state.

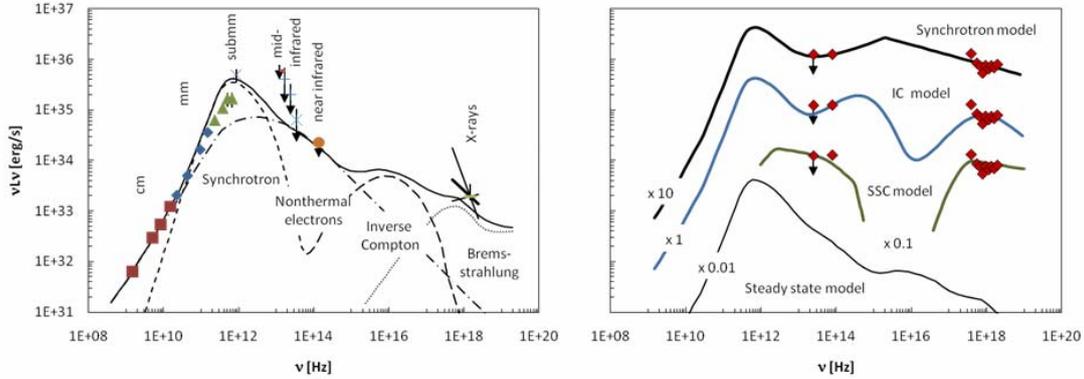

**Figure 7.1.1:** Spectral energy distribution of Sgr A*. All numbers are given for a distance of 8.3 kpc of the Galactic Center and are de-reddened for interstellar absorption (infrared and X-rays) and scattering (X-rays). Left: steady state. The Sgr A* radio spectrum follows roughly a power-law $\nu L_\nu \sim \nu^{4/3}$. The observed peak flux at submm wavelengths is about $5 \times 10^{35}$ erg/s. The spectrum then steeply drops towards infrared wavelengths down to less than the detection limit of about $2 \times 10^{34}$ erg/s at 2 μm. The only other unambiguous detection of Sgr A* in its steady state is at X-rays with energies from 2 - 10 keV with a flux of about $2 \times 10^{33}$ erg/s. The figure shows a compilation of data from ■ Zhao et al. 2001, ◆ Falcke et al. 1998, ▲ Zylka et al. 1995, × Serabyn et al. 1997, − Cotera et al. 1999, + Gezari 1999, ✶ Schödel et al. 2007b, ● Hornstein et al. 2002, − Baganoff et al. 2003. Overplotted is a model of the quiescent emission (from Yuan et al. 2003): the radio spectrum is well described by synchrotron emission of thermal electrons (short dashed line). The flattening of the radio spectrum at low frequency is modeled by the additional emission from a non-thermal power-law distribution of electrons, which carry about 1.5% of the total thermal energy (dashed-dotted line). The quiescent X-ray emission arises from thermal Bremsstrahlung from the outer parts of the accretion flow (dotted line). The secondary maximum (long-dashed line) at frequencies of about $10^{16}$ Hz is the result of the inverse Compton up-scattering of the synchrotron spectrum by the thermal electrons. Right: SED during a simultaneous X-ray and infrared flare: while the total energy in the radio-emission is largely unaffected during a flare, the IR and X-ray fluxes increase by factors of ten to hundred, respectively (from Dodds-Eden et al. 2009). The near-infrared emission results from synchrotron radiation of transiently heated electrons. Several emission mechanisms can account for the X-ray flares. Top: synchrotron model, in which both X-ray and infrared emission is synchrotron radiation from a population of ultra-relativistic electrons following a power law energy distribution. Middle: Inverse Compton model, in which the near-infrared emitting electrons up-scatter the sub-millimeter seed photons to X-ray energies (synchrotron and IC model from Dodds-Eden et al. 2009). Bottom: Synchrotron Self Compton model, in which the near-infrared emitting electrons transfer their energy to the synchrotron photons emitted from the same electron population (from Sabha et al. 2010).

The SED of Sgr A* in its steady state is shown in Figure 7.1.1. The steady state radio emission is linearly polarized at a level of 2 to 9 % at submm and millimeter wavelengths (Aitken et al. 2000, Marrone et al. 2006, 2007, Macquart et al. 2006). The centimeter emission is weakly circularly polarized (~0.3%), with a linear polarization of < 0.2% (Bower et al. 1999a, b, c, Sault & Macquart 1999). The emitted flux increases from 20 cm to 350 μm (Zhao et al. 2001, Falcke et al. 1998, Zylka et al. 1995, Serabyn et al. 1997), approximately with a power-law *($\nu L_\nu \sim \nu^{4/3}$)* and with a peak luminosity of ~ $5 \times 10^{35}$ erg/s (Serabyn et al. 1997). Simultaneous observations reveal that the power-law index increases slightly with frequency (Falcke et al. 1998, Melia and Falcke 2001 and references therein). This upturn of the spectrum is usually referred to as the *'submm bump'*. The SED at > 20 cm and < 350 μm is unknown, because here the limited angular resolution and sensitivity of the available observations make it impossible to disentangle the emission from Sgr A* from surrounding sources. In any case, the Sgr A* spectrum must drop off steeply between the submm and the mid-infrared. Sgr A* is not detected at mid-infrared wavelengths down to 8.6 μm, with upper limits on its flux of ~ 6, 4, 2, 0.6 × $10^{35}$ erg/s at 25, 18, 12, 8.6 μm, respectively (Cotera et al. 1999, Gezari et al. 1999, Schödel et al. 2007b). Sgr A* is frequently detected at near-infrared wavelengths between 5 and 1.5 μm, but the luminosity falls below about $2 \times 10^{34}$ erg/s at 2.2 μm in its steady state (Hornstein et al. 2002, Schödel et al. 2007b, Do et al. 2009b, Dodds-Eden et al. 2010b). The steady state X-ray emission, including 6.7 keV He-like Fe K line emission, is ~ $2 \times 10^{33}$ erg/s (Xu et al. 2006), and follows a moderately soft power-law spectrum $\nu L_\nu \sim \nu^{-2 \ldots 0.2}$ (Baganoff et al. 2003). Higher energy γ-ray observations again suffer strongly from comparably poor angular resolution, which makes an unambiguous identification with Sgr A* currently impossible (Aharonian et al. 2004, Kosack et al. 2004, Albert et al. 2006).

The observed spectral energy distribution, polarization, and compactness (§ 4.4) of the Sgr A* radio source directly constrain a number of fundamental properties of the steady state emitting region without the need for detailed spatial and dynamical modeling (e.g. Loeb & Waxman 2007). According to Loeb & Waxman the radio-emission is predominantly somewhat optically thick **synchrotron radiation from relativistic thermal electrons.** Its peak flux originates from within 10 $R_S$, the electron temperature and density are a few $10^{10}$ K (γ~10) and $10^6$ e⁻/cm³, respectively, and the magnetic field strength is about 10 to 50 Gauss.

The relative faintness of Sgr A* is difficult to understand, and has triggered the development of theoretical models with **radiatively inefficient accretion flows**. Among the early models is the advection dominated accretion flow (ADAF: Narayan et al. 1995, Narayan & Yi 1995), in which the low luminosity is explained by the combination of a high ratio of radial to tangential gas velocities, and the decoupling of hot protons and cold electrons in low density gas. However, the detection of linear polarization and the low electron densities estimated from the Faraday rotation measure rules out the large accretion rate of the standard ADAF model. This led to the development of convection dominated accretion flow models (CDAFs: Ball et al. 2001, Quataert & Gruzinov 2000, Narayan et al. 2002), which favor lower accretion rates and shallower density profiles. The latest sets of models are radiative inefficient accretion flow models with substantial mass loss (RIAFs: Yuan et al. 2003, 2004). Related models are the advection-dominated inflow-outflow solutions (ADIOS:

Blandford & Begelman 1999), jet models (Falke & Markoff 2000, Yuan et al. 2002), and Bondi-Hoyle models (Melia & Falcke 2001). All these models can be tuned to fit the steady-state spectrum of Sgr A*. As far as the spectral modeling is concerned, most of them are hybrid models (Özel et al. 2000), involving a spatially modeled thermal component, which is responsible for the bulk of the radio emission with its peak at (sub)mm wavelengths, plus a minor non-thermal component to explain the slight change of the power-law index across the radio wavelength range. The faint quiescent emission in the X-ray band is explained by thermal Bremsstrahlung originating from the transition region between the ambient medium and the accretion flow (Quataert 2002, Xu et al. 2006).

## 7.2 Variable emission

Sgr A* is 'variable' at all wavelengths but the degree of variation changes drastically across the electromagnetic spectrum. Figure 7.2.1 shows characteristic light-curves and the observed level of variability as a function of wavelength.

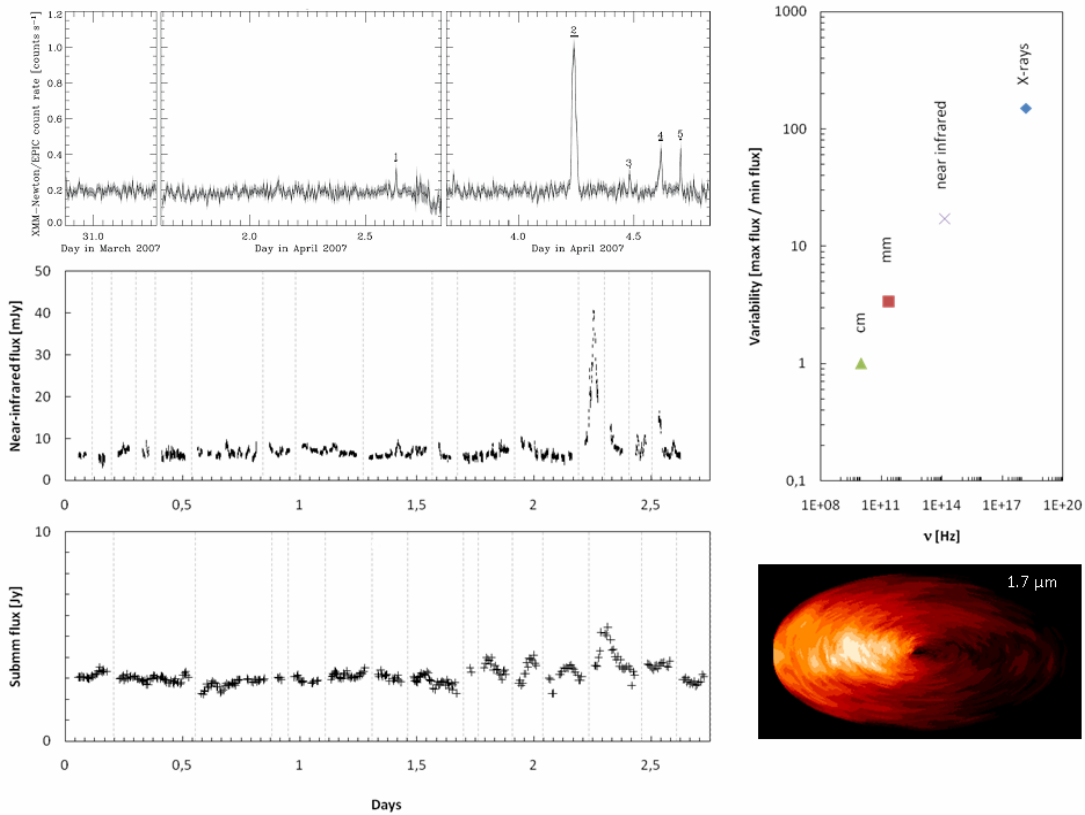

**Figure 7.2.1:** Variability of Sgr A*, in several non-simultaneous data sets. Top: X-ray light curve of Sgr A* (from Porquet et al. 2008) spanning 2¾ days in April 2007. The 2 − 10 keV peak luminosity of the bright flare is about $2.6 \times 10^{35}$ erg/s. Middle: near-infrared light curve of Sgr A* as observed in 2008 (adopted from Dodds-Eden et al. 2010b), the time axis also covering 2¾ days, collected from 16 nights (separated by dotted vertical lines) spread over several months. The flux at low levels is dominated by the star S17 close to the line of sight to Sgr A* in 2008. The intrinsic flux of Sgr A* in its minimum is below $1 − 2 \times 10^{34}$ erg/s (Hornstein et al. 2002, Schödel et al. 2007, Do et al. 2009, Dodds-Eden et al. 2010b). The peak flux of the bright flare is about 34 mJy, corresponding to a luminosity $\nu L_\nu \sim 10^{35.6}$ erg/s. Bottom: light curve at submm wavelengths as observed in 2007 − 2009 (adopted from Haubois et al., Yusef-Zadeh et al., Trap et al.; all in prep.). Again, the time axis covers 2¾ days, and the data is collected from several nights (separated by vertical dashed lines). Note

that the X-ray, infrared and submillimeter data are not simultaneous, nor continuous. They are meant to give a pictorial impression of the amplitude and time scales of variability on scales of days. Top right: flux variation of Sgr A* across the electromagnetic spectrum: the ratio between the observed peak and minimum flux strongly depends on the wavelength. While Sgr A* fluctuates only up to a factor few at radio wavelengths, the brightest flares observed at infrared and X-rays peak a factor 20 and 160 above the quiescent steady-state, respectively The centimeter variability was adopted from Macquart & Bower 2006; the millimeter variability from Zhao et al. 2001, 2003; near infrared = brightest K-band flare observed so far (Dodds-Eden et al. 2010b) divided by upper limit on steady state (Hornstein et al. 2002, Schödel et al. 2007b, Do et al. 2009b, Dodds-Eden et al.2010b); X-ray = brightest X-ray flare observed so far (Porquet et al. 2003) divided by quiescent state (Baganoff et al. 2003). Bottom right: snapshot from a magnetohydrodynamic simulation of the accretion disk in Sgr A* observed at 1.7 μm (taken from Chan et al. 2009).

For almost 30 years Sgr A* has been known to be variable at radio wavelengths (Brown & Lo 1982). The amplitude of the intensity modulation at wavelengths between 20 cm and 7 mm (Zhao et al. 1992, 2001, Herrnstein et al. 2004, Falcke et al. 2004) is 30 to 40% (Macquart & Bower 2006), with a clear trend that the spectral index of the centimeter emission becomes larger when the flux increases (Falcke et al. 1999). There is also a trend that the amplitude of variability increases at millimeter wavelengths (Zhao et al. 2001, 2003, Miyazaki et al. 2004, Mauerhan et al. 2005, Yusef-Zadeh et al. 2006a, Li et al. 2009), which can be as large as a factor few. The typical time scale of these millimeter fluctuations is about 1.5 to 2.5 hours. A similar variability, both in terms of amplitude and typical timescale is observed at submillimeter wavelengths (Eckart et al. 2006a, Yusef-Zadeh 2006a, 2008a, 2009, Marrone et al. 2006, 2008, Kunneriath et al. 2008). The luminosity fluctuations of Sgr A* at near-infrared wavelengths are about one order of magnitude larger than at radio wavelengths. Typically a few times per day Sgr A*'s near-infrared flux rises by factors of a few above the stellar background and confusion. The highest near-infrared flux observed at 2.2 μm was about 20 times above its minimum (Hornstein et al. 2002, Schödel et al. 2007b, Do et al. 2009b, Dodds-Eden et al. 2010b). These brightest flares frequently come with simultaneous X-ray emission, large polarization, and with strong, and perhaps quasi-periodic, substructure (§ 7.3). The largest flux variations are seen in the X-ray band. Here the brightest X-ray flare observed so far (Porquet et al. 2003) is about 160 times brighter than Sgr A* in its quiescent state (Baganoff et al. 2003). The statistics of the X-ray variability is less certain than that at near-infrared- and radio-wavelengths, but again it is clear that the X-ray variability is yet another order of magnitude above what is observed at near-infrared wavelengths.

The variable emission at radio wavelengths and at low and moderate near-infrared flux levels is well described by a featureless, red power-law, noise distribution (Mauerhan et al. 2005, Macquart & Bower 2006, Do et al. 2009b). This red power law distribution can naturally be explained by stochastic fluctuations in the accretion disk, perhaps triggered by magnetohydrodynamic turbulence (Chan et al. 2009, Dexter et al. 2009). Figure 7.2.1 (bottom right) shows a snapshot from magnetohydrodynamic simulations of the accretion disk in Sgr A* observed at 1.7 μm (taken from Chan et al. 2009). The level of X-ray variability of Sgr A* at low flux levels is unknown. The current X-ray telescopes do not provide the required angular resolution and sensitivity for such observations.

## 7.3 Flares

A key issue in the interpretation of the observed variability is whether the brightest variable emission from Sgr A* at near-infrared and X-ray wavelengths are statistical fluctuations from the probability distribution at low flux, or flare '*events*' with distinct properties. The event interpretation is suggested by the distinct peaks in the light curves of the brightest X-ray and infrared flares (Figure 7.2.1).

Up to mid 2009 about a dozen X-ray flares have been reported from observations with the XMM and Chandra telescopes (Baganoff et al. 2001, Baganoff 2003, Goldwurm et al. 2003, Porquet et al. 2003, 2008, Eckart et al. 2004, 2006, 2008, Belanger et al. 2005, Aharonian et al. 2008, Marrone et al. 2008). The X-ray light curves of Sgr A* clearly delineate well-separated flares (Figure 7.2.1). The brightest observed X-ray flare had a 2 − 10 keV peak-luminosity of ~ $3.9 \times 10^{35}$ erg/s. The X-ray spectrum of Sgr A* during the two brightest flares follows a power-law $\nu L_\nu \sim \nu^{-0.6\ldots 0.1}$ (Porquet et al. 2008), which is compatible with that of the quiescent state $\nu L_\nu \sim \nu^{-2\ldots 0.2}$ (Baganoff et al. 2003). The average power-law of all reported flares is slightly harder with $\nu L_\nu \sim \nu^{0.5}$, but the spectral indices for fainter X-ray flares are weakly constrained (typically to only ± 0.6), and the significance of a harder spectral index for fainter flares is under discussion. The available data are also inconclusive on the question whether the brightest X-ray light curves exhibit quasi-periodic oscillations (Belanger et al. 2006, Aschenbach et al. 2004).

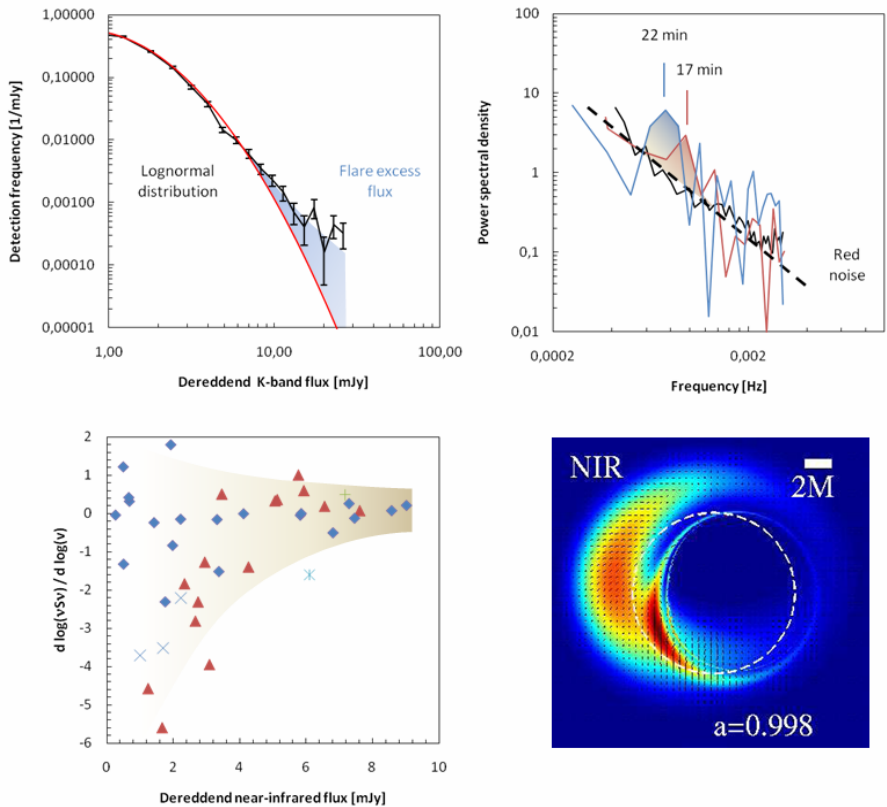

**Figure 7.3.1** Near infrared flares from Sgr A*. Top left: near-infrared (2.2 µm) flux density distribution function at the position of Sgr A* (adopted from Dodds-Eden et al. 2010b). The flux density distribution is well described by a log-normal distribution (red curve) at low flux levels and an additional high flux tail (blue shaded area). Top right: representative power spectral densities of Sgr A*'s near infrared flux. Black solid line: average power spectrum for 12 hours observing time (from Do et al. 2009b) with no significant quasi-periodic substructure, and fully consistent with a red-power law noise (dashed line). Red and blue: power spectra of infrared flares (red: from Genzel et al. 2003b, Trippe et al. 2007; blue: from Dodds-Eden et al. 2009), exhibiting quasi-periodic substructure of about 20 minutes on top of a red power spectrum. The near-infrared flux density distribution function and the variable strength of its quasi-periodic substructure can perhaps be explained by a two component model: at low flux levels the "variable" infrared emission from Sgr A* is characterized by a red-power-law noise process (Meyer et al. 2008, Do et al. 2009b), with a log-normal amplitude distribution function (Dodds-Eden et al. 2010b). At highest flux levels, which cannot be explained by the same lognormal flux-distribution, "flares" can show quasi-periodic substructure, are strongly polarized, and are often associated with X-ray flares. Bottom left: near-infrared spectral index $\alpha$ $(\nu L_\nu \sim \nu^\beta)$ of Sgr A* (crosses: Eisenhauer et al. 2005, plus: Ghez et al. 2005a, triangles: Gillessen et al. 2006, asterix: Krabbe et al. 2006, diamonds: Hornstein et al. 2007). The near-infrared spectrum is close to flat with $\nu L_\nu \sim \nu^{-1 \ldots 1}$ at high flux levels above about 5 mJy, the measurements at low fluxes are inconclusive, photometric observations favoring a constant spectral index irrespective of flux (Hornstein et al. 2007), and spectroscopic observations indicating a red spectrum with $\nu L_\nu \sim \nu^{-3 \ldots -1}$ (Eisenhauer et al. 2005, Gillessen et al. 2006). Bottom right: orbit-averaged near-infrared image from a simulation of a hot-spot orbiting the Galactic Center black hole on the last-stable orbit (from Broderick & Loeb 2006).

In contrast to the time period prior to 2003, near-infrared flares from Sgr A* are now routinely detected, perhaps owing to a combination of the much improved detection capabilities with AO-cameras on 8-10 m-class telescopes and the fact that S2 has moved out of the line of sight toward Sgr A* (Genzel et al. 2003b, Ghez et al. 2004, 2005a, Clenet et al. 2004, Eisenhauer et al. 2005, Eckart et al. 2006a,b, Gillessen et al. 2006, Krabbe et al. 2006, Yusef-Zadeh et al. 2006b, 2009, Meyer et al. 2006, 2007, 2008, Hornstein et al. 2007, Trippe et al. 2007, Hamaus et al. 2009, Do et al. 2009b, Nishiyama et al. 2009, Dodds-Eden et al. 2009, 2010b). The brightest observed near-infrared flares had a peak luminosity of $\nu L_\nu \sim 10^{35.6}$ erg/s (Dodds-Eden et al. 2010b). Figure 7.3.1 shows the measured near infrared flux-distribution at the position of Sgr A* from $\approx$ 12,000 images in 117 separate observation nights collected between 2004 and 2009 (Dodds-Eden et al. 2010b). The flux distribution of Sgr A* is well characterized by a two-component model, consisting of the 'variable state' log-normal power-law distribution dominating to about 5 mJy in the K-band, in combination with a significant "flare" tail at higher flux densities (Dodds-Eden et al. 2010b).

The transition from the log-normal distribution at low-flux levels to the tail of high fluxes may also explain the apparent mismatch between the detection vs. non-detection of *quasi-periodic substructures* in different near-infrared light curve studies. Several investigations have reported such quasi-periodic substructure during flares, with periods of 17 to 40 minutes (Genzel et al. 2003b, Eckart et al. 2006b, Meyer et al. 2006, Trippe et al. 2007, Hamaus et al. 2009, Figure 7.3.1), while other observations of Sgr A* are found to be fully consistent with a red-noise source without any periodic component (Meyer et al. 2008, Do et al. 2009b, top right panel in Figure 7.3.1). One reason for this apparent discrepancy is the question whether the observer is allowed to select certain parts of the light-curve to justify the significance of the quasi-periodic substructure of 'flares', which is otherwise washed out in the noise-like 'variable state' of longer time series. Such selection criteria are for example the simultaneous observations of a strong change in polarization (Zamaninasab et al.

2010), or an exceptional high peak flux (Trippe et al. 2007). The two-component flux-distribution described above may offer a natural explanation for this controversy: at low flux levels below about 5 mJy the 'variable' infrared emission from Sgr A* is characterized by a log-normally distributed red-power law noise (Meyer et al. 2008, Do et al. 2009b, Dodds-Eden et al. 2010b). With increasing flux an additional 'flare' component starts to dominate the emission. This excess cannot be explained by the same log-normal flux-distribution, and perhaps exhibits quasi-periodic substructure with varying strength. The high-flux flare state is associated with bright X-ray emission. The 'event' nature of strong flares is further supported by their strong and variable near-infrared linear polarization (Eckart et al. 2006b, 2008b, Meyer et al. 2006, Trippe et al. 2007, Nishiyama et al. 2009). The degree of near-infrared polarization is up to about 40%. The polarization changes by typically 20% on timescales comparable to the flux variations of few to several 10 minutes. The position angle of the polarization changes on similar time scales by about 70%.

Several groups have measured the spectral index (Figure 7.3.1 bottom left) of Sgr A*, either directly from K-band (2.2 μm) spectra obtained with imaging spectrographs (Eisenhauer et al. 2005, Krabbe et al. 2006, Gillessen et al. 2006), or with quasi-simultaneous multi-band photometry at 1.65 to 5 μm (Ghez et al. 2005a, Hornstein et al. 2007). The near-infrared SED is consistently found to be close to flat ($\nu L_\nu \sim \nu^{0\pm1}$) at high flux levels above about 5 mJy. The results at low flux levels below about 3 mJy are inconclusive because of the difficult and uncertain background subtraction. Several spectroscopic observations have found a correlation between the infrared color and flux of Sgr A*, with a significantly redder spectrum ($\nu L_\nu \sim \nu^{-3...-1}$) at low flux levels (Eisenhauer et al. 2005, Gillessen et al. 2006, Krabbe et al. 2006). The photometric observations, however, yield a constant spectral index irrespective of the flux of Sgr A* (Hornstein et al. 2007).

Following the discovery of X-ray and infrared flares of Sgr A* a number of research groups have conducted simultaneous multi-wavelength observations covering all accessible frequencies from the radio to γ-rays (Eckart et al. 2004, 2006a, 2008a, Yusef-Zadeh et al. 2006b, 2008a, 2009, Kunneriath et al. 2008, Marrone et al. 2008, Aharonian et al. 2008, Dodds-Eden et al. 2009). These multi-wavelength observations reveal the salient characteristics and the broad-band SED of Sgr A*'s variable emission. Every X-ray flare is accompanied by a near-infrared flare, but not every near-infrared flare is seen at X-rays. X-ray and near-infrared flares occur simultaneously, with no significant delay. Strong X-ray/near-infrared flares sometimes appear to be accompanied by a subsequent flare at (sub)mm wavelengths, with a typical delay of 100 min to 3 hours. No mid-infrared (8 to 12 μm) emission Sgr A* has been detected during an X-ray/near-infrared flare (see Marrone et al. 2008, Dodds-Eden et al. 2009, Yusef-Zadeh et al. 2009), even during bright near-infrared flares. The mid- to near-infrared spectral index of the brightest flares thus is flat or positive ($\nu L_\nu \sim \nu^\alpha$, $\alpha > 0$). Figure 7.1.1 (right) shows the spectral energy distribution of the currently best studied multi-wavelength flare (from Dodds-Eden et al. 2009). While the total energy in the radio-emission at centimeter wavelengths is largely unaffected during a typical flare, the IR and X-ray fluxes increase by factors few to few tens, respectively. There is still a turnover at submm/far-infrared wavelengths, and probably even a minimum between the submm and near-infrared bands. These suggest that the energy distribution of electrons in flares is distinct from that in the 'quiescent' state, and not just a modest increase in the high-energy tail of the quiescent distribution.

In contrast to the quiescent emission of Sgr A* (§ 7.1), the physics of the flare emission of Sgr A* is much less certain. However, there are a few properties we can directly conclude from the observations. First, infrared flares are due to optically thin synchrotron emission, as indicated by the high polarization of up to 40% (Eckart et al. 2006b, 2008b, Meyer et al. 2006, Trippe et al. 2007, Nishiyama et al. 2009). This infrared flare emission must originate from a population of highly relativistic electrons ($\gamma \sim 10^{2...3}$), if the magnetic field strength is within reasonable limits set by the steady state (§ 7.1). Since the infrared spectral energy distribution is flat for bright flares (Ghez et al. 2005a, Gillessen et al. 2006, Hornstein et al 2007), this highly relativistic electron population must be drawn from a power law distribution $n(\gamma) \sim \gamma^{-1...-5}$, or alternatively from a thermal distribution with a temperature of $10^{12...13}$ K. The situation is more complex for X-ray flares. The favored two main models are synchrotron emission from yet higher energetic electrons (Yuan et al. 2003, 2004, Dodds-Eden et al. 2009), or Inverse Compton (IC) scattering from the electrons, which are also responsible for the submm/infrared emission. In the latter model the electrons either transfer their energy via Synchrotron Self Compton (SSC) scattering to the synchrotron photons emitted from the same electron population (Markoff et al. 2001, Eckart et al. 2004, 2006a, Sabha et al. 2010), or scatter up the photons from the radio-submm steady-state emission (Yusef-Zadeh et al. 2006b). Considering the fairly large uncertainties in near- to mid-infrared and X-ray spectral indexes, all these models are able to broadly match the observed characteristics of the flare spectral energy distribution (right panel of Figure 7.1.1). One clear conclusion from these observations and associated modeling is that X-ray/near-infrared flares probably arise as a result of the ***acceleration of a small fraction of the particle distribution in the innermost accretion zone***, perhaps as the result of ***magnetic reconnection*** and ***conversion of magnetic energy into the energy density of relativistic particles*** (Sharma et al. 2007, Dodds-Eden et al. 2010a, Yuan 2010).

While the observed spectral energy distribution gives insights into the emission processes and physical conditions in the flare, it provides only few constraints on the source geometry. The three most plausible scenarios are a ***jet with blob(s) of ejected material, hot spot(s) orbiting the black hole, and statistical fluctuations*** in the accretion flow. The latter scenario is supported from the natural explanation of the red noise variability at low and intermediate flux levels (Do et al. 2009b), in which flares arise from the superposition of statistical fluctuations throughout the accretion disk. The substructure of the lightcurves then results from the sum of independent, spatially and temporarily disjunct emitting regions, in plausible agreement with the log-normal flux distribution discussed above. In contrast, the simultaneous change of the linear polarization, the short-timescale substructure of the light curves, the delayed submm emission, and the spectral energy distribution point towards a compact flare region with a size comparable to the Schwarzschild radius. This interpretation of a compact flare region holds both for the jet model (Markoff et al. 2001) and for orbiting hot spots in the accretion disk (Genzel et al. 2003b, Broderick & Loeb 2006, Eckart et al. 2006b, Gillessen et al. 2006, Meyer et al. 2006, Trippe et al. 2007). As such these are not necessarily two distinct scenarios. As indicated by the two-component infrared brightness distribution (see above), there may also be a ***smooth transition from the superposition of statistical fluctuations in the accretion disk at moderate flux levels*** *to a single, dominant hot blob or spot for the brightest flares*. The delayed submm emission has been interpreted as the consequence of *adiabatic cooling in an expanding emission region* (Yusef-Zadeh et al. 2006a, 2008a, 2009, Eckart et al. 2008, Marrone et al. 2008). An alternative scenario is that the delayed submm 'flares' may not be flares but rather a 'dip recovery' corresponding to the re-establishment of the steady state, which follows a violent disruption of the inner accretion flow in the preceding infrared and X-ray flaring event (Dodds-Eden et al. 2010a).

## 7.4 Accretion onto the black hole

Sgr A* is surprisingly faint despite the rich gas reservoir (§ 3) in its immediate surroundings. This faintness can in part be understood from the drastic decrease of the effective mass accretion towards the black hole, which is summarized in Table 7.1.

**Table 7.1** Overview of the gas mass and its accretion rate as a function of distance from the Galactic Center black hole.

|  | radius [pc, $R_S$] | gas mass [$M_\odot$] | Mass accretion rate [$M_\odot$/yr] |
|---|---|---|---|
| GMCs | 10 – few 100 pc | $10^{5.5}$ to $10^{6.5}$ | $10^{-2}$ |
| CND | 1.7 – 7 pc | a few $10^4$ to $10^5$ | $10^{-3} - 10^{-4}$ |
| central cavity, minispiral & stellar cluster | < 1.7 pc | a few $10^2$ | $10^{-3} \ldots 10^{-4}$ |
| stellar winds at Bondi radius | 0.05 pc = $10^5 R_S$ |  | a few $10^{-6}$ |
| outer accretion zone | $10^2 - 10^3 R_S$ |  | $< 10^{-6}$ |
| inner accretion zone | a few – $10^2 R_S$ |  | a few $10^{-9} - 10^{-7}$ |

Radio observations have revealed in detail the distribution and dynamics of the cold and ionized gas flowing from the Galactic Bulge at a few hundred parsecs down to the mini-spiral at distances less than one parsec (Morris & Serabyn 1996, Mezger et al. 1996, and references within). The interstellar medium of the nuclear bulge (~ 300 pc) is concentrated in a narrow layer of predominantly molecular gas with a height of about 30 − 50 pc (the 'central molecular zone'). The total gas mass is about $5 − 10 \times 10^7 M_\odot$. About half of the gas resides in compact Giant Molecular Clouds at radii between ten and 200 parsec with typical masses of $10^{5.5}$ to $10^{6.5} M_\odot$. Dynamical models of the cloud kinematics indicate an average in-fall rate of order $10^{-2} M_\odot$/yr at these radii (Von Linden et al. 1993). Some of this in-falling gas settles on the circum-nuclear disk at 1.7 − 5 pc radius (Jackson et al. 1993). The current mass accretion rate within the CND ranges between $10^{-3}$ and $10^{-4} M_\odot$/yr (Vollmer & Duschl 2002). The gas masses and motions in the ionized mini-spiral filaments at 0.1 − 2 pc imply similar accretion rates (Lacy et al. 1991, Zhao et al. 2009). The O/WR-stars in the central parsec eject stellar winds with a total mass loss rate of about $10^{-3} M_\odot$/yr (Martins et al. 2007), which currently dominate the Bondi-like accretion onto the black hole.

With ever improved observations of the stellar population (§ 2) the models for the gas accretion onto the Bondi radius ($R_{Bondi}$ ~ 1" ~ $10^5 R_S$) have evolved from a simple bow shock model of gas from the most nearby IRS 16 complex (Melia 1992), to early hydrodynamic models from pseudo-randomly distributed point-sources (Coker & Melia 1997), to spherically symmetric hydrodynamic models (Quataert 2004), and to 3D numerical simulations of stellar wind dynamics using the observed orbits and wind properties of individual stars (Cuadra et al. 2008a). The accretion rate is dominated by a few close, slow-wind ($v_w \leq 750$ km/s) stars, and is of the order few $\times$ $10^{-6}$ M$_\odot$/yr. This accretion at the Bondi-radius is traced by the extended X-ray emission (Baganoff et al. 2003; Xu et al. 2006). This emission can be explained by thermal Bremsstrahlung (Quataert 2002) from a hot plasma with $T \sim 4 \times 10^7$ K. The remaining gas is thermally driven out of the central star cluster in a wind.

At $\sim 10^2 - 10^3$ Schwarschild radii, the detection of linear polarization of Sgr A* (Aitken et al. 2000, Marrone et al. 2006, Macquart et al. 2006) directly indicates that the mass accretion rate must be very low ($< 10^{-6}$ M$_\odot$/yr), because otherwise extreme Faraday rotation gradients would depolarize the emission (Quataert & Gruzinov 2000; Agol 2000). The accretion rate at scales of a few to $10^2$ Schwarzschild radii is constrained from the observed Faraday rotation at submm-wavelengths (Marrone et al. 2006, 2007). Depending on the actual radius at which the electrons become relativistic and depending on the detailed magnetic field configuration, the accretion rate must be within $2 \times 10^{-9}$ to $2 \times 10^{-7}$ M$_\odot$/yr. Only about $10^{-6}$ of the gas entering the central molecular zone actually accretes onto the black hole. Theoretically this remarkably *low accretion (and radiation) efficiency* is now believed to originate from a combination of *inefficient angular momentum transport at all radii*, *relatively poor coupling between electrons and protons at the low gas densities* characteristic for SgrA*, and *strong outflows* and *convection in the innermost accretion zone* (Blandford & Begelman 1999, Hawley & Balbus 2002, Quataert 2003, Sharma et al. 2007b, Mościbrodzka et al. 2009, Yuan 2010).

This extremely low accretion efficiency is what also typifies other low-luminosity active galactic nuclei. As such the Galactic Center may be rather typical for 'normal' galactic nuclei in the local Universe. Similar to the Galactic Center black hole, these low luminosity active galactic nuclei are by no means fuel-starved. Instead the broad-line region and obscuring torus disappears in some of the faintest sources, and the optically thick accretion disk is thought to transform into a three-component structure consisting of an inner radiatively inefficient accretion flow, a truncated outer thin disk, and a jet or outflow (Ho 2008). In the future, pericenter passages of stars close to Sgr A* might be used to probe the inner accretion flow. The stellar radiation field provides seed photons that are Compton up-scattered by the hot electrons of the accretion flow, possibly leading to an observable increase of the X-ray flux (Nayakshin 2005b).

Even though the Galactic Center is currently one of the least luminous supermassive black holes, this may also be just a transient phase, and the gas accretion and luminosity of Sgr A* may have been much higher in the recent past. There is indication that Sgr A*'s past X-ray emission is echoed in the Compton scattered and reprocessed radiation emitted by the Sgr B2 giant molecular cloud. While there are other plausible scenarios for this transient irradiation from Sgr B2 (Fryer et al. 2006, Capelli et al. in prep.), the "echo" scenario suggests that Sgr A*'s luminosity must have been a few $10^5$ times higher a few hundred years ago (Sunyaev et al. 1993, Koyama et al. 1996, Revnivtsev et al. 2004, Muno et al. 2007, Ponti et al. 2010). Numerical simulations of the stellar wind accretion (Cuadra et al. 2008a) indeed show that Sgr A*'s X-ray luminosity could have varied by several orders of magnitude in the recent past, although this strong luminosity variation requires a highly non-linear dependence on mass accretion rate, as for example indicated by the 'fundamental plane of black hole activity' (Merloni et al. 2003). Still, the simulations fail to explain the putative extreme X-ray outburst a few hundred years ago, and the observational evidence is not yet conclusive on the question whether Sgr A* was indeed much more luminous in the resent past.

# 8. Concluding Remarks and Outlook

The progress in Galactic Center research over the last two decades and continuing until the time of writing this article has been astounding, with many unexpected discoveries and fundamental results. A number of the phenomena studied in the central parsec of our Milky Way are of broad relevance to other galactic nuclei. This success is largely based on the rapid advances in high-resolution observations across a broad range of wavelengths and, in the last decade, to ever more powerful computer simulations of dense star clusters, star formation, and accretion near/onto black holes.

Based on precise measurements of stellar orbits and of the central radio source SgrA*, the empirical evidence for the existence of a central massive black hole of about 4 million solar masses is compelling. The dense star cluster near the black hole has surprising properties, most of which were not anticipated. Massive O/WR-stars have been forming there recently, deep in the sphere of influence of the central black hole, at a high rate and probably with high efficiency. Similar 'starburst events' near the hole may have occured from time to time throughout the entire ~ 10 Gyr evolution of the Galactic Center. A very compact cluster of B-stars is centered on the massive black hole, with randomly oriented orbits on solar system scales. The lower mass, old stars that are traced by the current observations do not exhibit a concentration toward the central black hole, in contrast to basic theoretical predictions. Spatial distribution and dynamics of the O/WR- and B-stars are complex, indicating a number of processes at work that operate much faster than the classical two-body relaxation time scale. The massive black hole plays a central role in driving these processes.

Gas is streaming into the central parsecs at substantial rates, but the accretion into the event horizon at the present time is orders of magnitude lower than simple theoretical estimates had predicted. Theoretical work now suggests that this puzzling faintness of SgrA* (and of other low-luminosity AGNs) is due to a combination of a relatively low accretion rate at the Bondi radius, inefficient angular momentum transport, outflows and low radiation efficiency. There is tantalizing evidence that Sgr A* was much brighter in the recent past. Accretion onto the Galactic Center thus appears to be variable and chaotic, controlled by local processes near the black hole. Is this true for most other galaxies with a central black hole?

Where might the journey go in the next decade? Assuming that stellar orbit studies continue over this time scale, steady progress in the number and quality of the orbital determinations of the central star cluster should provide stringent tests of the different scenarios of the formation and evolution of the star disk(s) and the central S-star cluster that we have discussed extensively in this review. Detection and characterization of additional binary stars would be very helpful for the same purpose. Extending deep searches for young, massive stars to the region outside the central parsec would explore evidence for recent star formation outside the central star disk(s), as well as for the presence of a 'sea' of moderately massive stars that is proposed in several of the theoretical scenarios. Even more fundamentally, if the two peri-passages of the stars S31 (S08) in 2013 and S2 (S02) in 2018 are covered well by radial velocity measurements, there is a very good chance of a significant detection of the second-order ($O(v/c)^2$), post-Newtonian (transverse-Doppler effect and gravitational redshift) terms of Special and General Relativity in these orbits (Fragile

& Matthews 2000, Rubilar & Eckart 2001, Zucker et al. 2006, Weinberg et al. 2005, Angelil & Saha). Spectroscopic detection and tracking of faint stars even closer to Sgr A* would be extremely exciting and would open new prospects for future, still higher resolution measurements. As a by-product of these studies the accuracy of the distance determination to Sgr A* can be expected to reach ~ 1 to 2%. Continuing studies of the multi-wavelength properties, SED, polarization state and temporal behavior of Sgr A*'s emission are needed for a robust discrimination between the different emission scenarios discussed in this review, and for a deeper exploration of the nature of the flares. High quality light curves of a sufficient number of flares would settle the question whether or not quasi-periodic temporal substructures are present in these flares. As for many other questions in the Galactic Center, such multi-wavelength observations of the emission from Sgr A*, while time consuming and not always successful, will be enormously helpful for a deeper understanding of the important and wide-spread phenomenon of radiatively inefficient accretion flows in other normal galaxy nuclei.

If the near-infrared interferometry experiments now under development at the VLT(I) (PRIMA: Deplancke 2008, GRAVITY: Eisenhauer et al. 2008, Gillessen et al. 2010) and Keck (ASTRA: Pott et al. 2008) achieve the planned 10 to 100 μas astrometric accuracy in combination with mas imaging, the dynamics of gas and stars within a few hundred times the event horizon of the central black hole will become accessible for study. Likewise the next generation, extremely large optical/infrared telescopes (such as the European Extremely Large Telescope (E-ELT), the Giant Magellan Telescope (GMT), and the Thirty Meter Telescope (TMT)) will also be extremely valuable by combining superb sensitivity with 50 to 100 μas astrometric accuracy (Weinberg et al 2005, Davies et al. 2010b). VLBI imaging at ≤ 1 mm, with sub-mas resolution, will be another powerful tool for exploring the gas distribution in the central accretion zone (Fish & Doeleman 2010, Broderick et al. 2009b, Dexter et al. 2009, 2010, Huang et al. 2009). These new observational tools will probably give us the answer whether the Sgr A* flares originate from a jet, from hot spots, or from global fluctuations in the central accretion zone. The VLBI experiments may detect the effects of strong light-bending (the 'shadow') by the black hole (Falcke, Melia & Agol 2000). The near-infrared interferometry may be able to trace the motions of hot spots or a jet within 5 to10 $R_S$ (Eisenhauer et al. 2008, Hamaus et al. 2009). If there are stars sufficiently close to the central black hole, the Schwarzschild precession term, and perhaps even the Lense-Thirring precession term due to the spin of the hole might be detectable (Kraniotis 2007, Gillessen et al. 2008, Will 2008, Merritt et al. 2010). If we are lucky the Galactic Center black hole may then become a test bed for probing General Relativity in the strong field limit (Falcke et al. 2000, Fish & Doeleman 2009, Hamaus et al. 2009). Success is not guaranteed but the goals are extremely rewarding and make the efforts highly worthwhile.

**Acknowledgements:** *We would like to thank Tal Alexander, Fred Baganoff, Hendrik Bartko, Ian Bonnell, Avery Broderick, Warren Brown, Katia Cunha, Mel Davies, Katie Dodds-Eden, Shep Doeleman, Chris Fragile, Tobias Fritz, Alister Graham, Alessia Gualandris, Paul Ho, G.V. Kraniotis, Avi Loeb, Dieter Lutz, Fabrice Martins, Maria Montero-Castano, Sergei Nayakshin, Thomas Ott, Hagai Perets, Oliver Pfuhl, Pawel Pietrukowicz, Delphine Porquet, Mark Reid, Bob Sanders, Rainer Schödel, Amiel Sternberg, Linda Tacconi, Scott Tremaine, Farhad Yusef-Zadeh and Qingjuan Yu for material, valuable comments and suggestions.*

A&A:   Astronomy & Astrophysics
AJ:    Astronomical Journal
AN:    Astronomische Nachrichten
ApJ:   Astrophysical Journal
ApJL:  Astrophysical Journal Letters
ApJS:  Astrophysical Journal Supplement Series
ARAA:  Annual Review of Astronomy and Astrophysics
NA:    New Astronomy